\documentclass[aps,prd,preprint,showpacs,longbibliography, nofootinbib]{revtex4-1}

\usepackage{amsthm}
\usepackage{amsmath}
\usepackage{mathtools}
\usepackage{tabularx}
\usepackage{accents}
\usepackage{bbold}
\usepackage{dsfont}
\usepackage{bm}

\newcommand{\Ref}[1]{(\ref{#1})}

\newcommand{\C}{{\mathbb C}}

\newcommand{\vv}{\varphi}
\newcommand{\K}{\mathcal{K}}
\newcommand{\V}{\mathcal{V}}
\newcommand{\F}{\mathcal{F}}

\newcommand{\SU}{\mathrm{SU}}

\newcommand{\U}{\mathrm{U}}

\newcommand{\be}{\begin{equation}}
\newcommand{\ee}{\end{equation}}
\newcommand{\bea}{\begin{eqnarray}}
\newcommand{\eea}{\end{eqnarray}}

\newcommand{\bit}{\begin{itemize}}
	\newcommand{\eit}{\end{itemize}}

\newcommand{\tr}{{\rm Tr}}

\def\p{\partial}

\newcommand{\ex}{\mathrm{e}}
\newcommand{\dd}{\mathrm{d}}

\newcommand{\bra}[1]{\left\langle {#1}\right|}
\newcommand{\ket}[1]{\left|{#1}\right\rangle}
\newcommand{\braket}[2]{\langle {#1}|{#2}\rangle}

  \newcommand{\g}{\gamma}

\begin{document} 
	\title{Group Field theory and Tensor Networks:\\ towards a Ryu-Takayanagi formula in full quantum gravity}
	\author{Goffredo Chirco}
	\email{goffredo.chirco@aei.mpg.de}
	\affiliation{Max-Planck-Institut f\"ur Gravitationsphysik (Albert-Einstein-Institut),\\Am M\"uhlenberg 1, 14476 Golm, Germany}
	\author{Daniele Oriti}
	\email{daniele.oriti@aei.mpg.de}
	\affiliation{Max-Planck-Institut f\"ur Gravitationsphysik (Albert-Einstein-Institut),\\Am M\"uhlenberg 1, 14476 Golm, Germany}
	\author{Mingyi Zhang}
	\email{mingyi.zhang@aei.mpg.de}
	\affiliation{Max-Planck-Institut f\"ur Gravitationsphysik (Albert-Einstein-Institut),\\Am M\"uhlenberg 1, 14476 Golm, Germany}
	\date{\today}
	
	\begin{abstract}
		We establish a dictionary between group field theory (thus, spin networks and random tensors) states and generalized random tensor networks. Then, we use this dictionary to compute the R\'{e}nyi entropy of such states and recover the Ryu-Takayanagi formula, in two different cases corresponding to two different truncations/approximations, suggested by the established correspondence.
	\end{abstract}
	\maketitle
	\begin{widetext}
		\tableofcontents
	\end{widetext}
	
	\newpage

\section{Introduction}
Background independent approaches to quantum gravity suggest a picture of the microstructure of the universe in which continuum spacetime and geometry disappear and are replaced by discrete and non-spatiotemporal entities. Among them, Loop Quantum Gravity (LQG) \cite{Ashtekar:2004eh,Rovelli:QG,Thiemann:QG,Perez:2012wv,Rovelli:2014ssa}, the modern incarnation of the canonical quantization programme for the gravitational field, together with its covariant counterpart (spin foam models), and Group Field Theory (GFT) \cite{Oriti:2014uga,Baratin:2011aa,Oriti:2011jm,Oriti:2009wn}, a closely related formalism sharing the same type of fundamental degrees of freedom, identify this microstructure with (superpositions of) spin networks, which are graphs labeled by group-theoretic data. More precisely, in GFT models of quantum gravity spin network states arise as many-body states in a 2nd quantised context, whose kinematics and dynamics are governed by a quantum field theory over a group manifold with quanta corresponding to \emph{tensor maps} associated to nodes of the spin network graphs. Random combinatorial structures, corresponding both to the elementary building blocks of quantum spacetime and to their interaction processes, become central. The same is true in the related context of random tensor models \cite{Gurau:2011xp,Gurau:2016cjo,Rivasseau:2016zco}, which, for our present purposes can be seen as a simplified version of GFTs, stripped down of the group-theoretic data, leaving only the combinatorial aspects. Indeed, the random tensors can be understood as the GFT fields considered for the special case of a finite group. For a more detailed account of these three quantum gravity formalisms, and for the many results obtained, we refer to the cited literature. In the following, we will provide more precise definitions of their main ingredients.

Tensor networks, in recent years, have attracted a lot of attention as powerful quantum information tools in the context of condensed matter and, more generally, quantum many-body systems (including quantum field theory). For recent reviews, see \cite{Orus:2013kga,Bridgeman:2016dhh}. Also in this case, we will give precise definitions in the following. Here it suffices to say that tensor networks encode the entanglement properties of many-body systems in their combinatorial structure, in which tensors are connected along a network pattern and identify (the coefficients, in a given basis, of the wave function corresponding to) quantum states of the given system. Born as convenient mathematical tools for numerical evaluations of many-body wavefunctions, which become translatable into graphical manipulations, tensor network techniques have found an amazing number of applications: from the classification of exotic phases of quantum matter (e.g. topological order) \cite{Wen:2004ym,Wen:2016ddy} to new formulation of the non-perturbative renormalization of interacting quantum field theories \cite{Cirac:2009zz,verstraete2008matrix,augusiak2012modern}, down to realizations of the AdS/CFT correspondence \cite{Ryu:2006bv,Swingle:2009bg,Pastawski:2015qua,Hayden:2016cfa}.   
  
Despite their disparate origin, it should be clear already from our sketchy description that the type of mathematical structures identified by quantum gravity approaches and used in the theory of tensor networks are very similar. And consequently, it is very natural to try to put the two frameworks in more direct contact. This is the main goal of the present article. Indeed, the structural similarity had been noted before \cite{vidal2008class,Singh:2009cd,evenbly2011tensor,Han2016}, and also exploited in the context of renormalization of spin foam models treated as lattice gauge theories \cite{Dittrich:2014mxa,Delcamp:2016dqo,Dittrich:2011zh,Dittrich:2016tys}. The last set of works, in particular, has already shown how fruitful tensor network techniques can be for quantum gravity models.

Before we start presenting our results, we want to offer some motivations for our work, both from the quantum gravity perspective and from the tensor network side.

From the quantum gravity point of view, the general motivation is clear. Tensor networks provide a host of tools and results that could find useful application in quantum gravity; in particular they may become central tools in the renormalization analysis of GFT models \cite{Carrozza:2013wda,Benedetti:2014qsa,Carrozza:2016tih,Carrozza:2016vsq,Geloun:2016qyb,Lahoche:2015tqa}, in addition to their mentioned role in the renormalization analysis of spin foams models \cite{Bahr:2012qj,Bahr:2016hwc,Bahr:2014qza}. And such renormalization analyses are, in turn, the main avenue for solving the crucial problem of the continuum limit in such formalism. 

More specifically, tensor networks are very effective in taking into account and controlling the entanglement properties of quantum states in many-body systems. This is exactly the language in which GFT deals with quantum gravity states; moreover, in GFT, the very connectivity of spin network states, encoded in the links of the underlying graphs, is associated with entanglement between the fundamental quanta constituting them (associated to nodes) \cite{withFabio}. One example of this type of application, as we show in this paper, is the computation of entanglement entropy in spin network states and relate LQG with holography, which was also the subject of a number of other works in the LQG/GFT literature \cite{Livine:2005mw,Livine:2006xk,Donnelly:2008vx,DiazPolo:2011np,Perez:2014ura,Bianchi:2012ev, Chirco:2014naa,Ghosh:2014rra,Bonzom:2015ans,Hamma:2015xla,Bianchi:2015fra,Han:2014xna,Han:2015gma,Oriti:2015rwa,Han2016}.

Further, the identification of the true (interacting) vacuum state of a quantum gravity theory, in absence of any space-time background or preferred notions of energy, is a difficult matter even at the purely conceptual level, leaving aside the formidable technical challenges. One possible criterion, suited to this context, is to look for states which maximize entanglement, by some measure (e.g. entanglement entropy). In this respect, to reformulate the kinematics and dynamics of GFT and LQG states in terms of tensor networks, and to do the same for their renormalization, seems a promising strategy.

Finally, recent results in the application of tensor networks to AdS/CFT \cite{Swingle:2009bg,Pastawski:2015qua,Hayden:2016cfa} suggest that this application would be fruitful even within the conventional perspective of canonical quantum gravity (including LQG). From this perspective, in fact, the task of quantum gravity is the construction of the space of quantum states of the gravitational field which satisfy the (quantum counterpart of the) Hamiltonian constraint encoding the dynamics of quantised GR. A number of results in AdS/CFT suggest that a static AdS space-time, which we expect to be one such state, at the quantum level, satisfies the Ryu-Takayanaki (RT) formula \cite{Ryu:2006bv} for the entanglement entropy, which is very efficiently computed (as we also show in this paper) via random tensor network techniques \cite{Hayden:2016cfa}. One is led to conjecture that this may be a general properties of physically interesting quantum states of the gravitational field, and so far no counterexample to this conjecture has been found. This prompts the search, by the same techniques, for similar states in canonical quantum gravity. 

From the perspective of the theory of tensor networks, one general good point of dwelling into the correspondence with quantum gravity states should also be obvious. This identifies a new domain of applications, of truly fundamental nature, for techniques and ideas which have already proven powerful in others. Indeed, we expect that a number of key results obtained via tensor network techniques, most notably holographic mappings and indications of new topological phases in many-body systems, can be reproduced in this new context, with deep implications. In perspective, it is here that one will be able to test the suggestion that quantum information has a truly foundational role to play in our understanding of physical reality.

More practically, a number of techniques have been developed, and many results obtained, concerning the dynamics of GFT and spin-network states, also thanks to the many related developments in the theory of random tensors, and our dictionary proves that the GFT formalism provides a natural definition of the dynamics of random tensor networks. Specifically, it means that the many results in GFT can help dealing with general (non-Gaussian) probability distributions over random tensor networks, as well as offering new takes on more standard problems, like entropy calculations, in tensor network theory. In fact, we offer some examples of these applications in the following.

In this paper, we do not target the more ambitious objective of a calculation of the RT formula for the entanglement entropy in the full quantum gravity formalism of group field theory. Having established the general dictionary between group field theory states and (generalized) random tensor networks, we content ourselves with reproducing the RT formula, along the lines of the derivation given in \cite{Hayden:2016cfa} in two new cases: for group field theory states corresponding to generalized tensor networks, but only using a group field theory dynamics in the simplest approximation and dealing only with averages over the tensor functions associated to the network nodes, rather than treating the full tensor network as a group field theory observable; for the simple truncation of group field theory states corresponding to spin networks with fixed spin labels. We leave a more complete and comprehensive analysis for forthcoming work.\\

The paper is organized as follows. In the next section, we summarize the basic elements of spin network states and of their embedding in the GFT formalism, as well as the definition of tensor networks. Having done so, we define the precise correspondence between GFT states and tensor networks, showing how the first generalizes and provides a Fock space setting for the second. In the following section, 
we derive the $N$th R\'{e}nyi entropy using GFT techniques, in the group representation and for a generalized tensor network, but without taking advantage of the full GFT formalism; next, we compute the same R\'{e}nyi entropy and derive the RT formula from a purely spin-network perspective, seen as a truncation of more general GFT states. This is meant to be a clear example of how the same problem can be fruitfully approached from both sides of the correspondence. Finally, in the last section, we discuss one key universality result from the theory of random tensors, which extends to GFTs, and which could have direct impact on the applications of random tensor networks. 
We end up with a summary of our results.

\section{Group Field theory and Tensor Networks }\label{II}
A d-dimensional GFT is a combinatorially non-local field theory living on (d copies of) a group manifold \cite{Oriti:2014uga,Baratin:2011aa,Oriti:2011jm,Oriti:2009wn}. Due to the defining combinatorial structure, the Feynman diagrams $\F$ of the theory are dual to cellular complexes, and the perturbative expansion of the quantum dynamics defines a sum over random lattices of (a prior) arbitrary topology. A similar lattice interpretation can be given to the quantum states of the theory. For GFT models where appropriate group theoretic data are used and specific properties are imposed on the states and quantum amplitudes, the same lattice structures can be understood in terms of simplicial geometries. The associated many-body description of such lattice states can be given in terms of a tensor network decomposition. The corresponding (generalized) tensor networks are thus provided with a field theoretic formulation and a quantum dynamics (and, in specific models, with additional symmetries). In this section, after a brief introduction to the GFT formalism, we detail this correspondence between GFT states and (generalized) tensor networks.

\subsection{Group Field Theory}

Let $G$ denote an arbitrary semi-simple Lie group; in the following, we assume for simplicity that $G$ is compact, but the framework can easily be generalized to the non-compact case. A \emph{group field} $\varphi$ is a complex function defined on a number of copies of the group manifold $G$:
\begin{eqnarray}
\varphi&:G^d&\rightarrow \mathbb{C} \\ \nonumber
& {g_i}& \mapsto \varphi({g}_i)
\end{eqnarray}
where we use the shorthand notation ${g}_i$ for the set of $d$ group elements $\{g_1,g_2,\cdots,g_d\}$. 

The GFT field can be also seen as an infinite-dimensional tensor, transforming under the action of some (unitary) group $U^{\times d}$, as: 
\begin{eqnarray}\nonumber
\varphi(g_1,..,g_d) \rightarrow \int [\dd g_i]\, U(g'_1, g_1)\cdots U(g'_d, g_d)\, \varphi(g_1,...,g_d), 
\end{eqnarray}
and 
\begin{eqnarray}\nonumber
&&\varphi^*(g_1,..,g_d) \rightarrow \int [\dd g_i]\, U^*(g_1, g'_1)\cdots U^*(g_d, g'_d)\, \varphi^*(g_1,...,g_d)\\ 
&& \quad \text{for} \quad \int \dd g_i\, U(g'_i,g_i)^* U(g_i, \tilde{g}_i) = \delta(g'_i, \tilde{g}_i). 
\end{eqnarray}

This requires the $d$ arguments of the GFT field to be {\it labeled and ordered}. We will see in the following how one can decompose the same field into finite-dimensional tensors; in this finite-dimensional case, the correspondence with tensor network formalism will be evident, and it will also be evident then in which sense GFTs provide a generalization of it.

The GFT dynamics is defined by an action, at the classical level, and a partition function at the quantum level. The combinatorial structure of the pairing of field arguments in the GFT interactions is part of the definition of a GFT model. An interesting class of models \cite{Oriti:2014uga,Baratin:2011aa,Oriti:2011jm,Oriti:2009wn,Gurau:2011xp,Gurau:2016cjo,Rivasseau:2016zco} is defined by the requirement that the interaction monomials are {\it tensor invariants}, i.e. that GFT fields are convoluted in such a way as to produce an invariant under the above mentioned (unitary) transformations\footnote{Such invariants are in one to one correspondence with {\it colored d-graphs} $\mathcal{B}$ constructed as follows: for each GFT field (resp. its complex conjugate) draw a white (resp. black) node with $d$ outgoing links each labeled by $d$ different colors, then connect all links in such a way that a white (resp. black) node is always connected to a black (resp. white) node and that only links with the same color can be connected. }. 

Another class of GFT models is instead based on the requirement that the Feynman diagrams of the theory are simplicial complexes, which in turn requires the interaction kernels to have the combinatorial structure of d-simplices. This class of models is also the one on which model building for 4d quantum gravity has focused on, producing models whose Feynman amplitudes have the form of simplicial gravity path integrals and spin foam models \cite{Oriti:2014uga,Baratin:2011aa,Oriti:2011jm,Oriti:2009wn}, and, more generally, lattice gauge theories. This involves an additional symmetry requirements on the GFT fields and interactions, which will play a crucial role in the following.

In this simplicial case, the GFT action has the general form
\begin{eqnarray}
S_d[\varphi]&=&\frac{1}{2} \int \dd g_i \dd g'_i \, \vv(g_i)\, \K(g_i g_i'^{-1})\, \vv(g_i')\,+\\ \nonumber
&+&\frac{\lambda}{d+1} \int \prod_{i\neq j =1}^{d+1} \dd g_{ij}\, \V(g_{ij} g_{ji}'^{-1})\, \vv(g_{1j})\cdots \vv(g_{d+1j}),
\end{eqnarray}
where $\dd g_i$ is an invariant measure on G and we use the notation $\vv(g_{1j} ) =\vv(g_{12},  \cdots, g_{1d+1})$. $\K$ is the kinetic kernel, $\V$ the interaction kernel, $\lambda$ a coupling constant for the $d+1$-degree homogeneous interaction. 
The two kernels satisfy the invariance properties
\begin{eqnarray}
&\,&\K(h\,g_i g_i'^{-1}\,h') = \K(g_i g_i'), \\ \nonumber
&\,& \V(h_i\,g_{ij} g_{ji}'^{-1},h_j^{-1}) = \V(g_{ij} g_{ji}'^{-1}) \qquad \forall h,h',h_i \in G.
\end{eqnarray}

\noindent
This implies that the action is invariant under the gauge transformations $\delta \vv(g_i) = \tilde{\vv}(g_i)$, where $\tilde{\vv}$ is any function satisfying
\begin{eqnarray}\label{gauge}
\int_G \dd h \, \tilde{\vv}(hg_1, \cdots, hg_d)=0.
\end{eqnarray}
This symmetry is gauge fixed if one restricts the field $\vv$ to satisfy 
\begin{eqnarray}\label{symme}
\vv(hg_i) = \vv(g_i).
\end{eqnarray} 
The action is also invariant under the \emph{global symmetry}
\begin{eqnarray}
\vv(g_1,\cdots ,g_d) \rightarrow \vv(g_1h,\cdots,g_dh). 
\end{eqnarray}

GFT's Feynman diagrams define cellular complexes $\mathcal{F}$ weighted by amplitudes assigned to the faces, edges and vertices of the dual two-skeleton otabularf a chosen triangulation of a d dimensional topological spacetime $\mathcal{M}_{\mathcal{F}}$. As mentioned, their Feynman diagram evaluations reproduce the associated amplitudes of a spin foam model, or, in different variables, of a simplicial gravity path integral \cite{Baratin:2010wi,Baratin:2011tx,Baratin:2011hp}, providing a generalisation of the lattice formulation of gravity  \`a la Regge, with an accompanying sum over lattices, generalising matrix models for 2d gravity to any dimension \cite{Oriti:2014uga,Baratin:2011aa,Oriti:2011jm,Oriti:2009wn,Gurau:2011xp,Gurau:2016cjo,Rivasseau:2016zco}.

\begin{figure}[t!]
\centering 
\includegraphics[width=.5\textwidth]{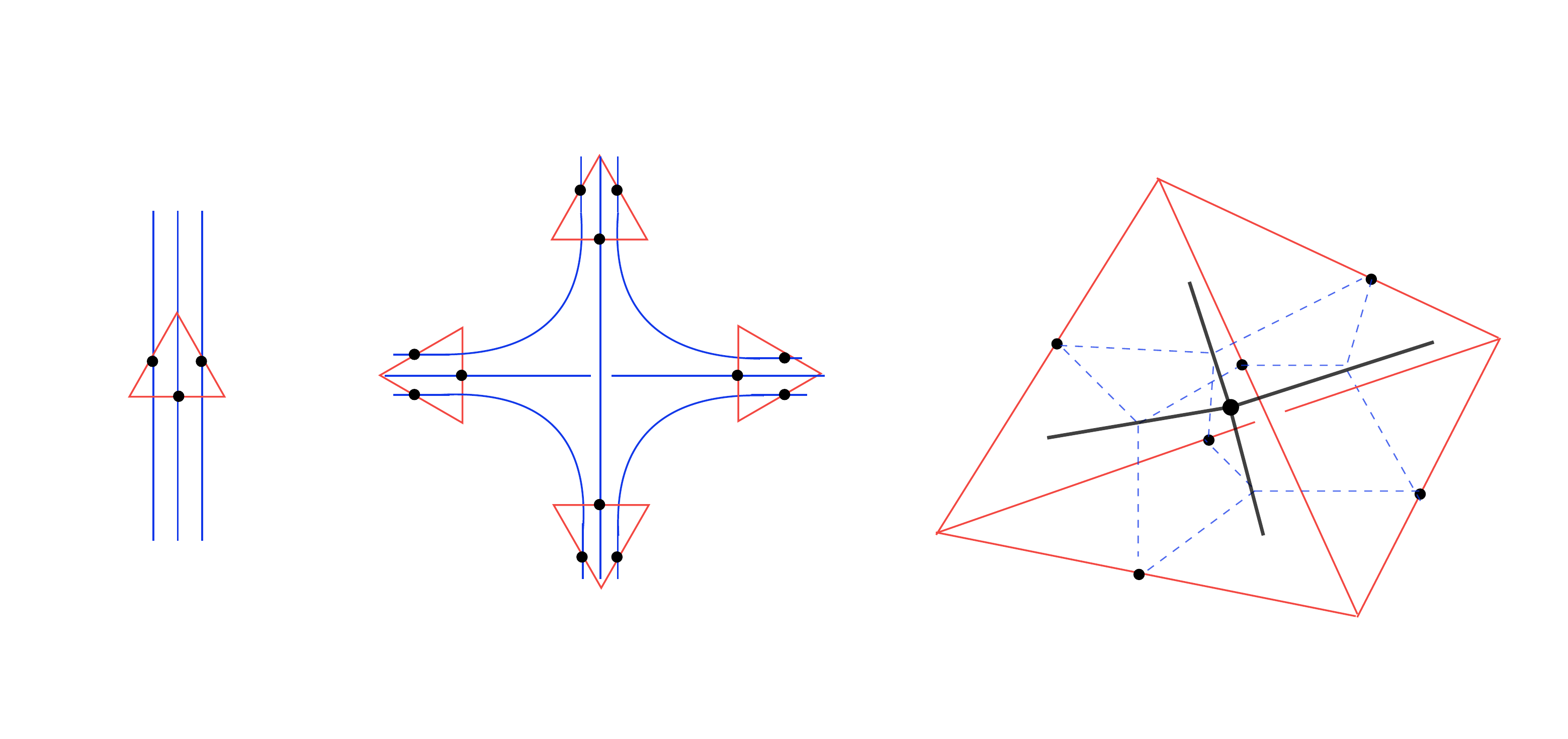}
\caption{Correspondence between Feynman diagram and triangulation: Each strand of the graph forms a closed loop which can be interpreted as the boundary of a 2d disk. These data are enough to reconstruct a topological 2d complex $\F$, the vertices and edges of this complex correspond to vertices and edges or the Feynman graph, the boundary of the faces of $\F$ correspond to the strands of the Feynman graph. }\label{fig1} 
\end{figure}

Let us give some more detail on the construction, to clarify the above points.
A specific theory, with a specific related Feynman cellular complex, is completely defined by the choice of the kernels. Lets consider the simplest case, consisting in the choice
\begin{eqnarray}
\K(g_i,g_i')&=&\int_G \dd h \prod_i \delta(g_ig_i'^{-1}\,h), \\ 
\V(g_{ij} g_{ji}'^{-1})&=&\int_G \prod_i \dd h_i \prod_{i < j} \delta(h_i\,g_{ij} g_{ji}'^{-1},h_j^{-1})\label{V}
\end{eqnarray}
where $\delta(\cdot)$ is the delta function on G and the integrals ensure the gauge invariance defined in (\ref{gauge}), and let us restrict to the case of dimension $d = 3$. To keep track of the combinatorics of field arguments in the kernels, it is useful to represent the Feynman diagram as a {\it stranded graph}. The field $\vv$ has three arguments, so each edge of a Feynman diagram comprises three strands running parallel to it. Four edges meet at each vertex and the form of the interaction $\V$ in (\ref{V}) forces the strands to recombine as in Figure \ref{fig1}. 

The three strands running along the edges can be understood to be dual to a triangle and the propagator $\K$ gives a prescription for the \emph{gluing} of two triangles. At the vertex, four triangles meet and their gluing via $\V$ form a tetrahedron. With this interpretation the Feynman diagram of a GFT is clearly dual to a triangulated 3d simplicial complex (which will be generically a singular pseudo-manifold) and this is true in any dimension \cite{DePietri:2000ke,DePietri:2000ii,DePietri:1999bx}.

The quantum states of the theory can be given a similar combinatorial characterization in terms of graphs and dual cellular complexes, as it should be already intuitive in the above example, in which GFT fields themselves are associated to triangles. We will not detail this aspect of the formalism.

\subsection{Fourier modes of the group field as tensor fields}\label{spin rep}

As a function on a group G, the field $\vv$ can be decomposed in terms of unitary irreducible representations $(\rho,V_{\rho})$ of G using the Peter-Weyl theorem, $L^2(G)\simeq \bigoplus_{\rho}V^{\rho} \otimes V^{*{\rho}}$, giving

\begin{eqnarray}
\vv(g)=\sum_{\rho} d_{\rho} \text{Tr}[\hat{\vv}_{ab}\, \rho^{ab}(g)]
\end{eqnarray}

Here, $d_\rho \in \mathbb{N}$ is the dimension of the representation $\rho : G\rightarrow \text{Aut} (V_{\rho})$, the indices $a,b = 1,\dots,d_\rho$ are matrix indices associated to the matrix $\rho(g)$ representing the group element $g$, and $\hat{\vv}_{\rho} \in V^{\rho} \otimes V^{*\rho}\simeq \text{End}(V_{\rho})$ is the matrix Fourier coefficient of the function $\vv$. In other words, each $\hat{\vv}_{\rho}$ is a rank $d_{\rho} = N$ matrix. 

Let us consider, as a specific example, the same decomposition for the case of $d=3$, with $G=SU(2)$. The unitary irreps of $SU(2)$, $\mathbb{V}^j$, are labeled by the spin $j\in \mathbb{N}/2$. Using the right invariance property of the field, one obtains the following decomposition 
\begin{eqnarray}\label{fourier}
&\,&\vv(g_1,g_2,g_3)= \sum_{\{j\}} \text{Tr}\left[ \vv^{\{j\}}_{m_1,m_2,m_3}\,\left(\prod_i \sqrt{d_{j_i} }D^{j_i}_{m_i,n_i}(g_i)\right)\,\bar{i}^{\{j\}}_{n_1,n_2,n_3}\right]
\end{eqnarray}
where $d_j$ is the dimension, $D^j(g) \in End(\mathbb{V}^j)$ the group matrix element and ${i}^{\{j\}}_{n_1,n_2,n_3} \in \text{Hom}_G(\mathbb{V}^{j_1}\otimes \mathbb{V}^{j_2} \otimes \mathbb{V}^{j_3}, \mathbb{C})$ is the three-valent intertwiner operator (related to the Clebsch-Gordan map $\Psi^{j_3}_{j_1 j_2}: \mathbb{V}^{j_1}\otimes \mathbb{V}^{j_2} \rightarrow \mathbb{V}^{j_3}$). We used the shorthand notation $\{j\}$ for the set of spin labels $(j_1,j_2,j_3)$.

The fields $\vv^{\{j\}}_{m_1,m_2,m_3}$ result from the contraction of the Fourier transformed GFT fields $\hat{\vv}_{\{j\}}$ with the intertwiner tensor imposing the gauge symmetry at the vertex. \footnote{This is the standard factorization of a symmetric tensor into a degeneracy tensor with all the degrees of freedom and a structural tensors (the Clebsch-Gordan coefficients) completely determined by the symmetry group G (Wigner-Eckart theorem) \cite{Singh:2009cd}.}
\begin{eqnarray}
\vv^{\{j\}}_{\{m\}}=\sum_{\{k\}}\hat{\vv}^{\,\{j\}}_{\{m\};\{k\}}\,i_{\{j\}}^{\{k\}}\,\prod_i \sqrt{d_{j_i} }.
\end{eqnarray}

The Fourier transformed fields depend on the (discrete) representation space labels of the Lie
group in question. Thus, generically Fourier transformed GFT fields are tensors of some rank d, $\vv_{\{m_j\}}$ with \emph{discrete} indices $\vec{m}_j=\{m_1,\dots, m_d\}$. \footnote{To regularize some quantities, especially at the dynamical level, it may be necessary to impose a (large) cut-off N in the range of the representation indices.}

In \eqref{fourier}, such tensors are contracted with the {\it spin network basis} tensors 
\begin{eqnarray}\label{see}
S^{\{j\}}_{\{m\}}=\left(\prod_i \sqrt{d_{j_i} }D^{j_i}_{m_i,n_i}(g_i)\right)\,\bar{i}^{\{j\}}_{\{n\}} ,
\end{eqnarray}
encoding the properties of the vertex of the spin network graph dual to the  (d-1)-dimensional triangulation that can be associated to the GFT states.

\subsection{Group Field Single Particle States}
Functions $\vv(g_i)$ can also be understood as \emph{single particle} wave functions for quanta corresponding to single open vertices of a spin network graph (in fact, they also label coherent states of the GFT field operator, which define the simplest condensate states of the theory \cite{Oriti:2015qva,Oriti:2015rwa,Oriti:2016acw}).

Let us define these \lq single-particle\rq quantum states as 
\begin{eqnarray}\label{gmod}
|\varphi \rangle=\int_{G^d} \dd{g_i}~\varphi({g_i})\ket{{g_i}}
\end{eqnarray}
where $\dd {g_i}\equiv \dd g_1\dd g_2\dots\dd g_d$ is the Haar measure on the group manifold $G^d$, invariant under the gauge transformation, and the vectors $\ket{g_1} \dots \ket{g_d}$  provide a basis on the respective infinite dimensional spaces $\mathbb{H}\simeq L^2[G]$.

The single particle state $\ket{\varphi}$ is then defined in $\mathbb{H}^{\otimes d}$. Moreover we require $\ket{\varphi}$ to be normalized (this is of course not the case for the classical GFT fields or the GFT condensate  wavefunctions):
\begin{eqnarray}
\braket{\varphi}{\varphi}=\int \dd {g_i} ~\overline{\varphi({g_i})}\varphi({g_i})=1.
\end{eqnarray}
Considering the case of $G=SU(2)$, we can decompose the basis $\ket{g}$ into the unitary irreducible representation of $SU(2)$ as
\begin{eqnarray}
\ket{g}\equiv \sum_{j,m,n}\sqrt{d_j} \overline{D^j_{mn}(g)}\ket{j,n,m^{\dag}}
\end{eqnarray}
and viceversa
\begin{eqnarray}
\ket{j,n,m^{\dag}} = \int_{SU(2)}\dd g \sqrt{d_j}D_{mn}^j(g)\ket{g}.
\end{eqnarray}
In particular, the tensor decomposition given in \eqref{fourier} holds at the quantum level, hence defining the quantum fields $\vv^{\{j\}}_{m_1,m_2,m_3}$ as actual  \emph{tensors states}.

Tensors in \eqref{see} defines the $SU(2)$-invariant single vertex spin network wave functions (in group representation)
\begin{eqnarray}
\psi_\chi (g_i)=\langle{\chi}|g_i\rangle=\left(\prod_i \sqrt{d_{j_i} }D^{j_i}_{m_i,n_i}(g_i)\right)\,\bar{i}^{\{j\}}_{\{n\}} ,
\end{eqnarray}
The basis vector $|\chi\rangle=|j, m, i\rangle$ denotes the standard $SU(2)$ spin network basis (labelled by spins and angular momentum projections associated to their d open edges, and intertwiner quantum numbers).

\subsection{Many-Body Description and Tensor Network States}\label{MBTNS}
We now describe the quantum states of the formalism, emphasizing their many-body structure, following \cite{Oriti:2013aqa}.

Consider a d-valent graph formed by V disconnected components, each corresponding to a single \emph{gauge invariant} d-valent vertex and d 1-valent vertices, thus having d edges.\footnote{One could work instead with the larger Hilbert spaces of non-gauge invariant states $L^2[G^{d\times V}]$ without imposing any gauge symmetry at the vertices of spin network graphs, and consider this condition as part of the dynamics. The above construction would proceed identically, with the same final result, but with the basis of single-vertex states now given by the above functions without the contraction of representation function with a G-intertwiner.} We refer to this type of disconnected components as \emph{open} spin network vertices.

\begin{figure}[t]
\centering 
\vspace{0cm}
\includegraphics[width=.36\textwidth]{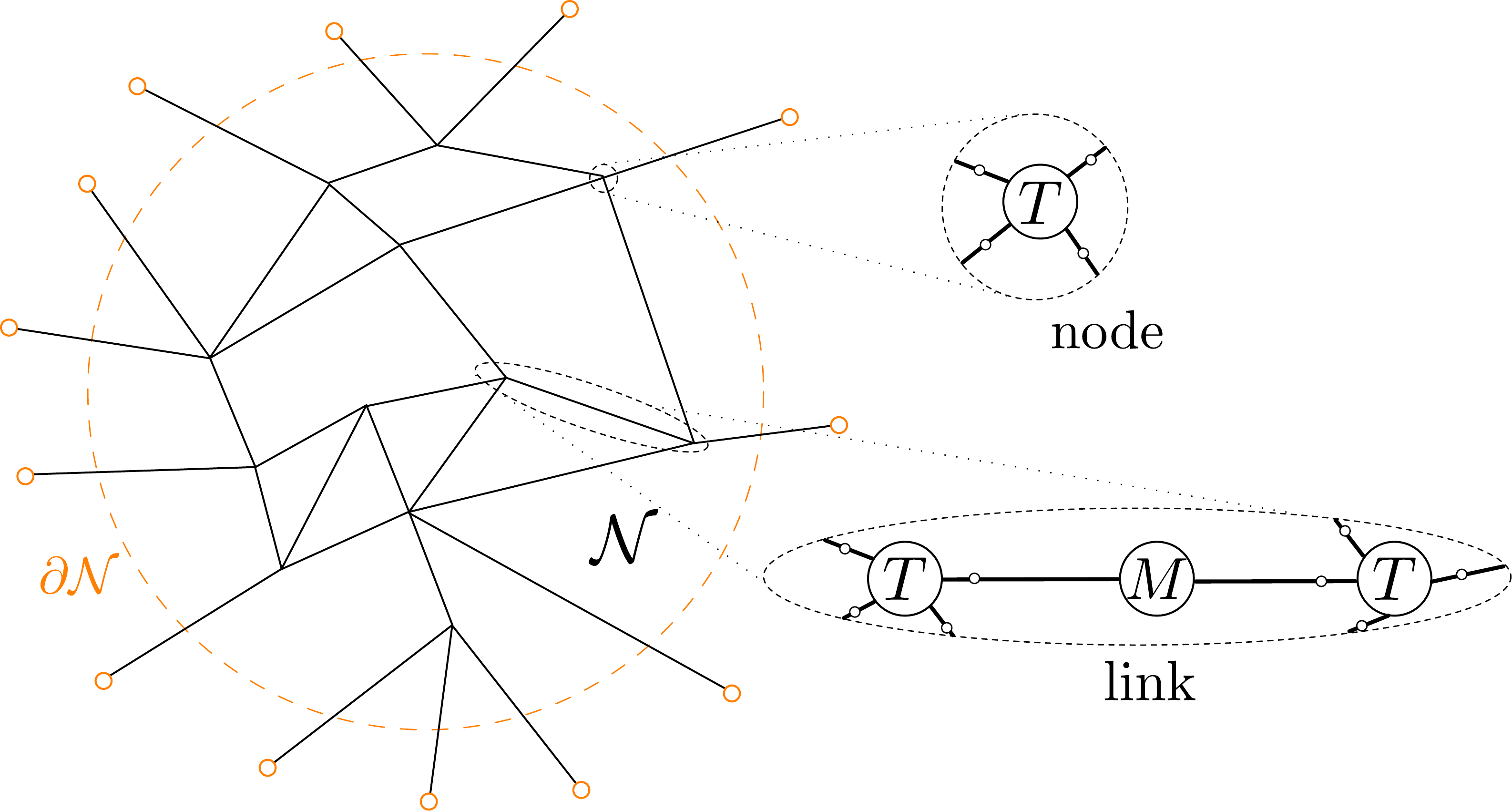}
\caption{\label{GFTN} A tensor network $\Gamma$  is a set of tensors whose indices are contracted according to a network pattern.
A network pattern can be always represented as a {\it graph}, given by a set of nodes $(n)$ and links $(\ell)$ connecting nodes. A link is called an {\it internal }link when it connects two \emph{different} nodes; while it is called a {\it boundary} link when it connects only one node. The number of links that connect to a node is called the valence of the node..}
\end{figure}
To such a graph we can associate a generic wavefunction given by a function of $d\times V$ group elements,
\begin{eqnarray}\label{multi}
&\Phi(g_a^i)=\Phi(g_1^1, ...,g_1^d, g_2^1, ...,g_2^d, \cdots, g_V^1, ...,g_V^d)  
\end{eqnarray}
defined on the group space ${G^{d \times V}}/{G^V}$ (V copies of $G^d$, quotiented by the isotropy group of the single particle function $\vv_{(v)}(g_i)$ at the each vertex); here the index $a$ runs over the set of vertices, while the index $i$ still runs over the links attached to each vertex).

These functions are exactly like many-particles wave functions for point particles living on the group manifold $G^d$, and having as classical phase space $(\mathcal{T}^*G)^d$ (which is also the classical phase space of a single open spin network vertex or polyhedron).

Accordingly, a state $|\Phi \rangle \in \mathbb{H}_V \simeq L^2[{G^{d \times V}}/{G^V}]$ can be conveniently decomposed into products of single-particle (single-vertex) states,
\begin{eqnarray}
\Phi(g_i^a)&=& \langle g_i^a |\Phi \rangle=\sum_{\chi_i, i=1...V} \vv^{\chi_1...\chi_V}\, \psi_{\chi_1} (g_i)\cdots \psi_{\chi_V} (g_i)
\end{eqnarray}

While the above decomposition is completely general, a special class of states can be constructed in direct association with a graph or network $\Gamma$. The association works as follows. Start from the d-valent graph with V disconnected components (open spin network vertices) to which a generic V-body state of the theory is associated. A partially connected d-valent graph can be constructed by choosing at least one edge $i$ in a vertex $a$ and {\it gluing} it  to one edge $j$ of the vertex $b$, i.e. joining the two edges along their 1-valent vertices. The final graph will be fully connected if  
all edges have been glued. Each pair of glued edges $\{a i, b j \}$ will identify a {\it link} $L$ of the resulting (partially) connected graph.
In the spin representation, i.e. in terms of the basis of functions $\psi_{\chi_1} (g_i)\cdots \psi_{\chi_V} (g_i)$, the gluing is implemented by the identification of the spin labels $j^a_i$ and $j_j^b$ associated to the two edges being glued and by the contraction of the corresponding vector indices $m_i^a$ and $m_j^b$. In other words, the corresponding wave functions for closed graphs can be decomposed in a basis of {\it closed spin network wave functions}, obtained from the general product basis by means of the same contractions:
\begin{eqnarray}
&&\Phi_\Gamma(g_i^a)= \langle g_i^a |\Phi_\Gamma \rangle =\sum_{\chi_a, a=1...V} \Phi_{\Gamma}^{j_i^1...j_i^V}\,\left[ \left(\prod_{L\in \Gamma} \delta_{j_i^a , j_j^b} \delta_{m_i^a , m_j^b}\right) \psi_{\chi_1} (g_i)\cdots \psi_{\chi_V} (g_i)\right]
\end{eqnarray}
 where the coefficients of the wave function can in turn be understood as the resulting of considering generic coefficients $\vv^{\chi_1...\chi_V}$ and contracting them with some choice of functions $M_{n_i^a n_j^b}^{j^a_i j^b_j} \delta_{j^a_i , j^b_j}$:

 \begin{equation}
 \Phi_\Gamma^{j_i^1...j_i^V} \, =\, \vv^{\chi_1...\chi_V} \, \left(\prod_{L\in \Gamma} \delta_{j_i^a , j_j^b}\, M^{j^a_i j^b_j}_{n_i^a , n_j^b}\right) \qquad , 
 \end{equation}
where the contraction is left implicit.

\noindent For fixed $\{j\}$, each resulting contraction scheme of tensors (each identified by a set of labels $\chi$) defines a \emph{tensor network state}.

\noindent In the group representation, the gluing amounts to considering wave functions with a specific symmetry under simultaneous group translation of the arguments associated to the edges being glued:
\begin{eqnarray}\label{multi}
&\Phi_\Gamma(g_a^i)=\Phi_\Gamma(g_1^1, ...,g_1^d h_{1V}^{d1}, g_2^1, ...,g_2^d, \cdots, g_V^1 h_{1V}^{d1}, ...,g_V^d)  \qquad .
\end{eqnarray}

\noindent In the end, given a tensor network with graph $\Gamma$, the $\Phi^{j_i^1...j_i^V}$  defined above will contain all the information about the combinatorics of the quantum geometry state.

\noindent A further special case corresponds to those states for which the coefficients  $\vv^{\chi_1...\chi_V}$ themselves have a product form, i.e. can be decomposed in terms of tensors. In this case, as it is for the spin network wave functions, the coefficients $\Phi^{j_i^1...j_i^V}$ can be obtained as a tensor trace
\begin{eqnarray}\label{tt}
\Phi^{j_i^1...j_i^V}\,= \text{Tr}[\bigotimes_L M \bigotimes_v \vv^{\{j\}\,(v)}_{\{m\}}] \qquad , 
\end{eqnarray}
again, in the case of fully connected graphs $\Gamma$ (otherwise, some angular momentum labels will remain on the left had side, corresponding to the edges that have not been glued). In lattice theory, we would say that the network $\Gamma$ (fixed $\{j\})$ provides a tensor network decomposition of the tensor state $\Phi^{j_i^1...j_i^V}$. 

The \emph{equivalence} of a special class of GFT states with the lattice tensor network states, and the sense in which GFT states generalise them, can be further elucidated by the following example.

\subsection{Link state as a gluing operation}
A tensor $\hat{T}$ is a multidimensional array of complex numbers $\hat{T}_{\lambda_1,\dots, \lambda_d} \in \mathbb{C}$. The rank of tensor $\hat{T}$ is the number $d$ of indices.  The size of an index $\lambda$, denoted $d_{|\lambda|}$, is the number of values that the index $\lambda \in \mathbb{N}$ takes \cite{singh2011tensor}.

Analogously, at the  quantum level, to each leg of the tensor one associates a Hermitian inner product space $\mathbb{H}_D$, with dimension $D$ given by the size of the indices $\lambda \in \{1,2,...,d_{|\lambda|}=D\}$.
Given an orthonormal basis $\ket{\lambda}$, in~$\mathbb{H}_D$, a  covariant tensor of rank $d$ is a~multilinear form on the Hilbert space of the vertex ${T} \colon \mathbb{H}_D^{\otimes d} \rightarrow \mathbb{C}$. 
Hence a tensor state is written as
\begin{eqnarray}
|T\rangle = \sum_{\lambda_1,\dots \lambda_d} \hat{T}_{\lambda_1\cdots \lambda_d}\ket{\lambda_1}\otimes\cdots\otimes\ket{\lambda_d} 
\end{eqnarray}
where $\hat{T}_{\lambda_1\cdots \lambda_d} \equiv { T} (\lambda_{1},\dots, \lambda_{d})$ denote the components in the canonical dual tensor product basis.

A tensor network is generally given by a set of $d$-valent vertices $v$, corresponding to rank $d$ tensors.
In particular,  a state corresponding to a set of unconnected vertices is written as a tensor product of individual vertex states
\begin{eqnarray}
\ket{\mathcal{T}_{\mathcal{N}}}\equiv \bigotimes_n \ket{T_n}
\end{eqnarray}

Individual vertex states are \emph{glued} by links. To each end of a link we associate a Hilbert space $\mathbb{H}_D$. The Hilbert space of the link $\ell$ is then $\mathbb{H}_{\ell}=\mathbb{H}_D^{\otimes 2}$ and a link state can be written as
\begin{eqnarray}\label{tlink}
\ket{M}=M_{\lambda_1\lambda_2}\ket{\lambda_1}\otimes\ket{\lambda_2}
\end{eqnarray}
where we choose to take the link states $\ket{M}$ to be generically entangled.\footnote{One can observe it by defining a density matrix $\rho_M\equiv \ket{M}\bra{M}$ and tracing out one of the Hilbert space, without losing generality, tracing out $\mathbb{H}_D$ of $\ket{\lambda_2}$, then computing the von Neumann entropy of the reduced density matrix $\rho_1\equiv \text{Tr}_2\rho_M= M^{\dag}M$. The entropy $S= \text{Tr} \rho_1\ln \rho_1$ is non-zero unless $M_{\lambda_1\lambda_2}$ can split as $M_{\lambda_1\lambda_2}= A_{\lambda_1}B_{\lambda_2}$.  For simplicity, in the next sections we will often assume that the link state is maximally entangled, i.e. 
\begin{eqnarray}
\ket{M}=\frac{1}{\sqrt{D}}\delta_{\lambda_1 \lambda_2}\ket{\lambda_1}\otimes\ket{\lambda_2}.
\end{eqnarray}}
In general, the entanglement of the links will encode the information on the connectivity of the graph. Two nodes are connected if their corresponding states contract with a link state,
\begin{eqnarray}
\hat{\mathcal{T}}_{12}&\equiv& \bra{M}\ket{T_1}\ket{T_2} =T^{(1)}_{\lambda_1\cdots\lambda_a\cdots\lambda_v}\overline{M}_{\lambda_a\lambda_b}T^{(2)}_{\lambda'_1\cdots\lambda_b\cdots\lambda_u'} \bigotimes_{i\neq a}^v\ket{\lambda_i}\otimes\bigotimes_{j\neq b}^u\ket{\lambda'_i}
\end{eqnarray}
Notice that if $\ket{M}$ was a non-entangled state, the connection would be trivial, i.e. the two nodes would be practically disconnected and the corresponding state could be written as a tensor product of two states,
\begin{eqnarray}\nonumber
\hat{\mathcal{T}}_{12}&=& T^{(1)}_{\lambda_a\lambda_1\cdots\lambda_v}\overline{A}_{\lambda_a}\bigotimes_{i=1}^v\ket{\lambda_i} \otimes \overline{B}_{\lambda_b}T^{(2)}_{\lambda_b\lambda'_1\cdots\lambda_u'}\bigotimes_{j=1}^u\ket{\lambda'_i}\\ 
&=&\ket{T'_1}\otimes\ket{T_2'}
\end{eqnarray}

Then given a network $\mathcal{N}$ with $N$ nodes and $L$ links, the corresponding state is
\begin{equation}\label{eq:TNS}
\ket{\Psi_{\mathcal{N}}}\equiv \bigotimes_\ell^L\bra{M_{\ell}}\bigotimes_n^N\ket{T_n}
\end{equation}
Because all links are contracted with nodes, $\ket{\Psi_{\mathcal{N}}}$ is then in the Hilbert space associated to the boundary links of the network, which is denoted as $\mathbb{H}_{\partial\mathcal{N}}$. $\ket{\Psi_{\mathcal{N}}}$ is a state in $\mathbb{H}_{\partial\mathcal{N}}$. 

The above structure can be identified also for the special GFT states mentioned at the end of the previous subsection, which are formed by generalised $L^2(G^d)$ functions associated to the nodes of the network.
In this case, the analogous of the generic link state in \eqref{tlink}, which is also the group counterpart of the gluing operators associated in the spin representation to the matrices $M$, can be defined in as the convolution functional
\begin{eqnarray}
\bra{M_{g_{\ell}}}\equiv \int\dd g_1\dd g_2~M(g_1^{\dag}g_{\ell}g_2)~\bra{g_1}\otimes\bra{g_2}~\in \mathbb{H}^{*\otimes 2},
\end{eqnarray}
where the functions $M(g)$ are assumed to be invariant under conjugation $M(g) = M(h g h^{-1})$.
When a link $\ell$ connects two nodes, say $a$ and $b$, the corresponding state $\bra{M_{g_{\ell}}}$ contracts with states $\ket{\varphi_a}$ and $\ket{\varphi_b}$ 
\begin{eqnarray}
&\,&\bra{M_{g_{\ell}}}\ket{\varphi_a}\ket{\varphi_b}=  \int\dd g_1\dd g_2\dd g_i^a\dd g_i^b~M(g_1^{\dag}g_{\ell}g_2)~\varphi_a(g_{1},g_i^{a})\varphi_b(g_{2},g_i^{b})\ket{g_i^{a}}|{g_i^{b}}\rangle \qquad , 
\end{eqnarray}
where we have singled out, among the arguments of the vertex wave functions $\vv$ the ones affected by the gluing operation.
In these terms, the open $d$-valent tensor network graph $\Gamma$ with $V$ vertices, can be written as
\begin{eqnarray}\label{TS}
\ket{\Phi_{\Gamma}^{g_\ell}} \equiv \bigotimes_{\ell \in \Gamma} \bra{M_{g_{\ell}}}\bigotimes_{n}^V \ket{\varphi_n}=\int \dd {g}_\p\,\Phi_{\Gamma} ({g}_{\ell},{g}_\p)\ket{{g}_{\p}}
\end{eqnarray}
where the $\{{g}_{\p}\}$ denote the group elements on the open links.

The role of the link state in tensor network, thus, is naturally generalised by the convolution function, defined for the group field variables. This is due to the fact that the group fields $\vv(g_i)$ on $G^d$ can be interpreted as rank $d$ tensors, with indices spanning the group space $G$, and associated Hilbert space (for each index) being $L^2(G)$.\footnote{The case of ordinary, finite-dimensional tensors is obtained if we pass from a Lie group to a discrete group. 
Let us consider, as a basic example, the case of a field defined on the discrete $n$th cyclic group $\mathbb{Z}_n$. Given the nonempty set 
\begin{eqnarray}
X=\{\vec{\lambda}\, | \,\vec{\lambda}=(\lambda_1,\dots,\lambda_d), \,\lambda_k \in \mathbb{Z}_n\}.
\end{eqnarray}
the field $\vv: X \rightarrow \mathbb{R} $ (or  $\mathbb{C}$) is a real or complex valued function on X and we indicate by
\begin{eqnarray}
\vv_{\vec{\lambda}}\equiv \vv(\vec{\lambda}).
\end{eqnarray} 
the value of $\vv$ on the set of d elements $\vec{\lambda}$. The function  $\vv_{\vec{\lambda}}$ can be interpreted as a tensor with d discrete indices $\vv_{\lambda_1,\dots, {\lambda_d}}$, where $\lambda \in \{1,2,\dots, |dim(\mathbb{Z}_n)|\}$.
}
The multiparticle state given in \eqref{multi} can then be interpreted as a tensor state with indices ${g}_\p$ and rank given by the number of open links of the spin network graph. 

\subsection{Link function in spin decomposition}
As showed in \ref{MBTNS}, many-body state can also be decomposed into spin representations. Suppose $M(g_1^\dag g_{\ell}g_2)$ can be written as
\begin{eqnarray}
M(g_1^\dag g_{\ell}g_2)=\sum_{jmn}d_jM_{ m n}^jD_{mn}^j(g_1^\dag g_{\ell}g_2)
\end{eqnarray}
Then, as a simple example, the state $\bra{M_{g_{\ell}}}\ket{\varphi_a}\ket{\varphi_b}$ can be written in terms of $\varphi_{\mathbf{k}\mathbf{n}}^{\mathbf{j}}$, $i_{\mathbf{m}}$ and $M_{m n}^j$ as\footnote{Notice that we are introducing the bold font for vectorial quantities, in order to shorten the notation in spin representation.}
\begin{eqnarray}\nonumber
&&\bra{M_{g_{\ell}}}\ket{\varphi_a}\ket{\varphi_b} = \int\dd g_1\dd g_2\dd {g}_i^a\dd {g}_i^b~M(g_1^{\dag}g_{\ell}g_2)\, \varphi_a(g_{1},{g}_i^{a})\varphi_b(g_{2}, {g}_i^{b})\ket{{g}_i^{a}}|{{g}_i^{b}}\rangle\\
&&= \sum_{jmnklpq}\sum_{i_ai_b}\overline{[i_{a}]_{\mathbf{p}}}[\varphi_{a}]_{\mathbf{p}\mathbf{n}_an}^{j\mathbf{j}_a}M_{nm}^j\overline{[i_{b}]_{\mathbf{q}}}[\varphi_{b}]_{\mathbf{q}\mathbf{m}_b(-m)}^{j\mathbf{j}_b} (-)^m \ket{\mathbf{j}_b,\mathbf{m}_b}\ket{\mathbf{j}_a,\mathbf{n_a}} [i_a]_{\mathbf{k}_ak}D_{kl}^j(g_{\ell})[i_b]_{\mathbf{l}_b(-l)} \times \nonumber\\ 
&&\times (-)^l ~|\mathbf{j}_a,\mathbf{k}_a^\dag\rangle |\mathbf{j}_b,\mathbf{l}_b^\dag \rangle
\end{eqnarray}

Graphically, the last line can be presented as
\begin{eqnarray}
\bra{M_{g_{\ell}}}\ket{\varphi_a}\ket{\varphi_b} =\sum_{jmnkl}\sum_{i_ai_b} ~~
\begin{minipage}[h]{0.4\linewidth}
	\vspace{10pt}
	\includegraphics[width=.6 \textwidth]{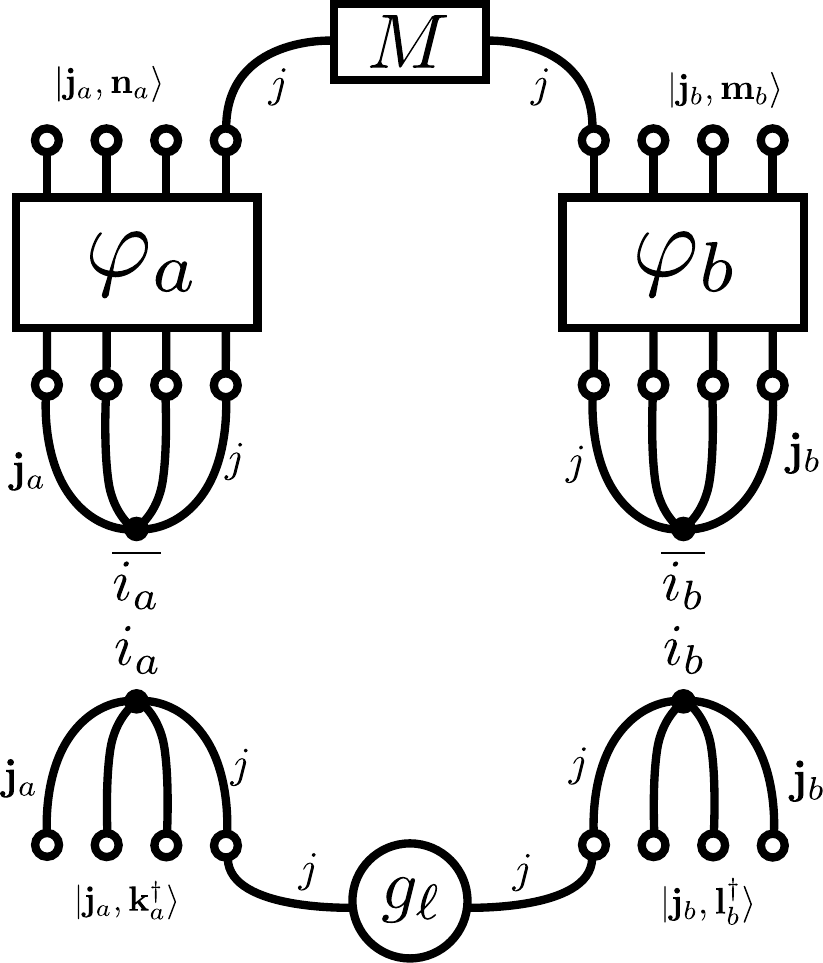}
	\vspace{20pt}
\end{minipage}
\end{eqnarray}
\smallskip
From the graphic equation, one can immediately observe that the upper part is an open tensor network $\ket{\Phi^{\mathbf{ji}}}$, given by the tensor trace of a collection of tensors 
\begin{eqnarray}
\phi_{\mathbf{m}}^{\,\mathbf{j}i}\equiv \sum_{\mathbf{n}}\overline{i_{\mathbf{n}}}\varphi_{\mathbf{nm}}^{\,\mathbf{j}}
\end{eqnarray}
for each node and matrices $M_{mn}^j$ for each link.

\subsection{Dictionary}

We summarize the established dictionary between group field theory states and generalized random tensor networks in terms of two synthetic tables. The correspondence between group field theory and tensor network description is summarized in Table A:
\begin{center}
\begin{tabularx}{1.1\textwidth}{ >{\setlength\hsize{.4\hsize}\centering}X >{\setlength\hsize{1\hsize}\centering}X|>{\setlength\hsize{1\hsize}\centering}X>{\setlength\hsize{.4\hsize}\centering}X } 
\hline
Table A	& Group Fields	& 	Tensors &
\tabularnewline \hline &&\tabularnewline	
group basis & $\ket{{g_i}\,} \in \mathbb{H} \simeq L^2[G] $ & $\ket{\lambda_i}$, $ \lambda_i=1,\dots ,D$ in~$\mathbb{H}_D$ & index basis\,
\tabularnewline &&\tabularnewline 
one particle state & $|\varphi \rangle=\int_{G^d} \dd{g_i}~\varphi({g_i})\ket{{g_i}}$ & $\ket{T_n} = \sum_{\{\lambda_i\}}T_{\{\lambda_i\}}|{{\lambda}_i}\rangle \in  \mathbb{H}_n=\mathbb{H}_D^{\otimes d}$ & tensor state
\tabularnewline &&\tabularnewline
gluing functional & $\bra{M_{g_{\ell}}}= \int\dd g_1\dd g_2~M(g_1^{\dag}g_{\ell}g_2)~\bra{g_1}\,\bra{g_2}$ \\ $\in \mathbb{H}^{*\otimes 2}$
& $\ket{M}=M_{\lambda_1\lambda_2}\ket{\lambda_1}\otimes\ket{\lambda_2} \in \mathbb{H}_{\ell}=\mathbb{H}_D^{\otimes 2} $&link state
\tabularnewline &&\tabularnewline
multiparticle\\ state &$\ket{\Phi_{\Gamma}} \in \mathbb{H}_V \simeq L^2[{G^{d \times V}}/{G^V}]$ & $\ket{\Psi_{\mathcal{N}}}$ & tensor network state
\tabularnewline &&\tabularnewline 
product state\\ convolution\\ \, &$\ket{\Phi_{\Gamma}^{g_\ell}} \equiv \bigotimes_{\ell \in \Gamma} \bra{M_{g_{\ell}}}\bigotimes_{n}^V \ket{\varphi_n}$\\$=\int \dd {g}_\p\,\Phi_{\Gamma} ({g}_{\ell},{g}_\p)\ket{{g}_{\p}}$ &$\ket{\Psi_{\mathcal{N}}}\equiv \bigotimes_\ell^L\bra{M_{\ell}}\bigotimes_n^N\ket{T_n} \in \mathbb{H}_{\partial\mathcal{N}}$ & tensor network\\ decomposition
\tabularnewline \hline &&\tabularnewline  
	{\bf randomness} & $\frac{1}{Z}d\nu(\vv)$\\ \footnotesize{field theory probability measure} &  $T^U_{\mu}\equiv (UT^0)_{\mu}$\\ \footnotesize{$T^0_\mu\equiv T^0_{\lambda_1\cdots \lambda_d} \in \mathbb{H}_T$, \,\,$U \in \U(\text{dim}(\mathbb{H}_T))$} &random tensor state\\ \,
\tabularnewline \hline
\end{tabularx}
\end{center}

The generalisation of tensor networks in terms of GFT states is evident in the spin-j decomposition of the latter  $\vv(g_i)= \sum_{\mathbf{j}} \text{Tr}[\, \vv^{\,\mathbf{j}}_{\{m\}}\,\left(\prod_i \sqrt{d_{j_i} }D^{j_i}_{m_i,n_i}(g_i)\right)\,\bar{i}^{\,\mathbf{j}}_{\{n\}}\, ]$. 

Once we turn off the sum over all possible $j$s, fix the representation labels and ask them to be equal, generically Fourier transformed GFT fields $\vv^{\,\mathbf{j}}_{\{m\}}$, are tensors of single rank d, with \emph{discrete} indices $m_i=\{m_1,\dots, m_d\}$ spanning a finite dimensional space. The equivalence is summarized in table B:

\begin{center}
\begin{tabularx}{1\textwidth}{ >{\setlength\hsize{0.5\hsize}\centering}X >{\setlength\hsize{1.2\hsize}\centering}X|>{\setlength\hsize{1.2\hsize}\centering}X|>{\setlength\hsize{0.8\hsize}\centering}X} 
\hline
Table B	&	GFT network	&	Spin Tensor Network	 &   Tensor Network
\tabularnewline \hline  &&\tabularnewline		

node	 &	$\vv(\vec{g})$\\ \footnotesize{$\equiv \vv(g_1,g_2,g_3,g_4)$}	& $\vv^{\,\mathbf{j}}_{\{m\}}$\\ \footnotesize{$\propto \sum_{\{k\}}\hat{\vv}^{\,\mathbf{j}}_{\{m\}\{k\}}\,i^{\mathbf{j} \{k\}}$} &	$T_{\{\mu\}}$	
\tabularnewline &&\tabularnewline

link	&	$M(g_1^\dag g_\ell g_2)$	& $M^j_{mn}$& $M_{\lambda_1\lambda_2}$
\tabularnewline &&\tabularnewline

sym & $\vv(h \vec{g})=\vv(\vec{g})$ & $\prod_s^v D^j_{m_sm'_s}(g)i_{m'_1\cdots m'_v}^i$\\$=i_{m_1\cdots m_v}^i$   & $\prod_{s}^v U_{\mu_s\mu'_s}T_{\mu'_1\cdots\mu'_v}=T_{\mu_1\cdots\mu_v}$
\tabularnewline &&\tabularnewline 

state &  $\ket{\Phi_{\Gamma}^{g_\ell}}\equiv \bigotimes_\ell\bra{M_{g_{\ell}}}\bigotimes_{n}\ket{\psi_n}$ & $|\Psi_{\Gamma}^{\mathbf{ji}}\rangle \equiv \bigotimes_\ell \langle{M^{j_\ell}}|\bigotimes_{n}|\phi_n^{~\mathbf{j}_ni_n}\rangle$ & $\ket{\Psi_{\mathcal{N}}}\equiv \bigotimes_\ell^L\bra{M_{\ell}}\bigotimes_n^N\ket{T_n} $
\tabularnewline &&\tabularnewline

indices	& $g_i \in {G}$ , \quad $\ket{{g_i}\,}  \in \mathbb{H} \simeq L^2[G]$  &	$m_i\in\mathbb{H}_j$, $\SU(2)$ spin-$j$ irrep.	&	$\mu_i\in\mathbb{Z}_n$, $n$th cyclic group
\tabularnewline &&\tabularnewline

dim & $\infty$ & $\dim\mathbb{H}_j=2j+1$ & \qquad $\dim\mathbb{Z}_n=n$ \\ \,
\tabularnewline \hline
\end{tabularx}
\end{center} 

\

In the following sections, with the longer-term goal of a full understanding and computation of the Ryu-Takayanagi (RT) formula \cite{Ryu:2006bv} in the field-theoretic GFT context, we are going to use the inputs provided by the established dictionary to investigate the holographic RT formula for the case of networks of combinatorial tensor group fields described by means of the GFT  formalism and spin network techniques, along the lines proposed for the case of  Random Tensor Networks by \cite{Hayden:2016cfa}. 

In the tensor network generalisation of the gauge gravity duality \cite{swi}, the RT formula strongly supports a general relation between entanglement and geometry, in turn leading to the suggestion that the whole of spacetime geometry can be understood as {\it emergent} from (quantum information-theoretic) properties of non-spatiotemporal quantum building blocks. Of course, this last suggestion has a life on its own and it has been brought forward in many different contexts \cite{Oriti:2013jga,Oriti:2016acw,Cao:2016mst,Seiberg:2006wf,Koslowski:2007kh,Oriti:2007qd,Konopka:2006hu,Rivasseau:2011hm,Steinacker:2010rh}. In this sense, our work provides further steps towards the calculation of the RT formula within a complete quantum gravity setting, a concrete and general indication of the holographic character of gravity, which goes beyond the AdS/CFT gauge gravity duality framework.

\section{Ryu-Takayanaki formula for a GFT Tensor Networks}\label{sec:GFT-RT}
%


The starting point of our analysis is the state $\ket{\Psi_{\Gamma}}$, corresponding to an open network graph where each node is dressed with a group field generalised tensor. Because of the field theoretic description, we can see the network as a random tensor network and use the established correspondence to apply standard path integral formalism to evaluate the expectation values of entropies and other tensor observables. 
In particular, then, our goal consists in investigate  the holographic entanglement properties of the GFT network by means of techniques recently applied to the study of the holographic behaviour for Random Tensor Networks \cite{Hayden:2016cfa}, building on the dictionary we have established between the two languages. This calculation is not in the full GFT setup, i.e. the state $\ket{\Psi_{\Gamma}}$ is not treated, in the calculation of the averaging over random (generalised) tensors, as an $n$-point function of a given GFT. This more complete calculation is postponed to a future analysis. Still, we apply several techniques from GFT and generalized the calculations in \cite{Hayden:2016cfa} based on our dictionary:
\begin{enumerate}
\item Tensors are generalized to group fields, from a finite dimensional object to a square integrable $L^2$ function, mapping from group manifolds to the complex numbers $\C$.
\item A gauge symmetry of the group field associated to each vertex as a vertex wave function is introduced in order to fit our setup more to the context of the quantum gravity theory. 
\item The average over the $N$-replica of the wave functions (generalised tensors) associated to each network vertex is reinterpreted as a $N$-point correlation function of a (simple) GFT model, which turns the averaged R\'{e}nyi entropy into an amplitude in GFT. \end{enumerate}

The last point can be seen as an approximation of a more complete calculation in which the (average over the) whole tensor network is understood as a GFT N-point function, and computed as such. 
This more complete calculation based on the full GFT setup is being explored \cite{ChircoRT}. We believe that the leading term of the entropy, at least for the entanglement entropy, would not be changed.

\begin{figure}[t!]
\centering 
\includegraphics[width=.3\textwidth]{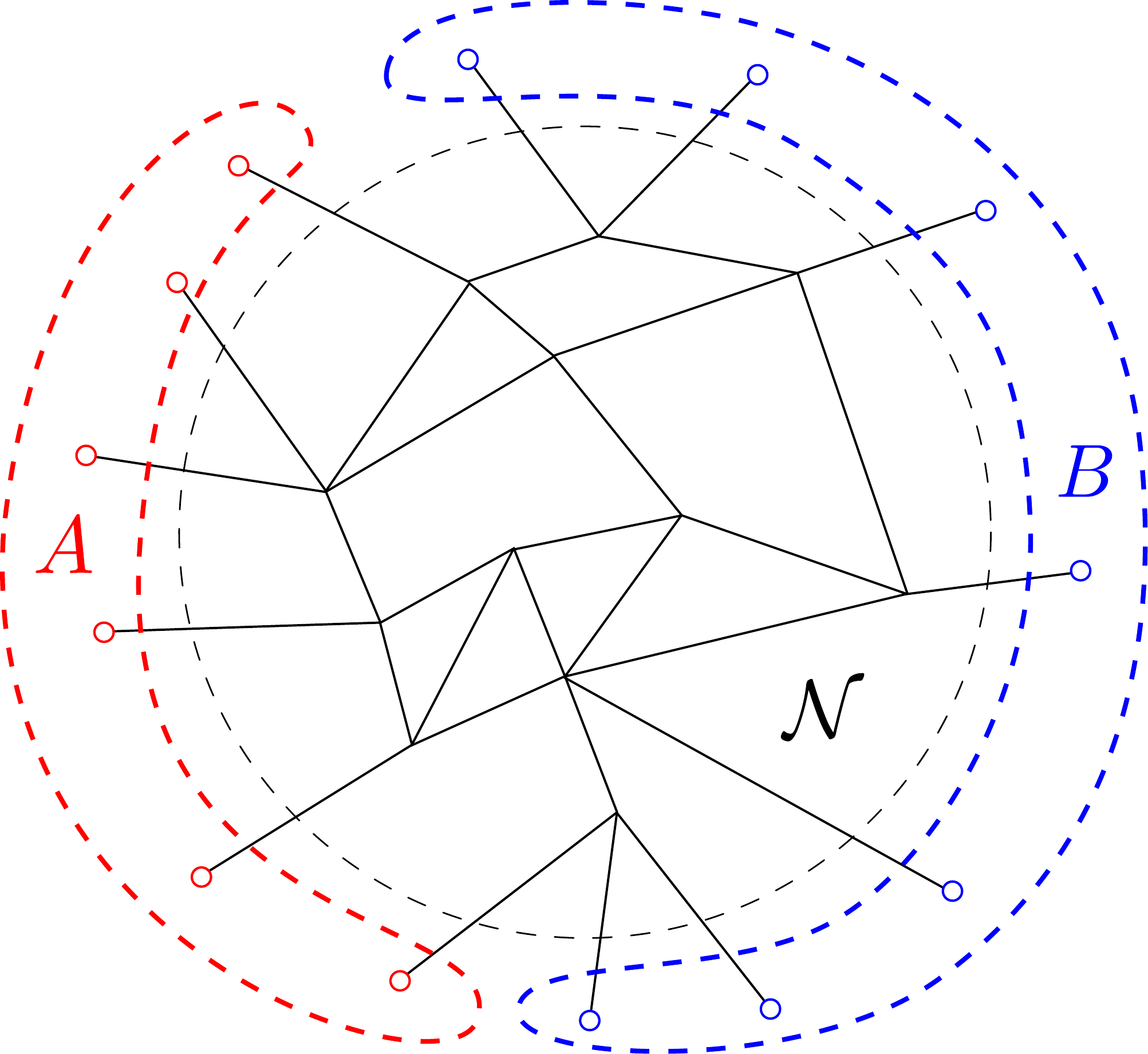}
\caption{\label{fig:AB} Boundary $\p\mathcal{N}$ of network $\mathcal{N}$ divided into two parts $A$ and $B$.}\label{AB}
\end{figure}

Given our tensor network state, $\ket{\Psi_{\mathcal{N}}}\equiv \bigotimes_\ell^L\bra{M_{\ell}}\bigotimes_n^N\ket{T_n} \in \mathbb{H}_{\partial\mathcal{N}}$, we start by considering  a bipartition of the boundary Hilbert space, 
\begin{equation}\label{bipart}
\mathbb{H}_{\partial\mathcal{N}}= \mathbb{H}_{A}\otimes \mathbb{H}_{B}
\end{equation}
associated to the definition of two -- a priori non adjacent -- subregions $A$ and $B$ of the boundary (see Figure \ref{AB}).
 
A measure of the entanglement between the two subsystems is given by the von Neumann entropy of the reduced density matrix of the subsystem, either $A$ or $B$, defined by partial tracing over the full system  Hilbert space. Focussing on subsystem $A$, for $ \rho \equiv \ket{\Psi_{\mathcal{N}}} \bra{\Psi_{\mathcal{N}}}
$, we have
\begin{eqnarray}
\rho_A = \text{Tr}_B (\rho) , 
\end{eqnarray}
and the entanglement entropy between $A$ and $B$ is given by the von Neumann entropy
\begin{eqnarray}
S_{\text{EE}}(A)=-\text{Tr} [\, \widehat{\rho}_A \ln \widehat{\rho}_A\,] , 
\end{eqnarray} 
where now
\begin{eqnarray}
\widehat{\rho}_A \equiv \frac{\rho_A}{\text{Tr}\rho}
\end{eqnarray}
is the normalized reduced density matrix.

In order to calculate $S_{\text{EE}}(A)$, due to the technical difficulty in computing the von Neumann entropy,  we need to make use of the standard replica trick. Contracting $N$ copies of the reduced density matrix $\rho_A$ and taking the logarithm of the trace of $\rho_A^N$, one obtains the $N$th-order R\'{e}nyi entropy
\begin{eqnarray}\label{eq:Sn}
S_N(A)=-\frac{1}{N-1}\ln\text{Tr}\widehat{\rho}_A^{\,N} \qquad .
\end{eqnarray}

The above formula is easier to compute and coincides with the von Neumann entropy of region $A$ in the limit $N \rightarrow 1$
\begin{eqnarray}
S_{\text{EE}}(A)=\lim_{N\rightarrow 1}S_N(A)
\end{eqnarray}

\subsection{$Nth$ R\'enyi entropy for a GFT random tensor network}

We focus now on the case of the $N$th R\'enyi entropy for a bipartite GFT state $\ket{\Psi_{\Gamma}}$ with support on a generic open graph $\Gamma$. We divide the boundary $\p\Gamma$ of the graph $\Gamma(V_\Gamma,L_{\Gamma},L_{\p\Gamma})$, with $V_{\Gamma}$ nodes, $L_{\Gamma}$ internal links and $L_{\p\Gamma}$ boundary links, into two parts, called $A$ and $B$. The $Nth$ R\'{e}nyi entropy between $A$ and $B$ is given by
\begin{eqnarray}
\ex^{(1-N)S_N}&=&\frac{Z_N}{Z_0^N}
\end{eqnarray}
with 
$\nonumber Z_0^{N}  \equiv (\text{Tr}\rho)^N$, 
 $Z_N  \equiv  \text{Tr}\rho_A^N = \text{Tr} [\rho^{\otimes N}\mathds{P}(\pi^0_A;N,d)]$ and the network density matrix $\rho$ defined as
\begin{equation}
\rho = \ket{\Psi_{\Gamma}}\bra{\Psi_{\Gamma}} = \tr_\ell\left[\bigotimes_{\ell}\ket{M_\ell}\bra{M_\ell}\bigotimes_n\ket{\psi_n}\bra{\psi_n}\right]\equiv \tr_\ell\left[\bigotimes_{\ell}\rho_\ell\bigotimes_n\rho_n\right] \qquad .
\end{equation}

Here, for convenience, we use the equivalence of the trace of the  reduced density with the result of the trace over the action of the permutation operator $\mathds{P}(\pi^0_A;N,d)$ on the full $\rho^N$\footnote{For $N=2$, e.g., the cyclic group $\mathcal{S}_n$ only has two elements: the identity $\mathds{1}$ and swap operator $\mathds{F}$, so that $\mathds{P}(\pi^0_A;2,d) \equiv \mathds{F}^{(A)}$. Then, $Z_2 = \tr[\rho^{\otimes 2}\mathds{F}^{(A)}]= \rho_{\overline{A}_1A_1\overline{B}_1B_1}\rho_{\overline{A}_2A_2\overline{B}_2B_2}\mathds{F}^{(A)}_{\overline{A}_1A_1\overline{A}_2A_2}\mathds{1}^{(B)}_{\overline{B}_1B_1\overline{B}_2B_2}$ and $Z_0^2= \tr[\rho^{\otimes 2}]=\rho_{\overline{A}_1A_1\overline{B}_1B_1}\rho_{\overline{A}_2A_2\overline{B}_2B_2}\mathds{1}^{(A)}_{\overline{A}_1A_1\overline{A}_2A_2}\mathds{1}^{(B)}_{\overline{B}_1B_1\overline{B}_2B_2}$.}, for
\begin{eqnarray}
\mathds{P}(\pi^0_A;N,d)=\prod_{s=1}^N\delta_{\mu_A^{([s+1]_D)}\mu_A^{(s)}}
\end{eqnarray}
with $d$ is the dimension of the Hilbert space in the same region $A$.

Given the random nature of the tensor network, we look for the typical value of the entropy.  Analogously to the case considered in~\cite{Hayden:2016cfa}, the variables $Z_N$ and $Z_0$ are easier to average than the entropy, since they are quadratic functions of the network density matrix $ff$. In particular, the entropy average can be expanded in powers of the fluctuations $\delta Z_N = Z_N - \mathbb{E}({Z_N})$ and $\delta Z_0^{N} = Z_0^{N} - \mathbb{E}({Z_0^{N}})$, so that
\begin{eqnarray}
\mathbb{E}({S_N(A)})&=&-\mathbb{E}\left({{\log\frac{\mathbb{E}({Z_N})+\delta  Z_N}{\mathbb{E}({Z_0^{N}})+\delta Z_0^N}}}\right)\\ \nonumber
&=&-\log \frac{\mathbb{E}({Z_N})}{\mathbb{E}({Z_0^{N}})}+\text{fluctuations}
\end{eqnarray}
As showed in \cite{Hayden:2016cfa},  for large enough bond dimensions $D$, as a direct consequence of the {concentration of measure phenomenon} \cite{hay2}, the statistical fluctuations around the average value are exponentially suppressed. Therefore, it is possible to approximate the entropy with high probability by the averages of $Z_N$ and $Z_0^{N} $,
\begin{eqnarray}
\ex^{(1-N)\mathbb{E}(S_N)}&\simeq&\frac{\mathbb{E}(\tr\rho_A^N)}{\mathbb{E}(\tr\rho)^N}=\frac{\mathbb{E}\tr [\rho^{\otimes N}\mathds{P}(\pi^0_A;N,d)]}{\mathbb{E}(\tr\rho)^N}\nonumber\\
&=& \frac{\tr \left[\bigotimes_{\ell}\rho^N_\ell\bigotimes_n\mathbb{E}(\rho^N_n)\mathds{P}(\pi^0_A;N,d)\right]}{\tr \left[\bigotimes_{\ell}\rho^N_\ell\bigotimes_n\mathbb{E}(\rho^N_n)\right]} \qquad .
\end{eqnarray}

In order to get the typical R\'enyi entropy one needs then to compute $\mathbb{E}({Z_N})$ and $\mathbb{E}({Z_0^{N}})$ separately.
The average over the tensor fields can be carried out before taking the partial trace, since the latter is a linear operation. Therefore, the key step consists in computing the quantity 
\begin{equation}
\mathbb{E}(\rho^N_n)=\mathbb{E}[(\ket{\psi_n}\bra{\psi_n})^N]=\mathbb{E}\left[\left(\int\prod_{a}^N\dd\mathbf{g}_a\dd\underline{\mathbf{g}_{a}}~\psi_n(\mathbf{g}_{a})\overline{\psi_n(\underline{\mathbf{g}_{a}})}\ket{\mathbf{g}_a}\bra{\underline{\mathbf{g}_{a}}}\right)\right] \qquad , 
\end{equation}
hence, eventually, the expectation value of $N$ copies of the network wavefunction,
\begin{equation}\label{eq:Npoint}
\mathbb{E}\left[\prod_a^N\psi_n(\mathbf{g}_{a})\overline{\psi_n(\underline{\mathbf{g}_{a}})}\right] \qquad .
\end{equation}
where $\dd\mathbf{g}\equiv \prod_i\dd g_i$, $\psi(\mathbf{g})\equiv\psi(g_1,\cdots,g_4)$ and $\underline{\mathbf{g}}$ is independent from $\mathbf{g}$, which denotes the arguments of $\overline{\psi}$. 

Now, we define the averaging operation $\mathbb{E}[\cdots]$ via the path integral of a generic group field theory model
\begin{equation}\label{eq:E}
\mathbb{E}\left[f[\psi,\overline{\psi}]\right]\equiv \int[\mathcal{D}\psi][\mathcal{D}\overline{\psi}]~f[\psi,\overline{\psi}]~\ex^{- S[\psi,\overline{\psi}]}
\end{equation}
where $S[\psi,\overline{\psi}]$ is the action of the given model of interest,
\begin{equation}
S[\psi,\overline{\psi}]=\int\dd\mathbf{g}\dd\underline{\mathbf{g}}~\overline{\psi(\mathbf{g})}\mathcal{K}(\mathbf{g},\underline{\mathbf{g}})\psi(\underline{\mathbf{g}})+ \lambda S_{\text{int}}[\psi,\overline{\psi}]+cc,
\end{equation}
the first term on the right hand side defining the kinetic term of the model. In the following calculation, we consider the particular case where 
\begin{equation}
\mathcal{K}(\mathbf{g},\underline{\mathbf{g}})=\delta(\mathbf{g}^\dag\underline{\mathbf{g}}) \qquad ,
\end{equation}
which thus implies  a free part of the action of the simple form \footnote{Notice that several GFT models of quantum gravity \cite{Oriti:2014uga,Baratin:2011aa,Oriti:2011jm,Oriti:2009wn} can be put in this form.}
\begin{equation}
S_0[\psi,\overline{\psi}]=\int\dd\mathbf{g}~\overline{\psi(\mathbf{g})}\psi(\mathbf{g})  \qquad .
\end{equation}
We further assume that the coupling constant $\lambda$ is much smaller than $1$, so the path integral $\mathbb{E}\left[f[\psi,\overline{\psi}]\right]$ can be perturbatively expanded in powers of $\lambda$
\begin{eqnarray}\label{eq:EExp}
\mathbb{E}\left[f[\psi,\overline{\psi}]\right] &=& \int[\mathcal{D}\psi][\mathcal{D}\overline{\psi}]~f[\psi,\overline{\psi}]~\ex^{- S_0[\psi,\overline{\psi}]}\left(1+\lambda S_{\text{int}}[\psi,\overline{\psi}]+\mathcal{O}(\lambda^2)\right)\nonumber\\
&\equiv& \mathbb{E}_0\left[f[\psi,\overline{\psi}]\right] + \mathcal{O}(\lambda) \qquad .
\end{eqnarray}
This is the regime of validity of the so-called spin foam expansion, seen from within the GFT formalism \cite{Oriti:2014uga,Baratin:2011aa,Oriti:2011jm,Oriti:2009wn,Ashtekar:2004eh,Rovelli:QG,Thiemann:QG,Perez:2012wv,Rovelli:2014ssa}.
In the following calculation, we will only focus on the leading term $\mathbb{E}_0\left[f[\psi,\overline{\psi}]\right]$\footnote{This, in turn, means that, from the point of view of the quantum gravity model, tthe quantum gravity dynamics is imposed only to the extent in which it is captured by the kinetic term in the GFT action.}.

Because of the gauge symmetry $\psi(h\mathbf{g})=\psi(\mathbf{g})$, the gauge equivalent paths in the above path integral have to be removed (via gauge fixing). In order to do so, we first introduce the following notation: if $\mathbf{g}=(g_1,g_2,g_3,g_4)$, then
\begin{equation}
[\mathbf{g}]\equiv g_1^{-1}\mathbf{g}=(\mathds{1},~g_1^{-1}g_2,~g_1^{-1}g_3,~g_1^{-1}g_4) \qquad .
\end{equation}
Then, we insert the delta functional $\delta[\psi(\mathbf{g})-\psi([\mathbf{g}])]$ constraint into the path integral, so that the average becomes
\begin{equation}
\mathbb{E}_0\left[f[\psi,\overline{\psi}]\right]\equiv \int[\mathcal{D}\psi][\mathcal{D}\overline{\psi}]~f[\psi,\overline{\psi}]~\delta[\psi(\mathbf{g})-\psi([\mathbf{g}])]~\ex^{-\int\dd\mathbf{g}~\overline{\psi(\mathbf{g})}\psi(\mathbf{g})} \qquad .
\end{equation}
Since this equation is simply the expectation value of $f[\psi,\overline{\psi}]$ in the free group field theory, we can immediately give the expectation value of \Ref{eq:Npoint} via Wick theorem:
\begin{eqnarray}\label{eq:gaugeReIntro}
\mathbb{E}_0\left[\prod_a^N\psi(\mathbf{g}_{a})\overline{\psi(\underline{\mathbf{g}_{a}})}\right]&=&\mathcal{C}\sum_{\pi\in\mathcal{S}_N}\prod_a^N\delta\left([\mathbf{g}_a][\underline{\mathbf{g}_{\pi(a)}}]^\dag\right)\nonumber\\ 
&=&\mathcal{C}\sum_{\pi\in\mathcal{S}_N}\int\prod_a^N\dd h_a \prod_a^N\delta\left(h_a \mathbf{g}_a\underline{\mathbf{g}_{\pi(a)}}^{\dag}\right) \qquad , 
\end{eqnarray}
where $\underline{\mathbf{g}}$ is independent from $\mathbf{g}$, $\delta([\mathbf{g}][\underline{\mathbf{g}}]^\dag)\equiv  \prod_{s=2}^4\delta\left(g_{1}^{\dag}g_s\underline{g_s}^{\dag}\underline{g_1}\right)$ and $\delta\left(h\mathbf{g}\underline{\mathbf{g}}\right)\equiv  \prod_{s=1}^4\delta\left(h g_s\underline{g_s}^{\dag}\right) $. 

In the second equality, we re-introduce the gauge symmetry by inserting integrals of $h_a\in \SU(2), N=1,2,\cdots N$ into the delta functions such that $g_{sa}$ on each leg of the node are on an equal footing, unlike $g_1=\mathds{1}$ in the gauge fixing procedure. So in the following calculation, the network is without gauge fixing, i.e. all integrals of $\mathbf{g}$ have to be performed. 

\noindent Denote now $\prod_a^N\delta\left(h_a \mathbf{g}_a\underline{\mathbf{g}_{\pi(a)}}^{\dag}\right)$ as
\begin{equation}\label{eq:Ph}
\mathds{P}_\mathbf{h}(\pi)\equiv \prod_a^N\delta\left(h_a \mathbf{g}_a\underline{\mathbf{g}_{\pi(a)}}^{\dag}\right)=\prod_{s=1}^4\prod_a^N\delta\left(h_a g_{sa}\underline{g_{s\pi(a)}}^{\dag}\right)\equiv \prod_s^4\mathds{P}^s_\mathbf{h}(\pi) \qquad , 
\end{equation}
where $\mathbf{h}$ denotes the set of $h_a$, $a=1,\cdots,N$. When $h_a=\mathds{1}$ for all $a$ from $1$ to $N$, 
\begin{equation}
\mathds{P}_\mathds{1}(\pi)=\prod_a^N\delta\left(\mathbf{g}_a\underline{\mathbf{g}_{\pi(a)}}^{\dag}\right)= \mathds{P}(\pi;N,D^4)=\prod_s^4 \mathds{P}^s(\pi;N,D^4)
\end{equation}
where $\mathds{P}(\pi;N,D^4)$ and $\mathds{P}^s(\pi;N,D^4)$ are the representations of $\pi\in\mathcal{S}_N$ on $\mathbb{H}^{\otimes 4}$ and $\mathbb{H}$, respectively. 

\noindent Then, $Z_N$ and $Z_0^N$ become
\begin{eqnarray}
Z_N & \approx & \mathcal{C}^{V_\Gamma}\sum_{\bm{\pi_n}\in\mathcal{S}_N}\int\prod_n\dd\mathbf{h}_n~\tr \left[\bigotimes_{\ell}\rho^N_\ell\bigotimes_n\mathds{P}_{\mathbf{h}_n}(\pi_n)\mathds{P}(\pi^0_A;N,d)\right]\nonumber\\
& \equiv & \mathcal{C}{V_\Gamma}\sum_{\bm{\pi_n}\in\mathcal{S}_N}\int\prod_n\dd\mathbf{h}_n~\mathcal{N}_A(\mathbf{h_n},\bm{\pi_n})\\
Z^N_0 & = & \mathcal{C}^{V_\Gamma}\sum_{\bm{\pi_n}\in\mathcal{S}_N}\int\prod_n\dd\mathbf{h}_n~\tr \left[\bigotimes_{\ell}\rho^N_\ell\bigotimes_n\mathds{P}_{\mathbf{h}_n}(\pi_n)\right]\nonumber\\
& \equiv & \mathcal{C}^{V_\Gamma}\sum_{\bm{\pi_n}\in\mathcal{S}_N}\int\prod_n\dd\mathbf{h}_n ~\mathcal{N}_0(\mathbf{h_n},\bm{\pi_n}) \qquad , 
\end{eqnarray}
which means that $Z_N$ and $Z_0^N$ correspond to summations of the networks $\mathcal{N}_A(\mathbf{h_n},\bm{\pi_n})$ and $\mathcal{N}_0(\mathbf{h_n},\bm{\pi_n})$ where at each node $n$ we have a contribution $\mathds{P}_{\mathbf{h}_n}(\pi_n)$ and at each link $\ell$ we have a contribution $\rho_\ell^N$. The only difference between these two networks is the boundary condition: where $Z_N$ is defined with $\mathds{P}(\pi^0_A;N,d)$ on $A$ of $\p\Gamma$ and $\mathds{P}(\mathds{1};N,d)$ on $\overline{A}$ of $\p\Gamma$, and $Z_0^N$ is defined with $\mathds{P}(\mathds{1};N,d)$ for all boundary region $\p\Gamma$. 

\noindent Since at each node $\mathds{P}_{\mathbf{h}_n}(\pi_n)$ is decoupled among the incident legs, because of \Ref{eq:Ph},  the value of the networks $\mathcal{N}_A(\mathbf{h_n},\bm{\pi_n})$ and $\mathcal{N}_0(\mathbf{h_n},\bm{\pi_n})$ can be written as products factorised over links: 
\begin{equation}
\mathcal{N}_A(\mathbf{h_n},\bm{\pi_n})=\prod_{\ell\in\Gamma} \mathcal{L_\ell}(\pi_{n},\pi_{n'};\mathbf{h}_{n},\mathbf{h}_{n'})\prod_{\ell\in A}\mathcal{L}_\ell(\pi_n,\pi_A^0;\mathbf{h}_n)\prod_{\ell\in\overline{A}}\mathcal{L}_{\ell}(\pi_n,\mathds{1};\mathbf{h}_n)
\end{equation}
\begin{equation}
\mathcal{N}_0(\mathbf{h_n},\bm{\pi_n})=\prod_{\ell\in\Gamma} \mathcal{L_\ell}(\pi_{n},\pi_{n'};\mathbf{h}_{n},\mathbf{h}_{n'})\prod_{\ell\in \p\Gamma}\mathcal{L}_{\ell}(\pi_n,\mathds{1};\mathbf{h}_n) \qquad .
\end{equation}
Because the $\mathcal{L}_{\ell}$ on the boundary are special cases of the $\mathcal{L}_{\ell}$ in the graph $\Gamma$, it is enough to calculate the $\mathcal{L}_{\ell}$ on the internal links. In general, $\mathcal{L}(\pi,\pi',\mathbf{h},\mathbf{h}')$ can be written as a trace of a modified representation of a permutation group element $\varpi\equiv(\pi')^{-1}\pi$ as
\begin{equation}
\mathcal{L}(\pi,\pi';\mathbf{h},\mathbf{h}')=\tr\left[\mathds{P}_{\mathbf{h}}(\pi)\rho_{\ell}^N\mathds{P}_{\mathbf{h}'}(\pi)\right] = \tr\left[\mathds{P}_{\mathbf{H}}\left((\pi')^{-1}\pi\right)\right]\equiv \tr\left[\mathds{P}_{\mathbf{H}}\left(\varpi\right)\right] \qquad, 
\end{equation}
where
\begin{equation}\label{eq:Hdef}
\mathbf{H}=\left\{H_a ~\big| ~ H_a\equiv \left(h'_{\varpi(a)}\right)^{\dag} h_a,~\forall a=1,\cdots,N\right\}
\end{equation} \qquad . 
\begin{figure}[htbp!]
\centering 
\includegraphics[width=.7\textwidth]{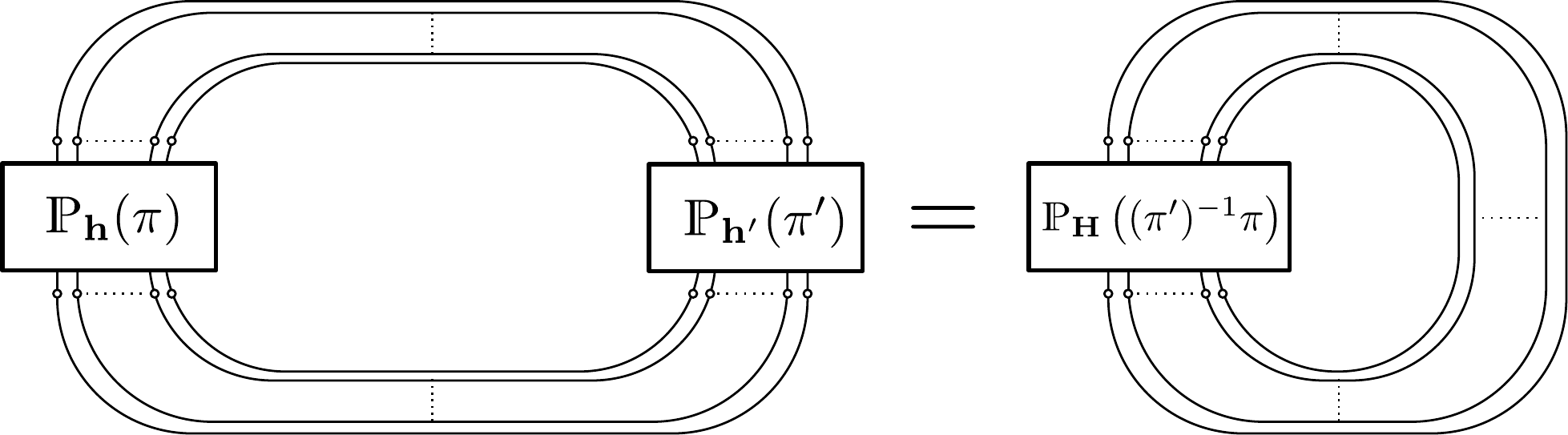}
\caption{\label{fig:L} $\mathcal{L}(\pi,\pi',\mathbf{h},\mathbf{h}')$}
\end{figure} 

\begin{figure}[htbp!]
\centering 
\includegraphics[width=.4\textwidth]{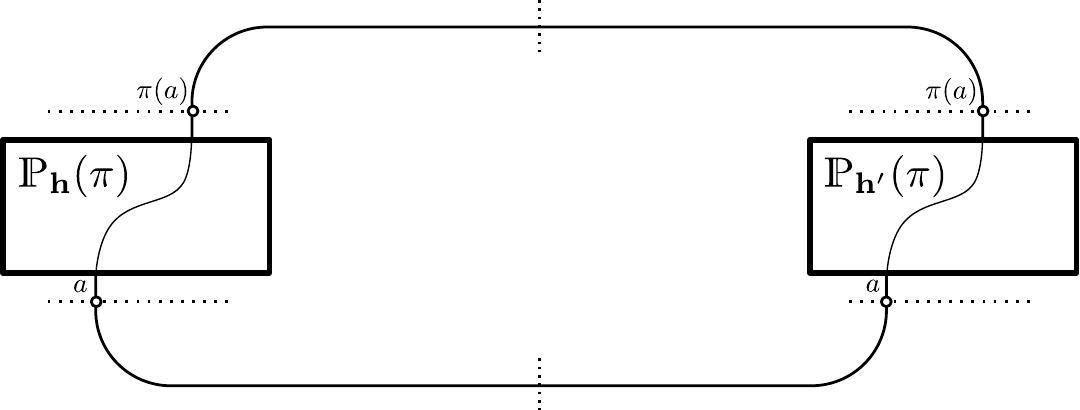}
\caption{\label{fig:Lp} $\mathcal{L}(\pi,\pi,\mathbf{h},\mathbf{h}')$}
\end{figure}

\noindent When $\pi=\pi'$, we have $\varpi=\mathds{1}$ and $\mathbf{H}=(\mathbf{h}')^\dag\mathbf{h}$, and then
\begin{eqnarray}\label{eq:LpipiD}
\mathcal{L}(\pi,\pi;\mathbf{h},\mathbf{h}') & = & \tr\left[\mathds{P}_{\mathbf{h}}(\pi)\rho_{\ell}^N\mathds{P}_{\mathbf{h}'}(\pi)\right]=\tr\left[\mathds{P}_{(\mathbf{h}')^\dag\mathbf{h}}\left(\mathds{1}\right)\right]\\
& = & \prod_a^N \int\dd g_a \dd g'_a\dd\underline{g_{\pi(a)}}\dd\underline{g'_{\pi(a)}}~\delta\left(h_{a}g_{a}\underline{g_{\pi(a)}}^\dag\right)\delta\left(\underline{g_{\pi(a)}}\underline{g'_{\pi(a)}}^{\dag}\right)\times\nonumber\\
&&\times\delta\left(h'_{a}g'_{a}\underline{g'_{\pi(a)}}^{\dag}\right)\delta\left(g'_{a}g_{a}^\dag\right)\nonumber\\
& = & \prod_a^N\delta\left((h'_a)^\dag h_a\right)= \prod_a^N\delta\left(H_a\right) \qquad , 
\end{eqnarray}
The above equation can be depicted graphically as in Fig.\ref{fig:Lp}

\noindent When $\pi\neq\pi'$, we have  
\begin{equation}
\mathcal{L}(\pi,\pi';\mathbf{h},\mathbf{h}')=\tr\left[\mathds{P}_{\mathbf{H}}\left(\varpi\right)\right] \qquad . 
\end{equation}
In order to perform the computation, it is necessary to use some facts about the permutation group $\mathcal{S}_N$, which we recall briefly, before proceeding. 
\begin{itemize}
\item Any element $\varpi\in\mathcal{S}_N$ can be expressed as the product of disjoint cycles $\mathcal{C}_i$
\begin{equation}
\varpi\equiv\prod_i^{\chi(\varpi)}\mathcal{C}_i
\end{equation}
where $1 \leq\chi(\varpi)\leq N$ is the number of cycles in $\varpi$, which is $1$ when $\varpi$ is a $1$-cycle and is $N$ only when $\varpi=\mathds{1}$. For instance, the permutation $\varpi=\{3241\}$ can be expressed as a product of two cycles $(134)(2)$, in which $\varpi(1)=3,\varpi(3)=4, \varpi(4)=1$ and $\varpi(2)=2$. $(132)$ is a $3$-cycle, because there are three elements in the cycle. 
We denote the number of elements in the cycle $\mathcal{C}_i$ as $r_i$, which is also called the length of the cycle. We also have $\sum_i r_i= N$.
Although the cycles $\mathcal{C}_i$ commute with each other, we order the cycles such that
\begin{equation}
1\leq \cdots \leq r_i \leq r_{i+1}\leq \cdots \leq N \qquad .
\end{equation}
We denote $a^i_k$, where $k$ is from $1$ to $r_i$, the elements of $\mathcal{C}_i$, and then we furthermore assume that 
\begin{equation}
\varpi(a^i_k)=a^i_{[k]_{r_i}+1} \qquad  .
\end{equation}
Thus, the cycle can be written as
\begin{equation}
\mathcal{C}_i=\left(a^i_1a^i_2\cdots a^i_{r_i}\right) \qquad . 
\end{equation}

\item The trace of $\mathds{P}_{\mathbf{H}}\left(\varpi\right)$ can be expressed as the product of the traces of the individual cycles $\mathcal{C}_i$
\begin{equation}
\tr\left[\mathds{P}_{\mathbf{H}}\left(\varpi\right)\right]=\prod_i\tr\left[\mathds{P}_{\mathbf{H}}\left(\mathcal{C}_i\right)\right] \qquad .
\end{equation}
Using the definition of $\mathds{P}_{\mathbf{H}}$, one can immediately obtain the trace of the cycle $\mathcal{C}_i$ as
\begin{equation}
\tr\left[\mathds{P}_{\mathbf{H}}\left(\mathcal{C}_i\right)\right] = \int\prod_{k=1}^{r_i}\dd g_{a^i_k} ~\delta\left(H_{a^i_k}g_{a^i_k}g^\dag_{a^i_{[k]_{r_i}+1}}\right)=\delta\left(\overleftarrow{\prod_{k=1}^{r_i}}H_{a^i_k}\right) \qquad , 
\end{equation}
where 
\begin{equation}
\overleftarrow{\prod_{k=1}^{r_i}}H_{a^i_k} \equiv H_{a^i_{r_i}}\cdots H_{a^i_2}H_{a^i_1} \qquad .
\end{equation}
Then the trace of $\mathds{P}_{\mathbf{H}}\left(\varpi\right)$ is
\begin{equation}
\mathcal{L}(\pi,\pi';\mathbf{h},\mathbf{h}')=\tr\left[\mathds{P}_{\mathbf{H}}\left(\varpi\right)\right]= \prod_i^{\chi(\varpi)}\delta\left(\overleftarrow{\prod_{k=1}^{r_i}}H_{a^i_k}\right) \qquad .
\end{equation}
\item On the boundary of $\mathcal{N}_0$ and $B$ of $\mathcal{N}_A$, $\mathcal{L}(\pi,\mathds{1};\mathbf{h})$ is a very special case of $\mathcal{L}(\pi,\pi';\mathbf{h},\mathbf{h}')$ where $\pi'=\mathds{1}$ and $\mathbf{h}'=\mathds{1}$
\begin{equation}
\mathcal{L}(\pi,\mathds{1};\mathbf{h})\equiv \mathcal{L}(\pi,\mathds{1};\mathbf{h},\mathds{1}) = \tr\left[\mathds{P}_{\mathbf{h}}\left(\pi\right)\right] = \prod_i^{\chi(\pi)}\delta\left(\overleftarrow{\prod_{k=1}^{r_i}}h_{a^i_k}\right)
\end{equation}
On the boudnary $A$ of $\mathcal{N}_A$, $\mathcal{L}(\pi,\pi_A^0;\mathbf{h})$ corresponds also to a special case of $\mathcal{L}(\pi,\pi';\mathbf{h},\mathbf{h}')$, where $\mathbf{h'}=\mathds{1}$ and $\pi'=\pi_A^0 = \mathcal{C}_0$, which is the $N$-cycle that for any integer $k$ from $1$ to $N$, $\mathcal{C}_0(k)=[k]_N+1$
\begin{equation}
\mathcal{L}(\pi,\pi_A^0;\mathbf{h})\equiv\mathcal{L}(\pi,\mathcal{C}_0;\mathbf{h},\mathds{1})=\tr\left[\mathds{P}_{\mathbf{h}}\left(\mathcal{C}_0^{-1}\pi\right)\right] = \prod_i^{\chi(\mathcal{C}_0^{-1}\pi)}\delta\left(\overleftarrow{\prod_{k=1}^{r_i}}h_{a^i_k}\right) \qquad .
\end{equation}
\end{itemize}
\noindent Altogether, for a given network $\mathcal{N}(\mathbf{h_n},\bm{\pi_n})$, defining the new variables $\varpi\equiv (\pi')^{-1}\pi$ and $\mathbf{H}$ given by \Ref{eq:Hdef} for each link, the corresponding link value is a product of $\chi(\varpi)$ delta function
\begin{equation}\label{eq:L}
\mathcal{L}(\pi,\pi';\mathbf{h},\mathbf{h}')\equiv \mathcal{L}(\varpi;\mathbf{H}) =\tr\left[\mathds{P}_{\mathbf{H}}\left(\varpi\right)\right]= \prod_i^{\chi(\varpi)}\delta\left(\overleftarrow{\prod_{k=1}^{r_i}}H_{a^i_k}\right)
\end{equation}
In particular, when 
$\pi=\pi'$,
the link value $\mathcal{L}(\pi,\pi;\mathbf{h},\mathbf{h}')$ is given by a product of $N$ delta functions as shown in \Ref{eq:LpipiD} and we re-present it here
\begin{equation}\label{eq:LpipiS}
\mathcal{L}(\pi,\pi;\mathbf{h},\mathbf{h}') =\prod_a^N\delta\left((h'_a)^\dag h_a\right)=\prod_a^N\delta\left(H_a\right) \qquad , 
\end{equation}
which is non-zero only when $\mathbf{h}=\mathbf{h}'$.

\noindent So in the end the network is divided into several regions, in each of which $\pi_n$ and $\mathbf{h}_n$ are the same. The links which connect different regions identify boundaries between each pair of different regions, called again domain walls. Corresponding to different domain walls and different assignments of permutation groups to each region, we have different patterns for the given network. We introduce pattern functions $\mathcal{P}_A(\bm{\pi_n})$ and $\mathcal{P}_0(\bm{\pi_n})$ such that
\begin{equation}
\mathcal{P}_A(\bm{\pi_n})\equiv \int\prod_n\dd\mathbf{h}_n~\mathcal{N}_A(\mathbf{h_n},\bm{\pi_n})
\end{equation}
\begin{equation}
\mathcal{P}_0(\bm{\pi_n})\equiv \int\prod_n\dd\mathbf{h}_n~\mathcal{N}_0(\mathbf{h_n},\bm{\pi_n}) \qquad .
\end{equation}
Given a set of $\{\pi_n\}$, $\mathcal{P}_A(\bm{\pi_n})$ and $\mathcal{P}_0(\bm{\pi_n})$ correspond to a certain network pattern with fixed boundary conditions, illustrated in the following figure.
\begin{figure}[t!]
\centering 
\includegraphics[width=.3\textwidth]{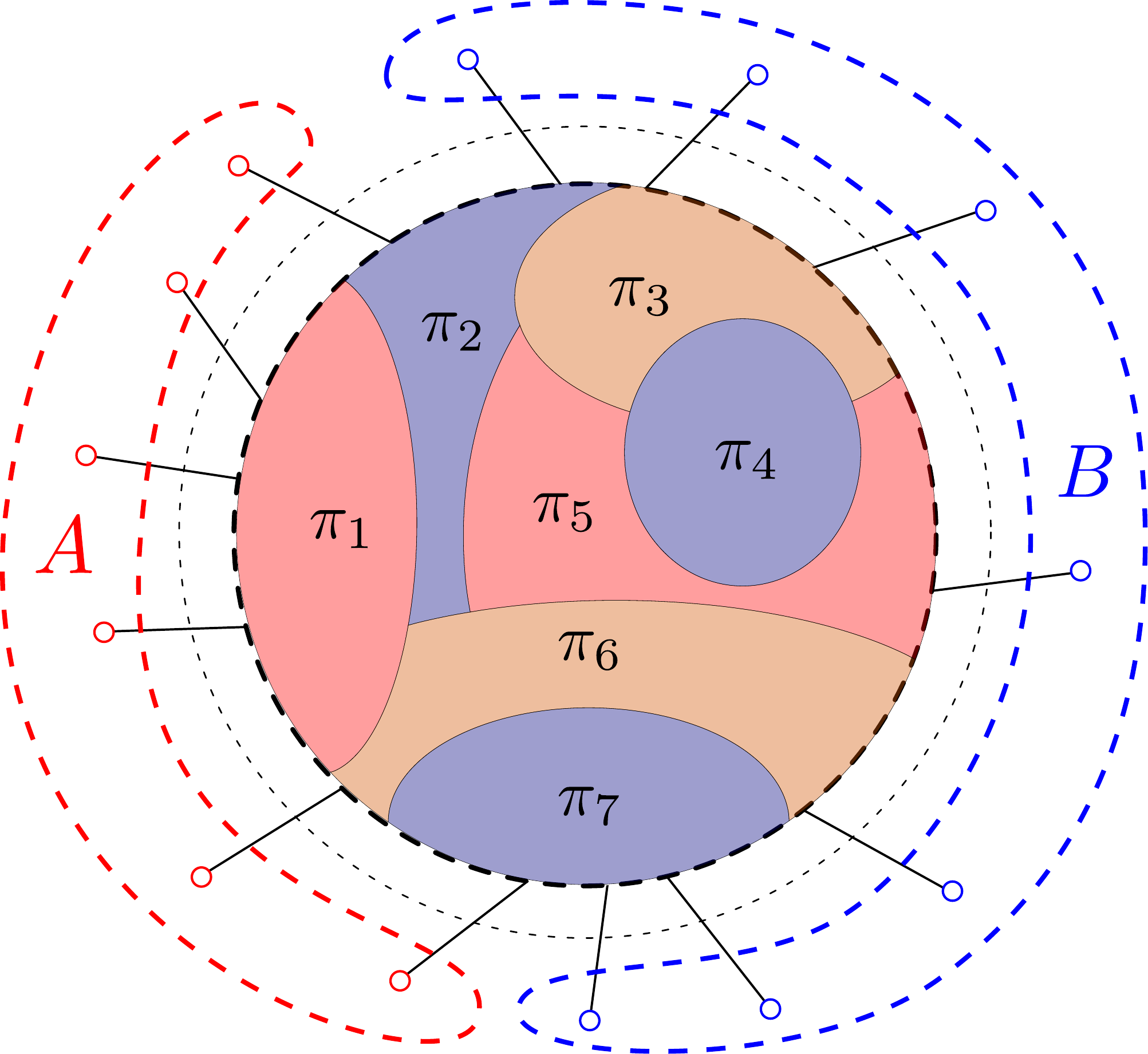}
\caption{\label{fig:PGFT} An example of pattern}
\end{figure}

\noindent More explicitly,
\begin{equation}
\mathcal{P}_A(\bm{\pi_n})=\int\prod_n\dd\mathbf{h}_n~\prod_{\ell\in\Gamma}\left[\prod_{i}^{\chi(\varpi_{\ell})}\delta\left(\overleftarrow{\prod_{k=1}^{r_i}}H_{\ell a^i_k}\right)\right]\prod_{\ell\in A}\left[\prod_i^{\chi(\mathcal{C}_0^{-1}\pi_{n\ell})}\delta\left(\overleftarrow{\prod_{k=1}^{r_i}}h_{\ell a^i_k}\right)\right]\prod_{\ell\in\overline{A}}\left[\prod_i^{\chi(\pi_{n\ell})}\delta\left(\overleftarrow{\prod_{k=1}^{r_i}}h_{\ell a^i_k}\right)\right]
\end{equation}
\begin{equation}
\mathcal{P}_0(\bm{\pi_n})=\int\prod_n\dd\mathbf{h}_n~\prod_{\ell\in\Gamma}\left[\prod_{i}^{\chi(\varpi_{\ell})}\delta\left(\overleftarrow{\prod_{k=1}^{r_i}}H_{\ell a^i_k}\right)\right]\prod_{\ell\in\p\Gamma}\left[\prod_i^{\chi(\pi_{n\ell})}\delta\left(\overleftarrow{\prod_{k=1}^{r_i}}h_{\ell a^i_k}\right)\right]
\end{equation}
They are exactly the amplitudes of a topological BF field theory, with given boundary condition, discretized on a specific 2-complex among the $N$ replica of networks, with each different pattern $\mathcal{P}$ corresponding to a different 2-complex. Each edge of the 2-complex is associated with a holonomy $h_{na}$ that is on node $n$ and the $a$th replica. The two ends of the holonomy are the vertices of the 2-complex. The $h_{na}$ inside a delta function form a loop holonomy, the corresponding edges of which form the face of the 2-complex. Then $Z_N$ and $Z_0^N$ are sum of BF amplitudes with different 2-complexes.
\begin{equation}
Z_N\equiv \mathcal{C}^{V_\Gamma}\sum_{\bm{\pi_n}\in\mathcal{S}_N}\mathcal{P}_A(\bm{\pi_n}),\quad Z_0^N\equiv \mathcal{C}^{V_\Gamma}\sum_{\bm{\pi_n}\in\mathcal{S}_N}\mathcal{P}_0(\bm{\pi_n})
\end{equation}
\noindent It is important to notice that this simple form of the various functions entering the calculation of the entropy, with the emergence of BF-like amplitudes, is {\it not} generic. It follows from the choice of GFT kinetic term, from the approximation used in the calculation of expectation values (neglecting GFT interactions) and from the special type of tensor network, in GFT language, that we have chosen (with simple delta functions associated to the links of the network). More involved, and interesting, cases could be considered.

\

\noindent What we are interested in is the leading term of $Z_N$ and $Z_0^N$, while the dimension $D$ of Hilbert space $\mathbb{H}$ is much larger than $1$. This leads us to seek the most divergent term of $\mathcal{P}_A(\bm{\pi_n})$ and $\mathcal{P}_0(\bm{\pi_n})$. In other words, we need to know the degree of divergence of $\mathcal{P}_A(\bm{\pi_n})$ and $\mathcal{P}_0(\bm{\pi_n})$. 
The divergence degree of BF amplitudes discretized on a lattice has been the subject of a number of works, both in the spin foam an GFT literature (see for example \cite{Freidel:2009hd, Magnen:2009at, Barrett:2006ru}), the most complete analysis being \cite{Bonzom:2010ar,Bonzom:2010zh,Bonzom:2011br}.

\noindent Let us first focus on a sub-region $R$ of the network such that $\pi_n=\pi$ for all nodes $n$ inside of $R$. Suppose that there are $L_i$ links inside $R$ and $L_e$ links connecting with other regions. Since we only consider 4-valent nodes, the number of nodes inside $R$ is
\begin{equation}
V \equiv \frac{1}{4} \left( 2 L_i + L_e \right) = \frac{L_i}{2} + \frac{L_e}{4}
\end{equation}
A minimum spanning tree (MST) $T$, which contains $\#_{T}=V-1$ links, can be found in $R$. 
\begin{equation}
T\equiv \{\ell|\ell\in \text{MST}\}
\end{equation}
According to \Ref{eq:LpipiS}, since $\pi_n=\pi$, there are $N$ delta functions on each link. The integrals over $\mathbf{h}_n$ would eliminate the $(V-1)N$ deltas associated to the MST and leave only one set of $N$ integrals over $\mathbf{h}=\{h_a\}$ and $(L_i/2 - L_e/4+1)N$ $\delta(\mathds{1})$'s. Here we keep indicating the divergent factor as the delta function evaluation originating it, but of course it should be understood more properly as a function of the cut-off used to regularize it. The pattern function of region $R$ is then
\begin{eqnarray}
\mathcal{P}_R(\pi) & \equiv & \int\prod_{n\in R}\dd \mathbf{h}_n ~ \prod_{\ell}^{L_i}\prod_a^N\delta(H_{\ell a})\prod_{\ell}^{L_e}\left[\prod_{i}^{\chi(\varpi_{\ell})}\delta\left(\overleftarrow{\prod_{k=1}^{r_i}}H_{\ell a^i_k}\right)\right]\nonumber\\
&=&\int\prod_{n\in R}\dd \mathbf{h}_n \prod_{\ell\in\text{MST}}\prod_a^N\delta(H_{\ell a})\prod_{\ell\notin\text{MST}}\prod_a^N\delta(H_{\ell a}) \prod_{\ell}^{L_e}\left[\prod_{i}^{\chi(\varpi_{\ell})}\delta\left(\overleftarrow{\prod_{k=1}^{r_i}}H_{\ell a^i_k}\right)\right]\nonumber\\
& = & \left[\delta(\mathds{1})\right]^{\left(\frac{L_i}{2}-\frac{L_e}{4}+1\right)N} \int\dd \mathbf{h} ~\prod_{\ell}^{L_e}\left[\prod_{i}^{\chi(\varpi_{\ell})}\delta\left(\overleftarrow{\prod_{k=1}^{r_i}}H_{\ell a^i_k}\right)\right] \qquad .
\end{eqnarray}
In the calculation, we have used 
\begin{equation}
\int\prod_{n\in R}\dd \mathbf{h}_n \prod_{\ell\in\text{MST}}\prod_a^N\delta(H_{\ell a}) = \int\dd \mathbf{h}
\end{equation}
and ($\mathbf{h}_n=\mathbf{h}$)
\begin{equation}
\prod_{\ell\notin\text{MST}}\prod_a^N\delta(H_{\ell a})=\left[\delta(\mathds{1})\right]^{\left(\frac{L_i}{2}-\frac{L_e}{4}+1\right)N} \qquad .
\end{equation}
The above calculation shows that we can coarse-grain the region $R$ into one single $L_e$-valent node which is colored by $\pi$ and $\mathbf{h}$. 
\begin{equation}\label{eq:CG}
\int\prod_{n\in R}\dd \mathbf{h}_n
\begin{minipage}[h]{0.15\linewidth}
	\vspace{0pt}
	\includegraphics[width=\textwidth]{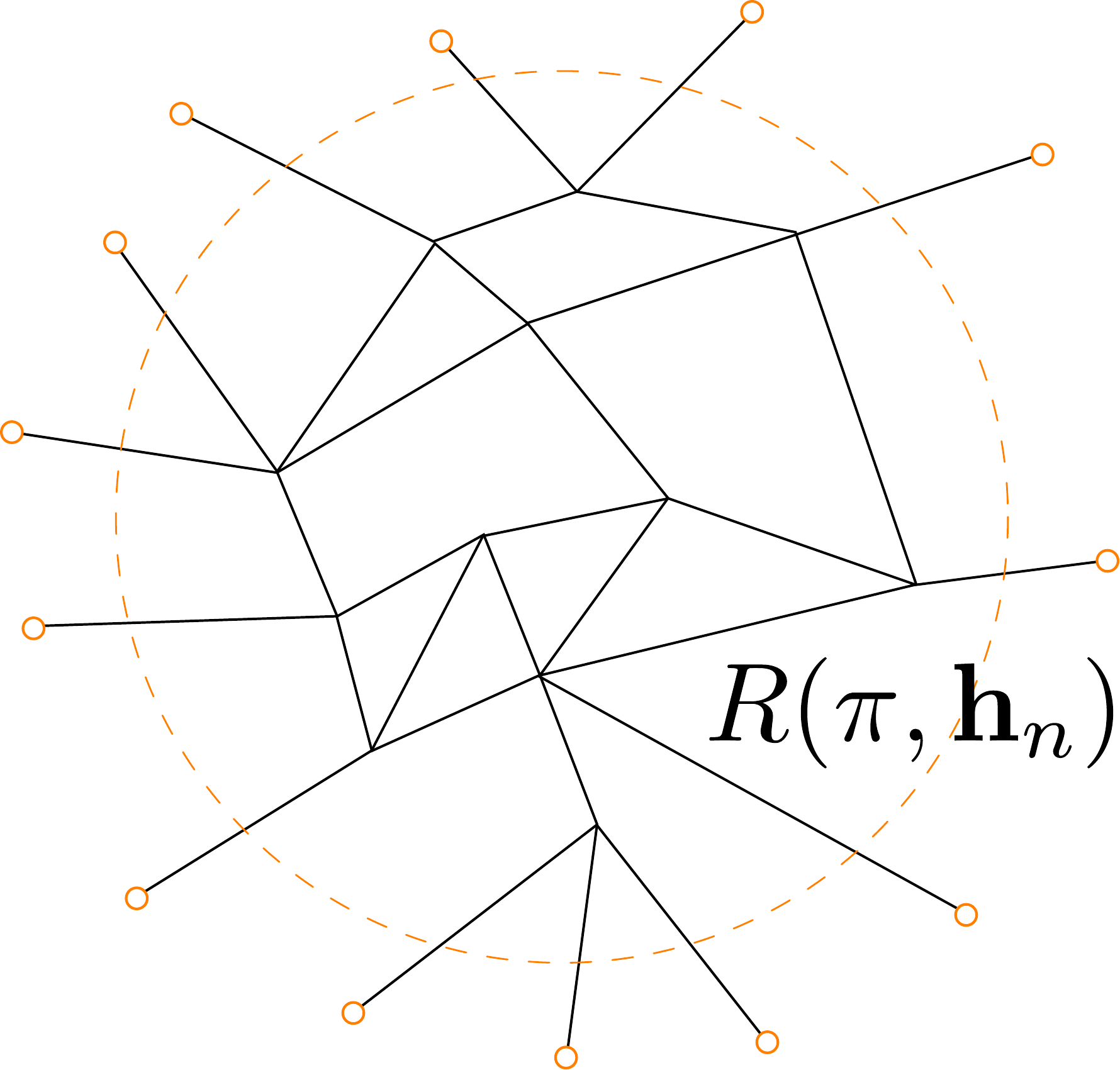}
\end{minipage}
= \int\prod_{n\in R}\dd \mathbf{h}_n
\begin{minipage}[h]{0.15\linewidth}
	\vspace{0pt}
	\includegraphics[width=\textwidth]{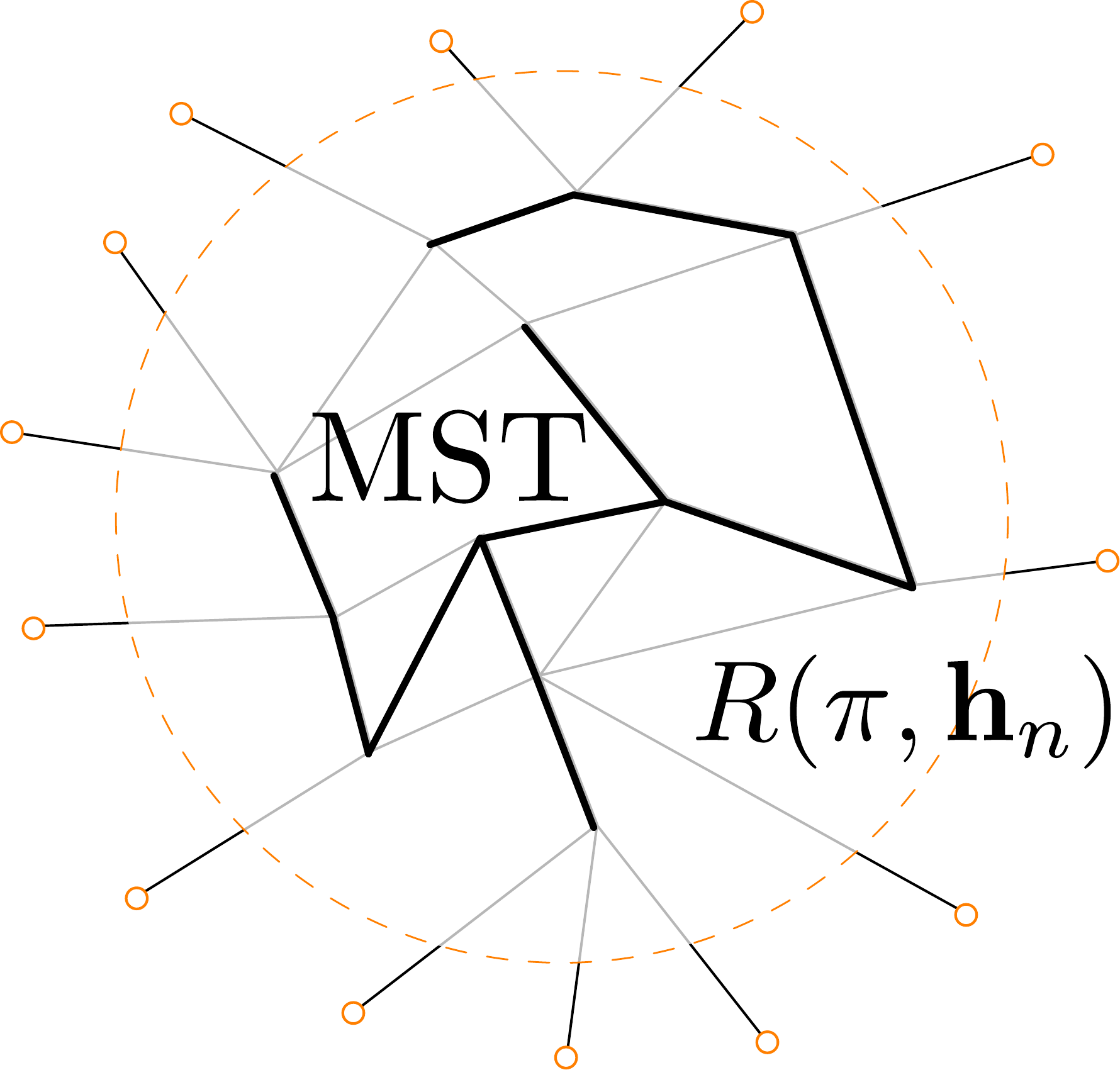}
\end{minipage}
= \int\dd \mathbf{h}
\begin{minipage}[h]{0.1\linewidth}
	\vspace{0pt}
	\includegraphics[width=\textwidth]{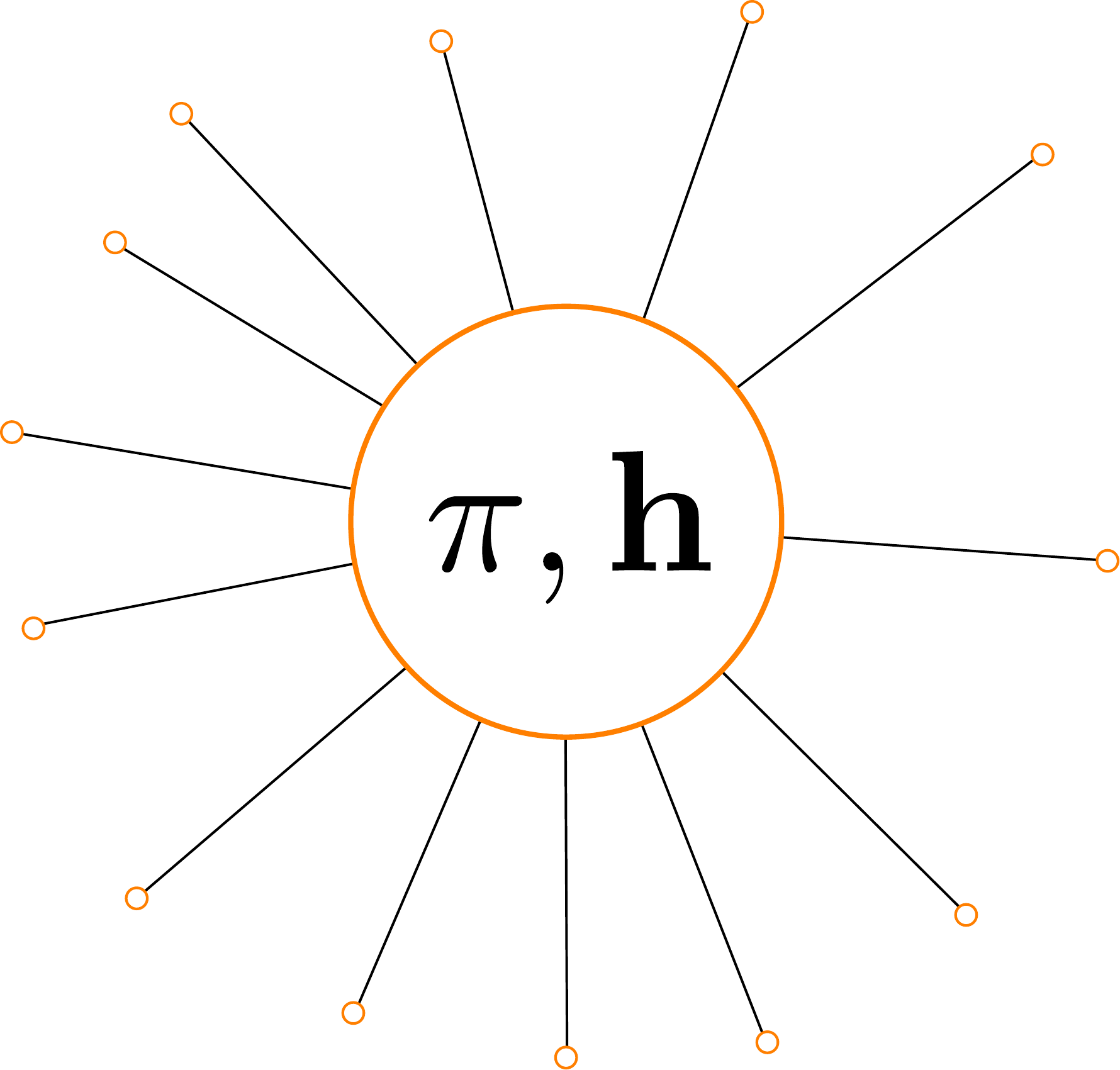}
\end{minipage}\left[\delta(\mathds{1})\right]^{\left(\frac{L_i}{2}-\frac{L_e}{4}+1\right)N}
\end{equation}
So the degree of divergence in region $R$ is: the number of internal links $\#_i=L_i$ subtracted the number of links in the MST $\#_{T}=V-1$, and then times the number of replica $N$,
\begin{eqnarray}\label{eq:DivPi}
\#_R = (\#_i-\#_{T})N= (L_i-V+1)N = \left(\frac{L_i}{2}-\frac{L_e}{4}+1\right)N \qquad .
\end{eqnarray}

\noindent Since the boundary condition of $\mathcal{N}_0$ is $\pi=\mathds{1}$ and $\mathbf{h}=\mathds{1}$, the boundary of $\mathcal{N}_0$ can be coarse-grained into a single node with $\pi=\mathds{1}$ and $\mathbf{h}=\mathds{1}$. The same consideration holds for $\mathcal{N}_A$: its boundary can be coarse-grained into two nodes, one of which corresponds to $A$ with $\pi=\mathcal{C}_0$, $\mathbf{h}=\mathds{1}$ and the other to $B$ with $\pi=\mathds{1}$ and $\mathbf{h}=\mathds{1}$. The corresponding closed graphs are denoted as $\Gamma_0$ and $\Gamma_{AB}$. A certain pattern $\mathcal{P}(\bm{\pi_n})$ divides $\Gamma_0$ and $\Gamma_{AB}$ into $M$ regions that can be coarse-grained into $M$ nodes, each of which is colored with permutation group $\pi_m$ and $N$ integrals over $\mathbf{h}_m$. Denote the graph with pattern $\mathcal{P}(\bm{\pi_n})$ as $\Gamma_0(\bm{\pi_m})$ and $\Gamma_{AB}(\bm{\pi_m})$, and denote the corresponding coarse-grained graphs as $\Gamma^{c}_0(\bm{\pi_m})$ and $\Gamma^c_{AB}(\bm{\pi_m})$.

\begin{figure}[t!]\label{fig:N0cg}
\centering 
\includegraphics[width=.6\textwidth]{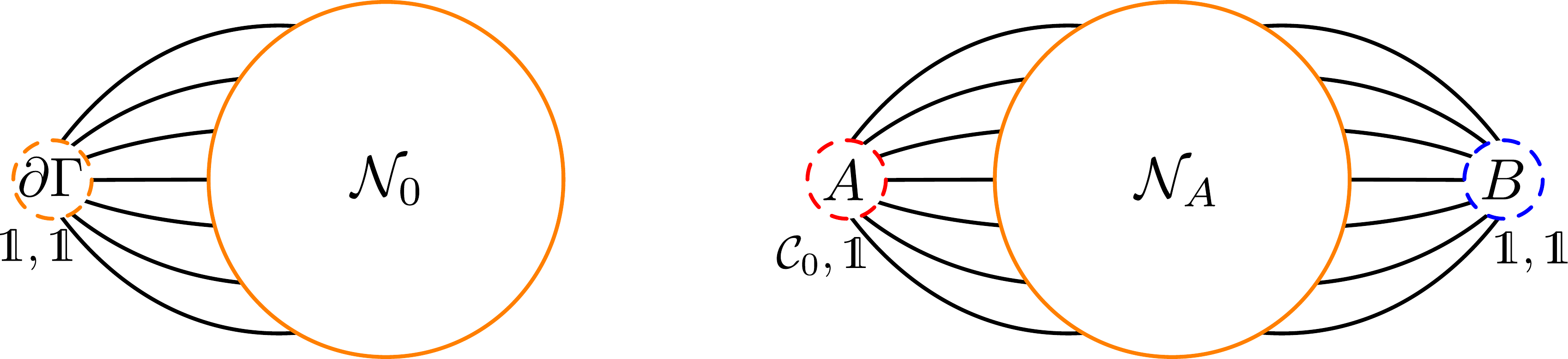}
\caption{The networks in which the boundaries are coarse-grained into nodes.}
\end{figure}

\noindent One can show that, for $\Gamma_0$, the pattern in which all nodes have assigned the same permutation group $\pi=\mathds{1}$ has the highest degree of divergence $\#_0$.
\begin{equation}
\#_0 = (\#_{\ell\in\Gamma_0}-\#_{T_{\Gamma_0}})N
\end{equation}
where $\#_{\ell\in\Gamma_0}$ is the number of links in graph $\Gamma_0$. Let us consider a coarse-grained graph $\Gamma^c_0(\bm{\pi_m})$. Denote the number of links in region $m$, between regions $m$ and $m'$, and between region $m$ and boundary $\p\Gamma$ are $L_m$, $L_{mm'}$ and $L_{m0}$, respectively. The proof goes as follows:
\begin{enumerate}
\item The permutation group on links between coarse-grained nodes $m$ and $m'$ is $\varpi_{mm'}\equiv \pi_m^{-1}\pi_{m'}$. As given by \Ref{eq:L}, the number of the delta functions on one of the links is the number of the disjoint cycles in $\varpi$, which is $\chi(\varpi_{mm'})<N$. Since all links between $m$ and $m'$ are identical, having the same link value, which is given by $\Ref{eq:L}$, when one integrate over $\mathbf{h}_m$ and $\mathbf{h}_{m'}$, only $\chi(\varpi_{mm'})$ deltas will be eliminated and left with $\delta(\mathds{1})$ to the order of $\chi(\varpi_{mm'})(L_{mm'}-1)$ and $2N-\chi(\varpi_{mm'})$ $\mathbf{h}$ integrals. In fact
\begin{eqnarray}\label{eq:step1}
\int \dd\mathbf{h}\dd\mathbf{h}'\left(\tr\left[\mathds{P}_{\mathbf{H}}\left(\varpi\right)\right]\right)^{L} & = & \int \dd\mathbf{h}\dd\mathbf{h}'\prod_i^{\chi(\varpi)}\left[\delta\left(\overleftarrow{\prod_{k=1}^{r_i}}H_{a^i_k}\right)\right]^{L}\nonumber\\
& = & \int\dd\mathbf{h} \prod_i^{\chi(\varpi)}\left[\delta\left(\overleftarrow{\prod_{k=1}^{r_i}}H_{a^i_k}\overrightarrow{\prod_{k=1}^{r_i}}H_{a^i_k}^\dag\right)\right]^{L-1}\nonumber\\
& = & \left[\delta \left(\mathds{1}\right)\right]^{\chi(\varpi)(L-1)}\int\dd\mathbf{h} 
\end{eqnarray}
\item MST can be chosen for $\Gamma_0(\bm{\pi_m})$, $\Gamma^c_0(\bm{\pi_m})$ and $M$ regions. It is obvious that, given a MST $T_m$ for each of the $M$ regions and a MST $T_{\Gamma^c_0(\bm{\pi_m})}$ for $\Gamma^c_0(\bm{\pi_m})$, rooting from the coarse-grained boundary node $\p\Gamma$, a MST $T_{\Gamma_0(\bm{\pi_m})}$ of $\Gamma_0(\bm{\pi_m})$ can be constructed. 
\begin{equation}
T_{\Gamma_0(\bm{\pi_m})} = \bigcup_m^M T_m\cup T_{\Gamma^c_0(\bm{\pi_m})}
\end{equation}
The number of branches of the trees is
\begin{equation}
\#_{T_{\Gamma_0(\bm{\pi_m})}} = \sum_m^M \#_{T_m}+ \#_{T_{\Gamma_0(\bm{\pi_m})}}
\end{equation}
\item The degree of divergence of region $m$ is given by \Ref{eq:DivPi}
\begin{equation}
\#_m = (L_m - \#_{T_m})N
\end{equation}
Similarly, for the divergence degree $\#_{\Gamma^c_0(\mathds{1})}$ of the pattern where all coarse-grained nodes have the same permutation $\pi_m=\mathds{1}$ is
\begin{equation}
\#_{\Gamma^c_0(\mathds{1})} = \left(\sum_{0\leq m<m'\leq M}L_{mm'}-\#_{T_{\Gamma^c_0(\bm{\pi_m})}}\right)N
\end{equation}
The degree of divergence of $\Gamma^c_0(\bm{\pi_m})$ is smaller than $\#_{\Gamma^c_0(\mathds{1})}$
\begin{equation}
\#_{\Gamma^c_0(\bm{\pi_m})} < \#_{\Gamma^c_0(\mathds{1})}
\end{equation}
This is because, after evaluating the delta functions on the MST $\Gamma^c_0(\bm{\pi_m})$ in accordance with \Ref{eq:step1}, there are still $MN-\sum_{(mm')\in T_{\Gamma^c_0(\bm{\pi_m})}}\chi(\varpi_{mm'})$ integrals over $\mathbf{h}$. Performing these integrals makes the degree of divergence of $\Gamma^c_0(\bm{\pi_m})$ not bigger than the following quantity
\begin{eqnarray}
\#_{\Gamma^c_0(\bm{\pi_m})} & \leq & \sum_{0\leq m<m'\leq M}L_{mm'}\chi(\varpi_{mm'})-\sum_{(mm')\in T_{\Gamma^c_0(\bm{\pi_m})}}\chi(\varpi_{mm'})\\
& = & \sum_{(mm')\notin T_{\Gamma^c_0(\bm{\pi_m})}}L_{mm'}\chi(\varpi_{mm'})\\ 
&& +\sum_{(mm')\in T_{\Gamma^c_0(\bm{\pi_m})}}(L_{mm'}-1)\chi(\varpi_{mm'})
\end{eqnarray}
which is definitely smaller than $\#_{\Gamma_0(\mathds{1})}$ because $\chi(\varpi_{mm'})<N$.
\item So the divergence degree of $\Gamma_0(\bm{\pi_m})$ is smaller than the divergence degree $\#_0$ for the pattern where all nodes have the same permutation.
\begin{equation}
\#_{\Gamma_0(\bm{\pi_m})}=\#_{\Gamma^c_0(\bm{\pi_m})}+\sum_m^M\#_m<\#_{\Gamma^c_0(\mathds{1})}+\sum_m^M\#_m = (\#_{\ell\in\Gamma_0}-\#_{T_{\Gamma_0}})N =  \#_0
\end{equation}
The leading term of $Z_0^N$ is $\mathcal{P}_0(\mathds{1})$, whose divergence degree is $\#_0$.
\begin{equation}
Z_0^N = \mathcal{C}^{V_{\Gamma}}[\delta(\mathds{1})]^{\#_0}\left[1+\mathcal{O}(\delta^{-1}(\mathds{1}))+\mathcal{O}(\lambda)\right] \qquad .
\end{equation}
\end{enumerate}

\noindent For $Z_N$, since the boundary is separated into two parts, the most divergent pattern 
$\mathcal{P}_A(\bm{\pi_n})$ is the one such that its corresponding coarse-grained graph has only two coarse-grained nodes $A$ and $B$, which are connected by the minimum number of links $\min(\#_{\ell\in\p_{AB}})$, whose divergence degree is
\begin{eqnarray}
\#_{AB} &=& \#_{A}+\#_{B}+\min(\#_{\ell\in\p_{AB}})\nonumber\\
&=& \left(\#_{\ell\in\Gamma_{AB}}-\min(\#_{\ell\in\p_{AB}})-\#_{T_{A}}-\#_{T_{B}}\right)N+\min(\#_{\ell\in\p_{AB}})\nonumber\\
& = & \left(\#_{\ell\in\Gamma_{AB}}-\#_{T_{A}}-\#_{T_{B}}\right)N+(1-N)\min(\#_{\ell\in\p_{AB}})\nonumber\\
& = & \#_0 + (1-N)\min(\#_{\ell\in\p_{AB}})
\end{eqnarray}
where the second equality is in terms of \Ref{eq:DivPi} and the forth equality is because $\#_{\ell\in\Gamma_{AB}}=\#_{\ell\in\Gamma_{0}}$ and $\#_{T_{A}}+\#_{T_{B}}=\#_{T_{\Gamma_0}}$\footnote{Since the boundary is coarse-grained into two nodes in $\Gamma_{AB}$, there are one more node in $\Gamma_{AB}$ than in $\Gamma_0$,
\begin{equation}
V_{\Gamma_{AB}} = V_{\Gamma_0} + 1
\end{equation}
Thus the number of the branches of the MST in $A$ and $B$ is equal to the number of the MST branches in $\Gamma_0$
\begin{equation}
\#_A + \#_B = (V_A - 1) - (V_B - 1) = V_{\Gamma_{AB}} - 2 = V_{\Gamma_0} - 1 = \#_{T_{\Gamma_0}}\quad .
\end{equation}
}. 

\noindent Let us consider a graph $\Gamma_{AB}(\bm{\pi_m})$ and its corresponding coarse-grained graph $\Gamma_{AB}^c(\bm{\pi_m})$. The divergence degree of $\Gamma_{AB}(\bm{\pi_m})$ is given as
\begin{equation}
\#_{\Gamma_{AB}(\bm{\pi_m})} = \#_{\Gamma_{AB}^c(\bm{\pi_m})} + \sum_{m = \{1,\cdots M,A,B\}}\#_m
\end{equation}
where $\#_m$ is given by \Ref{eq:DivPi}
\begin{equation}
\#_m = (L_m - \#_{T_m})N
\end{equation}
Adapting the same argument as for $Z_0^N$, because of the integral over $\mathbf{h_n}$, $\#_{\Gamma_{AB}^c(\bm{\pi_m})}$ should not be bigger than the following quantity
\begin{eqnarray}
\#_{\Gamma_{AB}^c(\bm{\pi_m})} & \leq &  \sum_{(mm')\notin T^A_{\Gamma_{AB}^c(\bm{\pi_m})},T^B_{\Gamma_{AB}^c(\bm{\pi_m})}}L_{mm'}\chi(\varpi_{mm'})\nonumber\\
& + & \sum_{(mm')\in T^A_{\Gamma_{AB}^c(\bm{\pi_m})}~ \text{or} ~T^B_{\Gamma_{AB}^c(\bm{\pi_m})}}(L_{mm'}-1)\chi(\varpi_{mm'})
\end{eqnarray}
where we assume $m<m'$ in order to avoid double counting, and $T^A_{\Gamma_{AB}^c(\bm{\pi_m})}$ and $T^B_{\Gamma_{AB}^c(\bm{\pi_m})}$ are the MST rooting from coarse-grained nodes $A$ and $B$, respectively. The right hand side of the above formula corresponds to the divergence degree of pattern $\mathcal{P}_A(\bm{\pi_m})$ on a graph $\overline{\Gamma_{AB}^c(\bm{\pi_m})}$ with all $\mathbf{h_n}=\mathds{1}$, which differs from $\Gamma_{AB}^c(\bm{\pi_m})$ by $T^A_{\Gamma_{AB}^c(\bm{\pi_m})}$ and $T^B_{\Gamma_{AB}^c(\bm{\pi_m})}$, i.e.
\begin{equation}
\overline{\Gamma_{AB}^c(\bm{\pi_m})} \equiv \Gamma_{AB}^c(\bm{\pi_m})\setminus \{T^A_{\Gamma_{AB}^c(\bm{\pi_m})},~T^B_{\Gamma_{AB}^c(\bm{\pi_m})}\}
\end{equation}
As presented in section 2, the major difference between \cite{Hayden:2016cfa,Han2016} and our paper is that we are considering the gauge transformation $\mathbf{h}_n$ on each node $n$. When all $\mathbf{h_n}$ are set to be the identity, our $Z_N$ and $Z_0^N$ simplify to the ones in \cite{Hayden:2016cfa,Han2016} up to overall normalization. In this case, as shown in \cite{Hayden:2016cfa,Han2016}, the patterns which gives only one domain wall for $\Gamma_{AB}$ have higher divergence degree than the divergent degree of multi-domain walls, which in our language means that the patterns whose corresponding coarse-grained graph contains only two coarse-grained nodes are more divergent than the patterns $\mathcal{P}_A(\bm{\pi_m})$, which give more than two coarse-grained nodes. So the divergence degree of the pattern $\mathcal{P}_A(\bm{\pi_m})$ on the graph $\overline{\Gamma_{AB}(\bm{\pi_m})}$ is not bigger than the pattern $\mathcal{P}_A(\mathds{1},\mathcal{C}_0)$. So we have 
\begin{eqnarray}
\#_{\Gamma_{AB}(\bm{\pi_m})} &=& \#_{\Gamma_{AB}^c(\bm{\pi_m})} + \sum_{m = \{1,\cdots M,A,B\}}\#_m\nonumber\\
& \leq & \#_{\overline{\Gamma_{AB}^c(\bm{\pi_m})}} + \sum_{m = \{1,\cdots M,A,B\}}\#_{\overline{m}}\nonumber\\
& \leq & \#_{A}+\#_{B}+\#_{\ell\in\p_{AB}}= \#_0 + (1-N)\#_{\ell\in\p_{AB}}\nonumber\\
& \leq & \#_{AB} = \#_0 + (1-N)\min(\#_{\ell\in\p_{AB}})
\end{eqnarray}
It follows that the amplitude $Z_N$ is
\begin{equation}
Z_N = \mathcal{C}^{V_{\Gamma}}[\delta(\mathds{1})]^{\#_0 + (1-N)\min(\#_{\ell\in\p_{AB}})}\left[1+\mathcal{O}(\delta^{-1}(\mathds{1}))+\mathcal{O}(\lambda)\right] \qquad .
\end{equation}

\noindent Finally, the $N$th order R\'{e}nyi entropy $S_N$ is then:
\begin{equation}
\ex^{(1-N)S_N} = \frac{Z_N}{Z_0^N} = [\delta(\mathds{1})]^{(1-N)\min(\#_{\ell\in\p_{AB}})}\left[1+\mathcal{O}(\delta^{-1}(\mathds{1}))+\mathcal{O}(\lambda)\right] \qquad .
\end{equation}
When $N$ goes to $1$, $S_N$ becomes the entanglement entropy $S_{\text{EE}}$. The leading term of the entanglement entropy $S_{\text{EE}}$ is therefore
\begin{equation}
S_{\text{EE}} = \min(\#_{\ell\in\p_{AB}})\ln\delta(\mathds{1}) \qquad ,
\end{equation}
which can be understood as the Ryu-Takayanagi formula in a GFT context. The minimal number of links $\min(\#_{\ell\in\p_{AB}})$ represents the minimal surface area which separates the bulk. 
\

\noindent Before moving on to a different derivation of the same result, we want to clarify the interpretation of this calculation. 

\noindent The definition of the expectation value \Ref{eq:E} in the GFT language shows that the exponential of $S_N$ can be interpreted as a GFT $2N$-point function, at least within the limits of the approximation made, focusing on the average over group field functions at each node, without recasting the whole generalized tensor network as a GFT correlation function. As shown in previous sections, the GFT amplitudes can in turn be written, by standard perturbative expansion, as a sum of Feynman amplitudes associated to Feynman diagrams, each of which corresponds to a different discretized \lq\lq space-time\rq\rq with fixed boundary, with the Feynman amplitude defining (for quantum gravity models) a lattice path integral for gravity discretised on the corresponding cellular complex. This allows a tentative (and partial) interpretation of the entropy formula we have derived, in geometric spatiotemporal terms. It implies, in fact, that, in the calculation of the entropy, not only the information of a time-slice of a space-time is considered, as encoded in a given network, but also its full quantum dynamics. This, at least, is true when the complete GFT partition function (for quantum gravity models) is employed in the computation of the entropy. The leading term, the free GFT amplitude, captures only a sector of that full quantum dynamics. With the specific (trivial) choice of kinetic term we have used, the quantum dynamics can at best correspond to (summing over) static space-times. When $N$ goes to $1$, in particular, the amplitude becomes the trivial propagation of GFT states, with any given network propagating to itself. This corresponds exactly to the context (static space-time) in which the Ryu-Takayanagi formula is usually derived. In other words, our calculation provides a realization of the Ryu-Takayanagi formula, at least in one extremely simple case, within the full dynamics of a non-perturbative approach to quantum gravity, the group field theory formalism, which can also be seen as a different definition of loop quantum gravity. Our result also shows that the same formalism allows to compute non-perturbative quantum gravity corrections to the Ryu-Takayanagi formula, by including the contributions from the GFT interaction term into the amplitude (as well as considering different choices for the GFT kinetic term).

\section{Ryu-Takayanaki formula for Spin-Network states}

We want now to perform a similar calculation of the Ryu-Takanayagi entropy using a different truncation of a generic GFT state, reformulated as a tensor network. We use a given linear combination of spin networks, corresponding to a specific assignment of spins to the links of the network, and thus to the tensors associated to its nodes.
 
\noindent As presented in Section 2, the spin representation of a GFT network is spin-network, in which each node is colored by a tensor $\phi^{\mathbf{j}i}_\mathbf{m}$
\begin{equation}
\phi^{\mathbf{j}i}_\mathbf{m} = \sum_{\mathbf{p}}\overline{i_{\mathbf{p}}}\psi^{\mathbf{j}}_{\mathbf{pm}},\qquad \ket{\phi^{\mathbf{j}i}}= \sum_{\mathbf{m}}\phi^{\mathbf{j}i}_\mathbf{m}\ket{\mathbf{j},\mathbf{m}} \in \bigotimes_{\ell}\mathbb{H}_{j_{\ell}}
\end{equation}
and each link is colored by matrix $M^j_{mm'}$
\begin{equation}
\ket{M^j} = \sum_{mm'}M^j_{mm'}\ket{j,m}\otimes\ket{j,m'}\in \mathbb{H}_j^{\otimes 2} \qquad , 
\end{equation}
where $\mathbb{H}_j$ is the spin-$j$ irreducible representation of $\SU(2)$. 

\noindent A spin-network has a clear geometric interpretation. The graph $\Gamma$ is the dual of a 3d cellular complex. When all nodes are 4-valent, the graph is dual to a 3d simplicial complex. Each node is dual to a tetrahedron and each link is dual to a triangle. The area of the triangle is given by the spin-$j$ irreducible representation associated with the dual link of the triangle. More precisely, the area $\mathcal{A}_{\ell}$ is
\begin{equation}
\mathcal{A}_\ell = 8\pi\gamma\sqrt{j_{\ell}(j_{\ell}+1)}\ell_p^2 \qquad ,
\end{equation}
where $\Gamma$ is the Barbero-Immirzi parameter and $\ell_p$ is the Planck length (while this results follows both from a canonical quantization of General Relativity in the continuum, and from the geometric quantization of simplicial geometries, the identification of the length scale with the Planck length is, of course, natural from the first perspective only). 

A detailed analysis (see e.g. \cite{Han:2013ina,Han:2013hna,Han:2013tap}) shows that the semi-classical regime of loop quantum gravity states, in which the Regge-Einstein gravity can be recovered, at least at the kinematical level, in the sense of approximating smooth geometries with simplicial ones, is at a scale intermediate between the Planck scale $\ell_p$ and the average background curvature scale $L_\Lambda$, which means that if we are working on this regime, area $\mathcal{A}_{\ell}$ of the triangle should be 
\begin{equation}
\ell_p^2\ll\mathcal{A}_\ell \ll L^2_\Lambda
\end{equation}\label{eq:semilow2}
Together with the relation $\mathcal{A}_\ell / L^2_\Lambda\sim \gamma^{-1}j^{1/2}\ll 1$ uncovered in \cite{Han:2013ina}, the above regime is equivalent to
\begin{equation}\label{eq:semilow}
\frac{1}{j}\ll\gamma\ll\frac{1}{j^{1/2}}
\end{equation}

\noindent In a semi-classical regime, then, one has $\mathcal{A}_{\ell} \approx \gamma j_{\ell}\ell^2_p$.

\noindent In \cite{Hamma:2015xla}, a special choice of $M^j_{mm'}$ 
\begin{equation}\label{eq:M}
M^j_{mm'} = \bra{j,m} n^\dag \ex^{-\pi\gamma L_z-\frac{\exp(1-2\pi\gamma L_z)}{4\pi\gamma}}n'\ket{j,m'}
\end{equation}
has been considered, with the property that the leading order of the entanglement entropy between the two $\mathbb{H}_j$ on a link is proportional to the same area $\mathcal{A}_{\ell} \approx \gamma j_{\ell}\ell^2_p$ in the semi-classical regime. In \Ref{eq:M}, $n$ and $n'$ are $\SU(2)$ elements; $L_z$ is the $SU(2)$ generator in $z$-axis. 
We use the same choice for $M^j_{mm'}$ in our calculation to obtain the Ryu-Takayanaki formula.

\noindent Considering the same graph $\Gamma$ as in the previous subsection, the spin-network state $\ket{\Psi^{\mathbf{ji}}}$ and its corresponding density matrix $\rho$ are given as
\begin{equation}
\ket{\Psi_{\Gamma}^{\mathbf{ji}}}\equiv \bigotimes_\ell\bra{M^{j_\ell}}\bigotimes_{n}\ket{\phi_n^{~\mathbf{j}_ni_n}}, \quad \rho\equiv \ket{\Psi_{\Gamma}^{\mathbf{ji}}}\bra{\Psi_{\Gamma}^{\mathbf{ji}}} \qquad .
\end{equation}
Just as in the previous calculation, we divide boundary $\p\Gamma$ into two parts $A$ and $B$. The $N$th R\'{e}nyi entropy is 
\begin{equation}
\ex^{(1-N)S_N} = \frac{Z_N}{Z_0^N}=\frac{\mathbb{E}\tr [\rho^{\otimes N}\mathds{P}(\pi^0_A;N,d)]}{\mathbb{E}(\tr\rho)^N} = \frac{\tr \left[\bigotimes_{\ell}\rho^N_\ell\bigotimes_n\mathbb{E}(\rho^N_n)\mathds{P}(\pi^0_A;N,d)\right]}{\tr \left[\bigotimes_{\ell}\rho^N_\ell\bigotimes_n\mathbb{E}(\rho^N_n)\right]} \qquad .
\end{equation}

\noindent The first key step is to calculate $\mathbb{E}(\rho^N_n)$. Because the gauge symmetry is already encoded in the intertwiner $\overline{i}$ for $\phi^{\mathbf{j}i}_\mathbf{m}$, $\phi^{\mathbf{j}i}_\mathbf{m}$ is not a gauge symmetric tensor, which is in the invariant space of $\mathbb{H}^{\otimes 4}$ as introduced in Section 2, but rather an ordinary tensor in $\otimes_\ell\mathbb{H}_{j_\ell}$. So the averge over $\rho^N_n$ can be performed in the same way as the one shown in \cite{Hamma:2015xla}:
\begin{equation}
\mathbb{E}(\rho^N_n) \equiv \int \dd\phi^{\mathbf{j}i} f(\phi^{\mathbf{j}i})\left(\ket{\phi^{\mathbf{j}i}}\bra{\phi^{\mathbf{j}i}}\right)^{\otimes N}\equiv \int_{\U(D)}\dd U f(\phi^{\mathbf{j}i})\left(U\ket{\phi_0^{\mathbf{j}i}}\bra{\phi_0^{\mathbf{j}i}}U^{\dag}\right)^{\otimes N} \qquad , 
\end{equation}
where $f(\phi^{\mathbf{j}i})$ is a distribution of $\phi^{\mathbf{j}i}$ and $U$ is the group element in the unitary group $\U(D)$, in which $D= \prod_{\ell\in n}d_{j_\ell}$. $f(\phi^{\mathbf{j}i})$ is invariant under the transformation of $\U(D)$ and in our following calculation we focus on either the uniform or the Gaussian distribution, which keep the main calculation unchanged up to an overall normalization that will be canceled in the final result.  

\noindent Because of Schur's lemma, $\mathbb{E}(\rho^N_n)$ is the invariant tensor in $\left(\otimes_\ell\mathbb{H}_{j_\ell}\right)^{\otimes N}$, which can be written as a sum of permutations
\begin{equation}
\mathbb{E}(\rho^N_n) = \mathcal{C}\sum_{\pi_n\in\mathcal{S}_N}\mathds{P}(\pi_n;N,D)=\mathcal{C}\sum_{\pi_n\in\mathcal{S}_N}\prod_{\ell\in n}\mathds{P}(\pi_n;N,d_{j_{\ell}})
\end{equation}
where $\mathcal{C}$ is an normalization factor which depends on the distribution. Then $Z_N$ and $Z_0^N$ can be written as a sum of different patterns $\mathcal{P}(\bm{\pi_n})$
\begin{equation}
Z_N = \mathcal{C}^{V_\Gamma}\sum_{\bm{\pi_n}\in\mathcal{S}_N}\mathcal{P}_A(\bm{\pi_n}),\quad Z_0^N = \mathcal{C}^{V_\Gamma}\sum_{\bm{\pi_n}\in\mathcal{S}_N}\mathcal{P}_0(\bm{\pi_n})
\end{equation}
where $\#$ is the number of nodes in $\Gamma$. $\mathcal{P}_A(\bm{\pi_n})$ and $\mathcal{P}_0(\bm{\pi_n})$ can be written as products of link values $\mathcal{L}(\pi_n,\pi'_n)$
\begin{equation}
\mathcal{P}_A(\bm{\pi_n})= \prod_{\ell\in\Gamma} \mathcal{L_\ell}(\pi_{n},\pi_{n'})\prod_{\ell\in A}\mathcal{L}_\ell(\pi_n,\pi_A^0)\prod_{\ell\in B}\mathcal{L}_{\ell}(\pi_n,\mathds{1})
\end{equation}
\begin{equation}
\mathcal{P}_0(\bm{\pi_n})= \prod_{\ell\in\Gamma} \mathcal{L_\ell}(\pi_{n},\pi_{n'})\prod_{\ell\in \p\Gamma}\mathcal{L}_{\ell}(\pi_n,\mathds{1})
\end{equation}
where $\mathcal{L_\ell}(\pi,\pi')$ is defined as
\begin{equation}\label{eq:Lsn}
\mathcal{L_\ell}(\pi,\pi') \equiv \tr[\mathds{P}(\pi;N,d_{j_{\ell}})\rho_\ell^N\mathds{P}(\pi';N,d_{j_{\ell}})]
\end{equation}

Suppose $\varpi\equiv (\pi')^{-1}\pi=\prod_i\mathcal{C}_i$, where $\mathcal{C}_i$ is an $r_i$-cycle, and impose \Ref{eq:M} into \Ref{eq:Lsn}. $\mathcal{L}(\pi,\pi')$ becomes
\begin{equation}\label{eq:Lchi}
\mathcal{L}(\pi,\pi') = \prod_{i=1}^{\chi(\varpi)}\chi_{j}\left(\ex^{-r_i2\pi\gamma L_z-r_i\frac{\exp(1-2\pi\gamma L_z)}{2\pi\gamma}}\right)
\end{equation}
In the semi-classical regime \Ref{eq:semilow}, the leading contribution of $\mathcal{L}(\pi,\pi')$ is obtained as
\begin{eqnarray}\label{eq:Lj}
\mathcal{L}(\pi,\pi') &\approx & \prod_{i=1}^{\chi(\varpi)}\frac{1}{r_i}\ex^{-1+(1-r_i)2\pi\gamma j-r_i\frac{\exp(1-2\pi\gamma j)}{2\pi\gamma}}\nonumber\\
&=& \ex^{-\chi(\varpi)+(\chi(\varpi)-N)2\pi\gamma j-N\frac{\exp(1-2\pi\gamma j)}{2\pi\gamma}}\prod_{i=1}^{\chi(\varpi)}\frac{1}{r_i}
\end{eqnarray}
A detailed calculation from \Ref{eq:Lchi} to \Ref{eq:Lj} can be found in the appendix. When $\varpi = \mathds{1}$, i.e. $\pi=\pi'$ and $\chi(\varpi)=N$, $\mathcal{L}(\pi,\pi)$ is then
\begin{equation}
\mathcal{L}(\pi,\pi) \approx \ex^{-N-N\frac{\exp(1-2\pi\gamma j)}{2\pi\gamma}}
\end{equation}
It is straightforward to check that $\mathcal{L}(\pi,\pi)\geq \mathcal{L}(\pi_1,\pi_2)$. In fact, because the sum of $r_i$ equals to $N$, $\mathcal{L}(\pi,\pi)$ can be rewritten as
\begin{equation}
\mathcal{L}(\pi,\pi) = \prod_{i=1}^{\chi(\varpi)}\ex^{-r_i-r_i\frac{\exp(1-2\pi\gamma j)}{2\pi\gamma}}
\end{equation}
Then the ratio between $\mathcal{L}(\pi_1,\pi_2)$ and $\mathcal{L}(\pi,\pi)$ is
\begin{equation}\label{eq:LL}
\frac{\mathcal{L}(\pi_1,\pi_2)}{\mathcal{L}(\pi,\pi)}=\prod_{i=1}^{\chi(\varpi)}\frac{\ex^{-1+(1-r_i)2\pi\gamma j-r_i\frac{\exp(1-2\pi\gamma j)}{2\pi\gamma}}}{r_i\ex^{-r_i-r_i\frac{\exp(1-2\pi\gamma j)}{2\pi\gamma}}}=\prod_{i=1}^{\chi(\varpi)}\frac{\ex^{(1-r_i)(2\pi\gamma j-1)}}{r_i} \leq 1 \qquad . 
\end{equation}
The last inequality holds because $r_i\geq 1$ and in the regime \Ref{eq:semilow} $\gamma j \gg 1$. The equality holds if and only if $\pi_1 = \pi_2$. 

\noindent If we assume that all $j_\ell$ are in the same order of magnitude, because of \Ref{eq:LL}, one can observe immediately that the leading term of $Z_0^N$ is $\mathcal{P}_0(\bm{\mathds{1}})$, i.e. the permutation group for all nodes is $\pi_n=\mathds{1}$. Suppose there are $L_i$ internal links and $L_e$ external links in $\Gamma$, then 
\begin{equation}
Z_0^N \approx \mathcal{C}^{V_\Gamma}\prod_\ell\ex^{-N-N\frac{\exp(1-2\pi\gamma j_\ell)}{2\pi\gamma}} \qquad .
\end{equation}
The $N$th order R\'{e}nyi entropy becomes
\begin{equation}
\ex^{(1-N)S_N}=\frac{Z_N}{Z_0^N} \approx \sum_{\bm{\pi_n}}\prod_{\ell}\prod_i^{\chi(\varpi_{\ell})}\frac{\ex^{(1-r_i)(2\pi\gamma j-1)}}{r_i} \qquad .
\end{equation}
As shown in \cite{Han2016}, in order for the single domain wall pattern to contribute the most to the R\'{e}nyi entropy, when three domain walls intersect, they should satisfy
\begin{equation}
\prod_i^{\chi(\varpi_{1})}\frac{\ex^{(1-r_i)(2\pi\gamma j-1)}}{r_i}\geq \prod_i^{\chi(\varpi_{2})}\frac{\ex^{(1-r_i)(2\pi\gamma j-1)}}{r_i}\prod_i^{\chi(\varpi_{3})}\frac{\ex^{(1-r_i)(2\pi\gamma j-1)}}{r_i} \qquad , 
\end{equation}
where $\varpi_1\varpi_2\varpi_3 = 1$. 
The above inequality can be simplified to
\begin{equation}
\ex^{[C(\varpi_2)+C(\varpi_3)-C(\varpi_1)](2\pi \g j - 1)}\frac{\prod_i^{\chi(\varpi_{2})}r_i\prod_i^{\chi(\varpi_{3})}r_i}{\prod_i^{\chi(\varpi_{1})}r_i}\geq 1 \qquad , 
\end{equation}
where $C(\varpi)$ is the Cayley weight of a permutation $\varpi$ which satisfies the triangular inequality $C(\varpi_1\varpi_2)\leq C(\varpi_1) + C(\varpi_2)$. In general, when $C(\varpi_1)< C(\varpi_2) + C(\varpi_3)$, the above inequality is satisfied because when $\gamma j \gg 1$ the exponential part of the inequality dominant. When $C(\varpi_1) = C(\varpi_2) + C(\varpi_3)$, one can check that the inequality is satisfied at least for $N \leq 3$\footnote{Using the geometric inequality, the left hand side of the above inequality becomes
\begin{eqnarray}
\frac{\prod_i^{\chi(\varpi_{2})}r_i\prod_i^{\chi(\varpi_{3})}r_i}{\prod_i^{\chi(\varpi_{1})}r_i} & \geq & \frac{(C(\varpi_2)+1)(C(\varpi_1)-C(\varpi_2)+1)}{\left(\frac{N}{N-C(\varpi_1)}\right)^{N-C(\varpi_1)}} \nonumber\\
& \geq &  (C(\varpi_1)+1)\left(\frac{N-C(\varpi_1)}{N}\right)^{N-C(\varpi_1)}
\end{eqnarray}
This simplification is very rough since one has to keep $\frac{N}{N-C(\varpi_1)}$ to be integer. Even in this approximate situation, we could find that it is bigger than 1 when $N$ is a bit smaller than $3$.}. Since we are only interested in the entropy while taking the limit $N\rightarrow 1$, this inequality is well satisfied. The R\'{e}nyi entropy $S_N$ for small $N$ is given as
\begin{equation}
\ex^{(1-N)S_N} \approx \prod_{\ell\in\p_{AB}}\exp\left((1-N)(2\pi\g j_{\ell} -1)-\ln N\right) \qquad .
\end{equation} 
When $N$ goes to zero, we have
\begin{equation}
S_{\text{EE}}\approx \sum_{\ell\in\p_{AB}}\left[2\pi\g j_{\ell} -1 -\lim_{N\rightarrow 1}\frac{\ln N}{1-N}\right] = \frac{\mathcal{A}_{\p_{AB}}}{4\ell_p^2} \qquad ,
\end{equation} 
which is exactly the Ryu-Takayanagi formula. Comparing with the calculation in \cite{Han2016}, we both reproduce the Ryu-Takayanagi formula from the spin-network state in the semi-classical regime \ref{eq:semilow2} of loop quantum gravity and GFT states. This gives further support to the expectation that a classical gravitational theory can be recovered in this formalism. Differently from \cite{Han2016}, however, our result directly relies on the fundamental degrees of freedom of the theory.

\section{Randomness and Universality}
The dictionary we have established between GFT states and (generalized) random tensor networks suggest the potential for useful cross-over of results across these two research areas. In particular, one can already envisage a direct application of results concerning the quantum dynamics of GFT models and the statistical properties of random tensor models to problems in statistical mechanics and condensed matter that can be formulated in terms of random tensor networks. 

Indeed, our path integral analysis generalises the statistical derivation given in \cite{Hayden:2016cfa}, where the random character of the tensors allowed to map the computation of typical R\'{e}nyi entropies to the evaluation of partition functions of generalized Ising models with inverse temperature $\beta \propto  log D$,  $D$ being the dimension of each leg of each tensor in the network. Interestingly, in the original work, the form of the averaged entropies was derived only in the \emph{large $D$ limit}, where the fluctuations of the partition functions are effectively suppressed. In the large $D$ (low temperature) limit, corresponding to the long-range ordered phase for the Ising models, the entropies of a boundary region can be directly related to the energy of a domain wall between different domains of the order parameter: the Ising action can be estimated by the lowest energy configuration and the minimal energy condition of the domain wall naturally leads to the RT formula. 

One set of results that appears immediately useful in this context concerns {\it universality} properties of probability distributions over random tensors, in the limit of large $D$ \cite{Gurau:2011kk}. They represent a generalization to tensor distributions of the central limit theorem for ordinary probability distributions. 

Indeed, a recently proved universality theorem for random tensor fields \cite{Gurau:2011kk} states that a rank-d random tensor whose entries are $N^d$ independent, identically distributed, complex random variables, and whose distribution is a trace invariant (of the type defining the interactions of {\it tensorial GFTs} as well), converges in distribution in the \emph{large $D$} limit to the distributional limit of a Gaussian tensor model, namely a Gaussian tensor field theory. This is already quite remarkable. However, a second, stronger, universality result \cite{Gurau:2011kk} states that under only the assumption that the joint probability distribution of tensor entries is invariant, assuming that the cumulants of this invariant distribution are uniformly bounded, the large $D$ limit the tensor distribution again converges to the distributional limit of a Gaussian tensor model.

We expect these theorems to have direct applicability to random tensor networks, and even to the generalized class corresponding to the infinite dimensional group fields, where the large $D$ limit refers to the regime in which any UV cut-off on group representations is removed.

The key point to be careful about is that such theorems generally apply to distributions of invariant tensor observables, constructed out of trace (bubble) invariants for bipartite d-colored graphs \cite{Gurau:2011xp}. Therefore, it does not directly apply to simple products of tensors as we have dealt with in this paper. However, one may wonder how much of such universal behavior survives for generic graphs  when distributions of generic tensor observable are considered, e.g. including polynomials made by contractions of tensors which leave some indices free, as for the case of a contracted tensor network state associated to an open graph.

Intuitively, if one randomizes tensors at the nodes independently of contractions, one can still rely on such results, to some extent, but the conclusions become much less solid, because contractions do affect the scaling of the tensors. Much more solid would be to treat the whole tensor network as an observable in a random tensor or GFT model; then, for tensor networks associated to d-colored graphs (trace invariants), the universality theorems would apply, thereby indicating a new direction for further characterizations of the tensor network states.
We postpone this type of evaluations to future work, alongside the complete reformulation of tensor network states and their statistical average within the 2nd quantized GFT framework.

\section{Conclusions}
Let us summarize our results in this paper. We have established a precise dictionary between GFT states and (generalized) random tensor networks. This dictionary also implies, under different restrictions on the GFT states, a correspondence between LQG spin network states and tensor networks, and a correspondence between random tensors models and tensor networks. Next, we have computed the R\'{e}nyi entropy and derived the RT entropy formula, for GFT and spin network techniques, first using a simple approximation to a complete definition of a random tensor network evaluation seen as a GFT correlation function, but still using a truly generalized tensor network seen as a GFT state, and then considering directly a spin network state as a random tensor network. This elucidates further the correspondence and its potential. Finally, we have discussed how universality theorems for random tensor models can be applied to tensor network states, as a first example of application of results from the theory of random tensors and GFT to tensor networks.
We are convinced that these results can be just the beginning of many further developments, made possible by the fertile meeting between tensor networks and fundamental quantum gravity, along the lines we have established.

\acknowledgments

The authors thank Razvan Gurau for comments on the universality theorems of random tensors. MZ also acknowledges the funding received from Alexander von Humboldt Foundation. 

\appendix

\section{From \Ref{eq:Lchi} to \Ref{eq:Lj}}
In this appendix we perform the calculation from \Ref{eq:Lchi} to \Ref{eq:Lj}. $\mathcal{L}(\pi,\pi')$ is given by \Ref{eq:Lchi}. Let us denote $2\pi\gamma$ as $c$ for simplicity, then $\mathcal{L}(\pi,\pi')$ can be written as
\begin{equation}
\mathcal{L}(\pi,\pi') = \prod_{i=1}^{\chi(\varpi)}\chi_{j}\left(\ex^{-r_i c L_z-r_i\frac{\exp(1-c L_z)}{c}}\right)\equiv \prod_{i=1}^{\chi(\varpi)}I_{r_i}
\end{equation}
$I_{r}$ can be written in terms of $\SU(2)$ coherent state as
\begin{eqnarray}
I_r &=&d_{j}\int \dd n\langle j,j|n^{\dag }\ex^{-r c L_{z}-r\frac{\exp
\left( 1-c L_{z}\right) }{c}}n|j,j\rangle  \nonumber\\
&=& d_{j}\sum_{k}^{\infty }\frac{\left( -\right) ^{k}r^{k}\ex^{k}}{k!c^{k}%
}\int \dd n\langle j,j|n^{\dag }\ex^{-\left( r+k\right) c
L_{z}}n|j,j\rangle  \\
&\equiv & d_{j}\sum_{k}^{\infty }\frac{\left( -\right) ^{k}r^{k}\ex^{k}}{k!c^{k}%
}\int \dd n~\ex^{S^{(k)}_{r}}\equiv d_{j}\int \dd n~\ex^{S_{r}}
\end{eqnarray}
where $S_r$ is the total action and $S_r^{(k)}\equiv 2j \ln \langle\uparrow|n^{\dag }e^{-c \left( r+k\right)
L_{z}}n|\uparrow\rangle$ and $\ket{\uparrow}\equiv\ket{\frac{1}{2},\frac{1}{2}}$. In the semi-classical regime of loop gravity, i.e. the large spin-$j$ regime, the leading contribution of $I_r$ is from the critical point of $S_r^{(k)}$, which is the solutions of the equations of motion
\begin{equation}
\delta_n S_r^{(k)} = 0 \quad \Rightarrow \quad  n^{\dag }\ex^{-c \left( r+k\right) L_{z}}n=e^{-\alpha L_{z}}
\end{equation}
One can obtain the solutions
\begin{equation}
n^{\dag }L_{z}n=\pm L_{z},\quad \alpha ^{\pm }=\pm c \left( r+k\right) 
\end{equation}
Bring the solutions back to $I_r$, we can get
\begin{equation}
I_r\sim d_{j}\sum_{\epsilon =\pm }\frac{e^{S_{r0}^{\epsilon }}}{\sqrt{\det
H_{r}^{\epsilon }}}\equiv \sum_{\epsilon =\pm }I_{r}^{\epsilon }
\end{equation}
where $S_{r0}^{\epsilon}$ is the total action $S_r$ on the critical point 
\begin{equation}
S_{r0}^{\epsilon} \equiv -\epsilon r c j-r\frac{\exp\left( 1-\epsilon c
j\right) }{\gamma }
\end{equation}
and $H_{r}^{\epsilon }$ is the Hessian matrix of $S_r$
\begin{equation}
H_{r}^{\epsilon } \equiv \frac{1}{2}\delta
_{n}^{2}S_r|_{\epsilon } 
\end{equation}
After perform the second derivation on $S_r$, one can obtain 
\begin{equation}
\det H^{\epsilon }=4j^{2}r^{2}\left( -c +\exp \left( 1-\epsilon c
j\right) \right) ^{2}
\end{equation}
In the semi-classical and low energy limit
\begin{equation}
\det H^{\epsilon}\sim 4j^{2}N^{2}\exp 2\left( 1-\epsilon c j\right) 
\end{equation}
Then $I_r^\epsilon$ becomes
\begin{equation}
I_{r}^{\epsilon} \sim \frac{\exp \left( -1+\epsilon c j (1-r)-r\frac{\exp\left( 1-\epsilon c j\right) }{c}\right) }{r} 
\end{equation}
One can observe that $I_r^+\gg I_r^-$ since when $\epsilon=-$, in the large spin regime $I_r^-$ goes to zero. $I_r$ thus becomes $I_r^+$, which is one of the term in the product of \Ref{eq:Lj}.
\begin{equation}
I_r \approx I_r^+ = \frac{\exp \left( -1+ c j (1-r)-r\frac{\exp\left( 1- c j\right) }{c}\right) }{r} 
\end{equation}


\begin{thebibliography}{87}%
	\makeatletter
	\providecommand \@ifxundefined [1]{%
		\@ifx{#1\undefined}
	}%
	\providecommand \@ifnum [1]{%
		\ifnum #1\expandafter \@firstoftwo
		\else \expandafter \@secondoftwo
		\fi
	}%
	\providecommand \@ifx [1]{%
		\ifx #1\expandafter \@firstoftwo
		\else \expandafter \@secondoftwo
		\fi
	}%
	\providecommand \natexlab [1]{#1}%
	\providecommand \enquote  [1]{``#1''}%
	\providecommand \bibnamefont  [1]{#1}%
	\providecommand \bibfnamefont [1]{#1}%
	\providecommand \citenamefont [1]{#1}%
	\providecommand \href@noop [0]{\@secondoftwo}%
	\providecommand \href [0]{\begingroup \@sanitize@url \@href}%
	\providecommand \@href[1]{\@@startlink{#1}\@@href}%
	\providecommand \@@href[1]{\endgroup#1\@@endlink}%
	\providecommand \@sanitize@url [0]{\catcode `\\12\catcode `\$12\catcode
		`\&12\catcode `\#12\catcode `\^12\catcode `\_12\catcode `\%12\relax}%
	\providecommand \@@startlink[1]{}%
	\providecommand \@@endlink[0]{}%
	\providecommand \url  [0]{\begingroup\@sanitize@url \@url }%
	\providecommand \@url [1]{\endgroup\@href {#1}{\urlprefix }}%
	\providecommand \urlprefix  [0]{URL }%
	\providecommand \Eprint [0]{\href }%
	\providecommand \doibase [0]{http://dx.doi.org/}%
	\providecommand \selectlanguage [0]{\@gobble}%
	\providecommand \bibinfo  [0]{\@secondoftwo}%
	\providecommand \bibfield  [0]{\@secondoftwo}%
	\providecommand \translation [1]{[#1]}%
	\providecommand \BibitemOpen [0]{}%
	\providecommand \bibitemStop [0]{}%
	\providecommand \bibitemNoStop [0]{.\EOS\space}%
	\providecommand \EOS [0]{\spacefactor3000\relax}%
	\providecommand \BibitemShut  [1]{\csname bibitem#1\endcsname}%
	\let\auto@bib@innerbib\@empty
	\bibitem [{\citenamefont {Ashtekar}\ and\ \citenamefont
		{Lewandowski}(2004)}]{Ashtekar:2004eh}%
	\BibitemOpen
	\bibfield  {author} {\bibinfo {author} {\bibfnamefont {A.}~\bibnamefont
			{Ashtekar}}\ and\ \bibinfo {author} {\bibfnamefont {J.}~\bibnamefont
			{Lewandowski}},\ }\href {\doibase 10.1088/0264-9381/21/15/R01} {\bibfield
		{journal} {\bibinfo  {journal} {Class. Quant. Grav.}\ }\textbf {\bibinfo
			{volume} {21}},\ \bibinfo {pages} {R53} (\bibinfo {year} {2004})},\ \Eprint
	{http://arxiv.org/abs/gr-qc/0404018} {arXiv:gr-qc/0404018} \BibitemShut
	{NoStop}%
	\bibitem [{\citenamefont {Rovelli}(2004)}]{Rovelli:QG}%
	\BibitemOpen
	\bibfield  {author} {\bibinfo {author} {\bibfnamefont {C.}~\bibnamefont
			{Rovelli}},\ }\href@noop {} {\emph {\bibinfo {title} {Quantum Gravity}}}\
	(\bibinfo  {publisher} {Cambridge University Press},\ \bibinfo {address}
	{London},\ \bibinfo {year} {2004})\BibitemShut {NoStop}%
	\bibitem [{\citenamefont {Thiemann}(2007)}]{Thiemann:QG}%
	\BibitemOpen
	\bibfield  {author} {\bibinfo {author} {\bibfnamefont {T.}~\bibnamefont
			{Thiemann}},\ }\href@noop {} {\emph {\bibinfo {title} {Modern canonical
				quantum general relativity}}},\ Cambridge Monographs on Mathematical Physics\
	(\bibinfo  {publisher} {Cambridge University Press},\ \bibinfo {address}
	{London},\ \bibinfo {year} {2007})\BibitemShut {NoStop}%
	\bibitem [{\citenamefont {Perez}(2012)}]{Perez:2012wv}%
	\BibitemOpen
	\bibfield  {author} {\bibinfo {author} {\bibfnamefont {A.}~\bibnamefont
			{Perez}},\ }\href {\doibase 10.12942/lrr-2013-3} {\bibfield  {journal}
		{\bibinfo  {journal} {Living Rev.Rel.}\ }\textbf {\bibinfo {volume} {16}},\
		\bibinfo {pages} {3} (\bibinfo {year} {2012})},\ \Eprint
	{http://arxiv.org/abs/1205.2019} {arXiv:1205.2019 [gr-qc]} \BibitemShut
	{NoStop}%
	\bibitem [{\citenamefont {Rovelli}\ and\ \citenamefont
		{Vidotto}(2014)}]{Rovelli:2014ssa}%
	\BibitemOpen
	\bibfield  {author} {\bibinfo {author} {\bibfnamefont {C.}~\bibnamefont
			{Rovelli}}\ and\ \bibinfo {author} {\bibfnamefont {F.}~\bibnamefont
			{Vidotto}},\ }\href
	{http://www.cambridge.org/mw/academic/subjects/physics/cosmology-relativity-and-gravitation/covariant-loop-quantum-gravity-elementary-introduction-quantum-gravity-and-spinfoam-theory?format=HB}
	{\emph {\bibinfo {title} {{Covariant Loop Quantum Gravity}}}},\ Cambridge
	Monographs on Mathematical Physics\ (\bibinfo  {publisher} {Cambridge
		University Press},\ \bibinfo {year} {2014})\BibitemShut {NoStop}%
	\bibitem [{\citenamefont {Oriti}(2014{\natexlab{a}})}]{Oriti:2014uga}%
	\BibitemOpen
	\bibfield  {author} {\bibinfo {author} {\bibfnamefont {D.}~\bibnamefont
			{Oriti}},\ }in\ \href
	{http://inspirehep.net/record/1312968/files/arXiv:1408.7112.pdf} {\emph
		{\bibinfo {booktitle} {to appear in ``Loop Quantum Gravity - 100 Years of
				General Relativity Series''}}},\ \bibinfo {editor} {edited by\ \bibinfo
		{editor} {\bibfnamefont {A.}~\bibnamefont {Ashtekar}}\ and\ \bibinfo {editor}
		{\bibfnamefont {J.}~\bibnamefont {Pullin}}}\ (\bibinfo {year} {2014})\
	\Eprint {http://arxiv.org/abs/1408.7112} {arXiv:1408.7112 [gr-qc]}
	\BibitemShut {NoStop}%
	\bibitem [{\citenamefont {Baratin}\ and\ \citenamefont
		{Oriti}(2012{\natexlab{a}})}]{Baratin:2011aa}%
	\BibitemOpen
	\bibfield  {author} {\bibinfo {author} {\bibfnamefont {A.}~\bibnamefont
			{Baratin}}\ and\ \bibinfo {author} {\bibfnamefont {D.}~\bibnamefont
			{Oriti}},\ }\bibfield  {booktitle} {\emph {\bibinfo {booktitle}
			{{Proceedings, International Conference on Non-perturbative / background
					independent quantum gravity (Loops 11): Madrid, Spain, May 23-28, 2011}}},\
	}\href {\doibase 10.1088/1742-6596/360/1/012002} {\bibfield  {journal}
		{\bibinfo  {journal} {J. Phys. Conf. Ser.}\ }\textbf {\bibinfo {volume}
			{360}},\ \bibinfo {pages} {012002} (\bibinfo {year} {2012}{\natexlab{a}})},\
	\Eprint {http://arxiv.org/abs/1112.3270} {arXiv:1112.3270 [gr-qc]}
	\BibitemShut {NoStop}%
	\bibitem [{\citenamefont {Oriti}(2011)}]{Oriti:2011jm}%
	\BibitemOpen
	\bibfield  {author} {\bibinfo {author} {\bibfnamefont {D.}~\bibnamefont
			{Oriti}},\ }in\ \href
	{http://inspirehep.net/record/942719/files/arXiv:1110.5606.pdf} {\emph
		{\bibinfo {booktitle} {{Proceedings, Foundations of Space and Time:
					Reflections on Quantum Gravity: Cape Town, South Africa}}}}\ (\bibinfo {year}
	{2011})\ pp.\ \bibinfo {pages} {257--320},\ \Eprint
	{http://arxiv.org/abs/1110.5606} {arXiv:1110.5606 [hep-th]} \BibitemShut
	{NoStop}%
	\bibitem [{\citenamefont {Oriti}(2009)}]{Oriti:2009wn}%
	\BibitemOpen
	\bibfield  {author} {\bibinfo {author} {\bibfnamefont {D.}~\bibnamefont
			{Oriti}},\ }\bibfield  {booktitle} {\emph {\bibinfo {booktitle}
			{{Proceedings, 25th Max Born Symposium: The Planck Scale: Wroclaw, Poland,
					June 29-July 3, 2009}}},\ }\href {\doibase 10.1063/1.3284386} {\bibfield
		{journal} {\bibinfo  {journal} {AIP Conf. Proc.}\ }\textbf {\bibinfo {volume}
			{1196}},\ \bibinfo {pages} {209} (\bibinfo {year} {2009})},\ \Eprint
	{http://arxiv.org/abs/0912.2441} {arXiv:0912.2441 [hep-th]} \BibitemShut
	{NoStop}%
	\bibitem [{\citenamefont {Gurau}\ and\ \citenamefont
		{Ryan}(2012)}]{Gurau:2011xp}%
	\BibitemOpen
	\bibfield  {author} {\bibinfo {author} {\bibfnamefont {R.}~\bibnamefont
			{Gurau}}\ and\ \bibinfo {author} {\bibfnamefont {J.~P.}\ \bibnamefont
			{Ryan}},\ }\href {\doibase 10.3842/SIGMA.2012.020} {\bibfield  {journal}
		{\bibinfo  {journal} {SIGMA}\ }\textbf {\bibinfo {volume} {8}},\ \bibinfo
		{pages} {020} (\bibinfo {year} {2012})},\ \Eprint
	{http://arxiv.org/abs/1109.4812} {arXiv:1109.4812 [hep-th]} \BibitemShut
	{NoStop}%
	\bibitem [{\citenamefont {Gurau}(2016)}]{Gurau:2016cjo}%
	\BibitemOpen
	\bibfield  {author} {\bibinfo {author} {\bibfnamefont {R.}~\bibnamefont
			{Gurau}},\ }\href {\doibase 10.3842/SIGMA.2016.094} {\bibfield  {journal}
		{\bibinfo  {journal} {SIGMA}\ }\textbf {\bibinfo {volume} {12}},\ \bibinfo
		{pages} {094} (\bibinfo {year} {2016})},\ \Eprint
	{http://arxiv.org/abs/1609.06439} {arXiv:1609.06439 [hep-th]} \BibitemShut
	{NoStop}%
	\bibitem [{\citenamefont {Rivasseau}(2016)}]{Rivasseau:2016zco}%
	\BibitemOpen
	\bibfield  {author} {\bibinfo {author} {\bibfnamefont {V.}~\bibnamefont
			{Rivasseau}},\ }\href {\doibase 10.3842/SIGMA.2016.069} {\bibfield  {journal}
		{\bibinfo  {journal} {SIGMA}\ }\textbf {\bibinfo {volume} {12}},\ \bibinfo
		{pages} {069} (\bibinfo {year} {2016})},\ \Eprint
	{http://arxiv.org/abs/1603.07278} {arXiv:1603.07278 [math-ph]} \BibitemShut
	{NoStop}%
	\bibitem [{\citenamefont {Orus}(2014)}]{Orus:2013kga}%
	\BibitemOpen
	\bibfield  {author} {\bibinfo {author} {\bibfnamefont {R.}~\bibnamefont
			{Orus}},\ }\href {\doibase 10.1016/j.aop.2014.06.013} {\bibfield  {journal}
		{\bibinfo  {journal} {Annals Phys.}\ }\textbf {\bibinfo {volume} {349}},\
		\bibinfo {pages} {117} (\bibinfo {year} {2014})},\ \Eprint
	{http://arxiv.org/abs/1306.2164} {arXiv:1306.2164 [cond-mat.str-el]}
	\BibitemShut {NoStop}%
	\bibitem [{\citenamefont {Bridgeman}\ and\ \citenamefont
		{Chubb}(2016)}]{Bridgeman:2016dhh}%
	\BibitemOpen
	\bibfield  {author} {\bibinfo {author} {\bibfnamefont {J.~C.}\ \bibnamefont
			{Bridgeman}}\ and\ \bibinfo {author} {\bibfnamefont {C.~T.}\ \bibnamefont
			{Chubb}},\ }\href@noop {} {\  (\bibinfo {year} {2016})},\ \Eprint
	{http://arxiv.org/abs/1603.03039} {arXiv:1603.03039 [quant-ph]} \BibitemShut
	{NoStop}%
	\bibitem [{\citenamefont {Wen}(2004)}]{Wen:2004ym}%
	\BibitemOpen
	\bibfield  {author} {\bibinfo {author} {\bibfnamefont {X.~G.}\ \bibnamefont
			{Wen}},\ }\href@noop {} {\emph {\bibinfo {title} {{Quantum field theory of
					many-body systems: From the origin of sound to an origin of light and
					electrons}}}}\ (\bibinfo {year} {2004})\BibitemShut {NoStop}%
	\bibitem [{\citenamefont {Wen}(2016)}]{Wen:2016ddy}%
	\BibitemOpen
	\bibfield  {author} {\bibinfo {author} {\bibfnamefont {X.-G.}\ \bibnamefont
			{Wen}},\ }\href@noop {} {\  (\bibinfo {year} {2016})},\ \Eprint
	{http://arxiv.org/abs/1610.03911} {arXiv:1610.03911 [cond-mat.str-el]}
	\BibitemShut {NoStop}%
	\bibitem [{\citenamefont {Cirac}\ and\ \citenamefont
		{Verstraete}(2009)}]{Cirac:2009zz}%
	\BibitemOpen
	\bibfield  {author} {\bibinfo {author} {\bibfnamefont {J.~I.}\ \bibnamefont
			{Cirac}}\ and\ \bibinfo {author} {\bibfnamefont {F.}~\bibnamefont
			{Verstraete}},\ }\href {\doibase 10.1088/1751-8113/42/50/504004} {\bibfield
		{journal} {\bibinfo  {journal} {J. Phys.}\ }\textbf {\bibinfo {volume}
			{A42}},\ \bibinfo {pages} {504004} (\bibinfo {year} {2009})}\BibitemShut
	{NoStop}%
	\bibitem [{\citenamefont {Verstraete}\ \emph {et~al.}(2008)\citenamefont
		{Verstraete}, \citenamefont {Murg},\ and\ \citenamefont
		{Cirac}}]{verstraete2008matrix}%
	\BibitemOpen
	\bibfield  {author} {\bibinfo {author} {\bibfnamefont {F.}~\bibnamefont
			{Verstraete}}, \bibinfo {author} {\bibfnamefont {V.}~\bibnamefont {Murg}}, \
		and\ \bibinfo {author} {\bibfnamefont {J.~I.}\ \bibnamefont {Cirac}},\
	}\href@noop {} {\bibfield  {journal} {\bibinfo  {journal} {Advances in
				Physics}\ }\textbf {\bibinfo {volume} {57}},\ \bibinfo {pages} {143}
		(\bibinfo {year} {2008})}\BibitemShut {NoStop}%
	\bibitem [{\citenamefont {Augusiak}\ \emph {et~al.}(2012)\citenamefont
		{Augusiak}, \citenamefont {Cucchietti},\ and\ \citenamefont
		{Lewenstein}}]{augusiak2012modern}%
	\BibitemOpen
	\bibfield  {author} {\bibinfo {author} {\bibfnamefont {R.}~\bibnamefont
			{Augusiak}}, \bibinfo {author} {\bibfnamefont {F.}~\bibnamefont
			{Cucchietti}}, \ and\ \bibinfo {author} {\bibfnamefont {M.}~\bibnamefont
			{Lewenstein}},\ }\href@noop {} {\bibfield  {journal} {\bibinfo  {journal}
			{Lect. Not. Phys}\ }\textbf {\bibinfo {volume} {843}},\ \bibinfo {pages}
		{245} (\bibinfo {year} {2012})}\BibitemShut {NoStop}%
	\bibitem [{\citenamefont {Ryu}\ and\ \citenamefont
		{Takayanagi}(2006)}]{Ryu:2006bv}%
	\BibitemOpen
	\bibfield  {author} {\bibinfo {author} {\bibfnamefont {S.}~\bibnamefont
			{Ryu}}\ and\ \bibinfo {author} {\bibfnamefont {T.}~\bibnamefont
			{Takayanagi}},\ }\href {\doibase 10.1103/PhysRevLett.96.181602} {\bibfield
		{journal} {\bibinfo  {journal} {Phys.Rev.Lett.}\ }\textbf {\bibinfo {volume}
			{96}},\ \bibinfo {pages} {181602} (\bibinfo {year} {2006})},\ \Eprint
	{http://arxiv.org/abs/hep-th/0603001} {arXiv:hep-th/0603001 [hep-th]}
	\BibitemShut {NoStop}%
	\bibitem [{\citenamefont {Swingle}(2012)}]{Swingle:2009bg}%
	\BibitemOpen
	\bibfield  {author} {\bibinfo {author} {\bibfnamefont {B.}~\bibnamefont
			{Swingle}},\ }\href {\doibase 10.1103/PhysRevD.86.065007} {\bibfield
		{journal} {\bibinfo  {journal} {Phys. Rev.}\ }\textbf {\bibinfo {volume}
			{D86}},\ \bibinfo {pages} {065007} (\bibinfo {year} {2012})},\ \Eprint
	{http://arxiv.org/abs/0905.1317} {arXiv:0905.1317 [cond-mat.str-el]}
	\BibitemShut {NoStop}%
	\bibitem [{\citenamefont {Pastawski}\ \emph {et~al.}(2015)\citenamefont
		{Pastawski}, \citenamefont {Yoshida}, \citenamefont {Harlow},\ and\
		\citenamefont {Preskill}}]{Pastawski:2015qua}%
	\BibitemOpen
	\bibfield  {author} {\bibinfo {author} {\bibfnamefont {F.}~\bibnamefont
			{Pastawski}}, \bibinfo {author} {\bibfnamefont {B.}~\bibnamefont {Yoshida}},
		\bibinfo {author} {\bibfnamefont {D.}~\bibnamefont {Harlow}}, \ and\ \bibinfo
		{author} {\bibfnamefont {J.}~\bibnamefont {Preskill}},\ }\href {\doibase
		10.1007/JHEP06(2015)149} {\bibfield  {journal} {\bibinfo  {journal} {JHEP}\
		}\textbf {\bibinfo {volume} {06}},\ \bibinfo {pages} {149} (\bibinfo {year}
		{2015})},\ \Eprint {http://arxiv.org/abs/1503.06237} {arXiv:1503.06237
		[hep-th]} \BibitemShut {NoStop}%
	\bibitem [{\citenamefont {Hayden}\ \emph {et~al.}(2016)\citenamefont {Hayden},
		\citenamefont {Nezami}, \citenamefont {Qi}, \citenamefont {Thomas},
		\citenamefont {Walter},\ and\ \citenamefont {Yang}}]{Hayden:2016cfa}%
	\BibitemOpen
	\bibfield  {author} {\bibinfo {author} {\bibfnamefont {P.}~\bibnamefont
			{Hayden}}, \bibinfo {author} {\bibfnamefont {S.}~\bibnamefont {Nezami}},
		\bibinfo {author} {\bibfnamefont {X.-L.}\ \bibnamefont {Qi}}, \bibinfo
		{author} {\bibfnamefont {N.}~\bibnamefont {Thomas}}, \bibinfo {author}
		{\bibfnamefont {M.}~\bibnamefont {Walter}}, \ and\ \bibinfo {author}
		{\bibfnamefont {Z.}~\bibnamefont {Yang}},\ }\href {\doibase
		10.1007/JHEP11(2016)009} {\bibfield  {journal} {\bibinfo  {journal} {JHEP}\
		}\textbf {\bibinfo {volume} {11}},\ \bibinfo {pages} {009} (\bibinfo {year}
		{2016})},\ \Eprint {http://arxiv.org/abs/1601.01694} {arXiv:1601.01694
		[hep-th]} \BibitemShut {NoStop}%
	\bibitem [{\citenamefont {Vidal}(2008)}]{vidal2008class}%
	\BibitemOpen
	\bibfield  {author} {\bibinfo {author} {\bibfnamefont {G.}~\bibnamefont
			{Vidal}},\ }\href@noop {} {\bibfield  {journal} {\bibinfo  {journal}
			{Physical review letters}\ }\textbf {\bibinfo {volume} {101}},\ \bibinfo
		{pages} {110501} (\bibinfo {year} {2008})}\BibitemShut {NoStop}%
	\bibitem [{\citenamefont {Singh}\ \emph {et~al.}(2010)\citenamefont {Singh},
		\citenamefont {Pfeifer},\ and\ \citenamefont {Vidal}}]{Singh:2009cd}%
	\BibitemOpen
	\bibfield  {author} {\bibinfo {author} {\bibfnamefont {S.}~\bibnamefont
			{Singh}}, \bibinfo {author} {\bibfnamefont {R.~N.~C.}\ \bibnamefont
			{Pfeifer}}, \ and\ \bibinfo {author} {\bibfnamefont {G.}~\bibnamefont
			{Vidal}},\ }\href {\doibase 10.1103/PhysRevA.82.050301} {\bibfield  {journal}
		{\bibinfo  {journal} {Phys. Rev.}\ }\textbf {\bibinfo {volume} {A82}},\
		\bibinfo {pages} {050301} (\bibinfo {year} {2010})},\ \Eprint
	{http://arxiv.org/abs/0907.2994} {arXiv:0907.2994 [cond-mat.str-el]}
	\BibitemShut {NoStop}%
	\bibitem [{\citenamefont {Evenbly}\ and\ \citenamefont
		{Vidal}(2011)}]{evenbly2011tensor}%
	\BibitemOpen
	\bibfield  {author} {\bibinfo {author} {\bibfnamefont {G.}~\bibnamefont
			{Evenbly}}\ and\ \bibinfo {author} {\bibfnamefont {G.}~\bibnamefont
			{Vidal}},\ }\href@noop {} {\bibfield  {journal} {\bibinfo  {journal} {Journal
				of Statistical Physics}\ }\textbf {\bibinfo {volume} {145}},\ \bibinfo
		{pages} {891} (\bibinfo {year} {2011})}\BibitemShut {NoStop}%
	\bibitem [{\citenamefont {Han}\ and\ \citenamefont {Hung}(2016)}]{Han2016}%
	\BibitemOpen
	\bibfield  {author} {\bibinfo {author} {\bibfnamefont {M.}~\bibnamefont
			{Han}}\ and\ \bibinfo {author} {\bibfnamefont {L.-Y.}\ \bibnamefont {Hung}},\
	}\href@noop {} {\  (\bibinfo {year} {2016})},\ \Eprint
	{http://arxiv.org/abs/1610.02134} {arXiv:1610.02134 [hep-th]} \BibitemShut
	{NoStop}%
	\bibitem [{\citenamefont {Dittrich}\ \emph
		{et~al.}(2016{\natexlab{a}})\citenamefont {Dittrich}, \citenamefont
		{Mizera},\ and\ \citenamefont {Steinhaus}}]{Dittrich:2014mxa}%
	\BibitemOpen
	\bibfield  {author} {\bibinfo {author} {\bibfnamefont {B.}~\bibnamefont
			{Dittrich}}, \bibinfo {author} {\bibfnamefont {S.}~\bibnamefont {Mizera}}, \
		and\ \bibinfo {author} {\bibfnamefont {S.}~\bibnamefont {Steinhaus}},\ }\href
	{\doibase 10.1088/1367-2630/18/5/053009} {\bibfield  {journal} {\bibinfo
			{journal} {New J. Phys.}\ }\textbf {\bibinfo {volume} {18}},\ \bibinfo
		{pages} {053009} (\bibinfo {year} {2016}{\natexlab{a}})},\ \Eprint
	{http://arxiv.org/abs/1409.2407} {arXiv:1409.2407 [gr-qc]} \BibitemShut
	{NoStop}%
	\bibitem [{\citenamefont {Delcamp}\ and\ \citenamefont
		{Dittrich}(2016)}]{Delcamp:2016dqo}%
	\BibitemOpen
	\bibfield  {author} {\bibinfo {author} {\bibfnamefont {C.}~\bibnamefont
			{Delcamp}}\ and\ \bibinfo {author} {\bibfnamefont {B.}~\bibnamefont
			{Dittrich}},\ }\href@noop {} {\  (\bibinfo {year} {2016})},\ \Eprint
	{http://arxiv.org/abs/1612.04506} {arXiv:1612.04506 [gr-qc]} \BibitemShut
	{NoStop}%
	\bibitem [{\citenamefont {Dittrich}\ \emph {et~al.}(2012)\citenamefont
		{Dittrich}, \citenamefont {Eckert},\ and\ \citenamefont
		{Martin-Benito}}]{Dittrich:2011zh}%
	\BibitemOpen
	\bibfield  {author} {\bibinfo {author} {\bibfnamefont {B.}~\bibnamefont
			{Dittrich}}, \bibinfo {author} {\bibfnamefont {F.~C.}\ \bibnamefont
			{Eckert}}, \ and\ \bibinfo {author} {\bibfnamefont {M.}~\bibnamefont
			{Martin-Benito}},\ }\href {\doibase 10.1088/1367-2630/14/3/035008} {\bibfield
		{journal} {\bibinfo  {journal} {New J. Phys.}\ }\textbf {\bibinfo {volume}
			{14}},\ \bibinfo {pages} {035008} (\bibinfo {year} {2012})},\ \Eprint
	{http://arxiv.org/abs/1109.4927} {arXiv:1109.4927 [gr-qc]} \BibitemShut
	{NoStop}%
	\bibitem [{\citenamefont {Dittrich}\ \emph
		{et~al.}(2016{\natexlab{b}})\citenamefont {Dittrich}, \citenamefont
		{Schnetter}, \citenamefont {Seth},\ and\ \citenamefont
		{Steinhaus}}]{Dittrich:2016tys}%
	\BibitemOpen
	\bibfield  {author} {\bibinfo {author} {\bibfnamefont {B.}~\bibnamefont
			{Dittrich}}, \bibinfo {author} {\bibfnamefont {E.}~\bibnamefont {Schnetter}},
		\bibinfo {author} {\bibfnamefont {C.~J.}\ \bibnamefont {Seth}}, \ and\
		\bibinfo {author} {\bibfnamefont {S.}~\bibnamefont {Steinhaus}},\ }\href
	{\doibase 10.1103/PhysRevD.94.124050} {\bibfield  {journal} {\bibinfo
			{journal} {Phys. Rev.}\ }\textbf {\bibinfo {volume} {D94}},\ \bibinfo {pages}
		{124050} (\bibinfo {year} {2016}{\natexlab{b}})},\ \Eprint
	{http://arxiv.org/abs/1609.02429} {arXiv:1609.02429 [gr-qc]} \BibitemShut
	{NoStop}%
	\bibitem [{\citenamefont {Carrozza}\ \emph {et~al.}(2014)\citenamefont
		{Carrozza}, \citenamefont {Oriti},\ and\ \citenamefont
		{Rivasseau}}]{Carrozza:2013wda}%
	\BibitemOpen
	\bibfield  {author} {\bibinfo {author} {\bibfnamefont {S.}~\bibnamefont
			{Carrozza}}, \bibinfo {author} {\bibfnamefont {D.}~\bibnamefont {Oriti}}, \
		and\ \bibinfo {author} {\bibfnamefont {V.}~\bibnamefont {Rivasseau}},\ }\href
	{\doibase 10.1007/s00220-014-1928-x} {\bibfield  {journal} {\bibinfo
			{journal} {Commun. Math. Phys.}\ }\textbf {\bibinfo {volume} {330}},\
		\bibinfo {pages} {581} (\bibinfo {year} {2014})},\ \Eprint
	{http://arxiv.org/abs/1303.6772} {arXiv:1303.6772 [hep-th]} \BibitemShut
	{NoStop}%
	\bibitem [{\citenamefont {Benedetti}\ \emph {et~al.}(2015)\citenamefont
		{Benedetti}, \citenamefont {Ben~Geloun},\ and\ \citenamefont
		{Oriti}}]{Benedetti:2014qsa}%
	\BibitemOpen
	\bibfield  {author} {\bibinfo {author} {\bibfnamefont {D.}~\bibnamefont
			{Benedetti}}, \bibinfo {author} {\bibfnamefont {J.}~\bibnamefont
			{Ben~Geloun}}, \ and\ \bibinfo {author} {\bibfnamefont {D.}~\bibnamefont
			{Oriti}},\ }\href {\doibase 10.1007/JHEP03(2015)084} {\bibfield  {journal}
		{\bibinfo  {journal} {JHEP}\ }\textbf {\bibinfo {volume} {03}},\ \bibinfo
		{pages} {084} (\bibinfo {year} {2015})},\ \Eprint
	{http://arxiv.org/abs/1411.3180} {arXiv:1411.3180 [hep-th]} \BibitemShut
	{NoStop}%
	\bibitem [{\citenamefont {Carrozza}\ and\ \citenamefont
		{Lahoche}(2016)}]{Carrozza:2016tih}%
	\BibitemOpen
	\bibfield  {author} {\bibinfo {author} {\bibfnamefont {S.}~\bibnamefont
			{Carrozza}}\ and\ \bibinfo {author} {\bibfnamefont {V.}~\bibnamefont
			{Lahoche}},\ }\href@noop {} {\  (\bibinfo {year} {2016})},\ \Eprint
	{http://arxiv.org/abs/1612.02452} {arXiv:1612.02452 [hep-th]} \BibitemShut
	{NoStop}%
	\bibitem [{\citenamefont {Carrozza}(2016)}]{Carrozza:2016vsq}%
	\BibitemOpen
	\bibfield  {author} {\bibinfo {author} {\bibfnamefont {S.}~\bibnamefont
			{Carrozza}},\ }\href {\doibase 10.3842/SIGMA.2016.070} {\bibfield  {journal}
		{\bibinfo  {journal} {SIGMA}\ }\textbf {\bibinfo {volume} {12}},\ \bibinfo
		{pages} {070} (\bibinfo {year} {2016})},\ \Eprint
	{http://arxiv.org/abs/1603.01902} {arXiv:1603.01902 [gr-qc]} \BibitemShut
	{NoStop}%
	\bibitem [{\citenamefont {Ben~Geloun}\ \emph {et~al.}(2016)\citenamefont
		{Ben~Geloun}, \citenamefont {Martini},\ and\ \citenamefont
		{Oriti}}]{Geloun:2016qyb}%
	\BibitemOpen
	\bibfield  {author} {\bibinfo {author} {\bibfnamefont {J.}~\bibnamefont
			{Ben~Geloun}}, \bibinfo {author} {\bibfnamefont {R.}~\bibnamefont {Martini}},
		\ and\ \bibinfo {author} {\bibfnamefont {D.}~\bibnamefont {Oriti}},\ }\href
	{\doibase 10.1103/PhysRevD.94.024017} {\bibfield  {journal} {\bibinfo
			{journal} {Phys. Rev.}\ }\textbf {\bibinfo {volume} {D94}},\ \bibinfo {pages}
		{024017} (\bibinfo {year} {2016})},\ \Eprint
	{http://arxiv.org/abs/1601.08211} {arXiv:1601.08211 [hep-th]} \BibitemShut
	{NoStop}%
	\bibitem [{\citenamefont {Lahoche}\ and\ \citenamefont
		{Oriti}(2017)}]{Lahoche:2015tqa}%
	\BibitemOpen
	\bibfield  {author} {\bibinfo {author} {\bibfnamefont {V.}~\bibnamefont
			{Lahoche}}\ and\ \bibinfo {author} {\bibfnamefont {D.}~\bibnamefont
			{Oriti}},\ }\href {\doibase 10.1088/1751-8113/50/2/025201} {\bibfield
		{journal} {\bibinfo  {journal} {J. Phys.}\ }\textbf {\bibinfo {volume}
			{A50}},\ \bibinfo {pages} {025201} (\bibinfo {year} {2017})},\ \Eprint
	{http://arxiv.org/abs/1506.08393} {arXiv:1506.08393 [hep-th]} \BibitemShut
	{NoStop}%
	\bibitem [{\citenamefont {Bahr}\ \emph {et~al.}(2013)\citenamefont {Bahr},
		\citenamefont {Dittrich}, \citenamefont {Hellmann},\ and\ \citenamefont
		{Kaminski}}]{Bahr:2012qj}%
	\BibitemOpen
	\bibfield  {author} {\bibinfo {author} {\bibfnamefont {B.}~\bibnamefont
			{Bahr}}, \bibinfo {author} {\bibfnamefont {B.}~\bibnamefont {Dittrich}},
		\bibinfo {author} {\bibfnamefont {F.}~\bibnamefont {Hellmann}}, \ and\
		\bibinfo {author} {\bibfnamefont {W.}~\bibnamefont {Kaminski}},\ }\href
	{\doibase 10.1103/PhysRevD.87.044048} {\bibfield  {journal} {\bibinfo
			{journal} {Phys. Rev.}\ }\textbf {\bibinfo {volume} {D87}},\ \bibinfo {pages}
		{044048} (\bibinfo {year} {2013})},\ \Eprint {http://arxiv.org/abs/1208.3388}
	{arXiv:1208.3388 [gr-qc]} \BibitemShut {NoStop}%
	\bibitem [{\citenamefont {Bahr}\ and\ \citenamefont
		{Steinhaus}(2016)}]{Bahr:2016hwc}%
	\BibitemOpen
	\bibfield  {author} {\bibinfo {author} {\bibfnamefont {B.}~\bibnamefont
			{Bahr}}\ and\ \bibinfo {author} {\bibfnamefont {S.}~\bibnamefont
			{Steinhaus}},\ }\href {\doibase 10.1103/PhysRevLett.117.141302} {\bibfield
		{journal} {\bibinfo  {journal} {Phys. Rev. Lett.}\ }\textbf {\bibinfo
			{volume} {117}},\ \bibinfo {pages} {141302} (\bibinfo {year} {2016})},\
	\Eprint {http://arxiv.org/abs/1605.07649} {arXiv:1605.07649 [gr-qc]}
	\BibitemShut {NoStop}%
	\bibitem [{\citenamefont {Bahr}(2014)}]{Bahr:2014qza}%
	\BibitemOpen
	\bibfield  {author} {\bibinfo {author} {\bibfnamefont {B.}~\bibnamefont
			{Bahr}},\ }\href@noop {} {\  (\bibinfo {year} {2014})},\ \Eprint
	{http://arxiv.org/abs/1407.7746} {arXiv:1407.7746 [gr-qc]} \BibitemShut
	{NoStop}%
	\bibitem [{\citenamefont {Chirco}\ \emph {et~al.}({\natexlab{a}})\citenamefont
		{Chirco}, \citenamefont {Mele}, \citenamefont {Oriti},\ and\ \citenamefont
		{Vitale}}]{withFabio}%
	\BibitemOpen
	\bibfield  {author} {\bibinfo {author} {\bibfnamefont {G.}~\bibnamefont
			{Chirco}}, \bibinfo {author} {\bibfnamefont {F.}~\bibnamefont {Mele}},
		\bibinfo {author} {\bibfnamefont {D.}~\bibnamefont {Oriti}}, \ and\ \bibinfo
		{author} {\bibfnamefont {P.}~\bibnamefont {Vitale}},\ }in\ \href@noop {}
	{\emph {\bibinfo {booktitle} {to appear}}}\BibitemShut {NoStop}%
	\bibitem [{\citenamefont {Livine}\ and\ \citenamefont
		{Terno}(2006{\natexlab{a}})}]{Livine:2005mw}%
	\BibitemOpen
	\bibfield  {author} {\bibinfo {author} {\bibfnamefont {E.~R.}\ \bibnamefont
			{Livine}}\ and\ \bibinfo {author} {\bibfnamefont {D.~R.}\ \bibnamefont
			{Terno}},\ }\href {\doibase 10.1016/j.nuclphysb.2006.02.012} {\bibfield
		{journal} {\bibinfo  {journal} {Nucl. Phys.}\ }\textbf {\bibinfo {volume}
			{B741}},\ \bibinfo {pages} {131} (\bibinfo {year} {2006}{\natexlab{a}})},\
	\Eprint {http://arxiv.org/abs/gr-qc/0508085} {arXiv:gr-qc/0508085 [gr-qc]}
	\BibitemShut {NoStop}%
	\bibitem [{\citenamefont {Livine}\ and\ \citenamefont
		{Terno}(2006{\natexlab{b}})}]{Livine:2006xk}%
	\BibitemOpen
	\bibfield  {author} {\bibinfo {author} {\bibfnamefont {E.~R.}\ \bibnamefont
			{Livine}}\ and\ \bibinfo {author} {\bibfnamefont {D.~R.}\ \bibnamefont
			{Terno}},\ }\href@noop {} {\  (\bibinfo {year} {2006}{\natexlab{b}})},\
	\Eprint {http://arxiv.org/abs/gr-qc/0603008} {arXiv:gr-qc/0603008 [gr-qc]}
	\BibitemShut {NoStop}%
	\bibitem [{\citenamefont {Donnelly}(2008)}]{Donnelly:2008vx}%
	\BibitemOpen
	\bibfield  {author} {\bibinfo {author} {\bibfnamefont {W.}~\bibnamefont
			{Donnelly}},\ }\href {\doibase 10.1103/PhysRevD.77.104006} {\bibfield
		{journal} {\bibinfo  {journal} {Phys. Rev.}\ }\textbf {\bibinfo {volume}
			{D77}},\ \bibinfo {pages} {104006} (\bibinfo {year} {2008})},\ \Eprint
	{http://arxiv.org/abs/0802.0880} {arXiv:0802.0880 [gr-qc]} \BibitemShut
	{NoStop}%
	\bibitem [{\citenamefont {Diaz-Polo}\ and\ \citenamefont
		{Pranzetti}(2012)}]{DiazPolo:2011np}%
	\BibitemOpen
	\bibfield  {author} {\bibinfo {author} {\bibfnamefont {J.}~\bibnamefont
			{Diaz-Polo}}\ and\ \bibinfo {author} {\bibfnamefont {D.}~\bibnamefont
			{Pranzetti}},\ }\href {\doibase 10.3842/SIGMA.2012.048} {\bibfield  {journal}
		{\bibinfo  {journal} {SIGMA}\ }\textbf {\bibinfo {volume} {8}},\ \bibinfo
		{pages} {048} (\bibinfo {year} {2012})},\ \Eprint
	{http://arxiv.org/abs/1112.0291} {arXiv:1112.0291 [gr-qc]} \BibitemShut
	{NoStop}%
	\bibitem [{\citenamefont {Perez}(2014)}]{Perez:2014ura}%
	\BibitemOpen
	\bibfield  {author} {\bibinfo {author} {\bibfnamefont {A.}~\bibnamefont
			{Perez}},\ }\href {\doibase 10.1103/PhysRevD.90.084015,
		10.1103/PhysRevD.90.089907} {\bibfield  {journal} {\bibinfo  {journal} {Phys.
				Rev.}\ }\textbf {\bibinfo {volume} {D90}},\ \bibinfo {pages} {084015}
		(\bibinfo {year} {2014})},\ \bibinfo {note} {[Addendum: Phys.
		Rev.D90,no.8,089907(2014)]},\ \Eprint {http://arxiv.org/abs/1405.7287}
	{arXiv:1405.7287 [gr-qc]} \BibitemShut {NoStop}%
	\bibitem [{\citenamefont {Bianchi}\ and\ \citenamefont
		{Myers}(2014)}]{Bianchi:2012ev}%
	\BibitemOpen
	\bibfield  {author} {\bibinfo {author} {\bibfnamefont {E.}~\bibnamefont
			{Bianchi}}\ and\ \bibinfo {author} {\bibfnamefont {R.~C.}\ \bibnamefont
			{Myers}},\ }\href {\doibase 10.1088/0264-9381/31/21/214002} {\bibfield
		{journal} {\bibinfo  {journal} {Class.Quant.Grav.}\ }\textbf {\bibinfo
			{volume} {31}},\ \bibinfo {pages} {214002} (\bibinfo {year} {2014})},\
	\Eprint {http://arxiv.org/abs/1212.5183} {arXiv:1212.5183 [hep-th]}
	\BibitemShut {NoStop}%
	\bibitem [{\citenamefont {Chirco}\ \emph {et~al.}(2015)\citenamefont {Chirco},
		\citenamefont {Rovelli},\ and\ \citenamefont {Ruggiero}}]{Chirco:2014naa}%
	\BibitemOpen
	\bibfield  {author} {\bibinfo {author} {\bibfnamefont {G.}~\bibnamefont
			{Chirco}}, \bibinfo {author} {\bibfnamefont {C.}~\bibnamefont {Rovelli}}, \
		and\ \bibinfo {author} {\bibfnamefont {P.}~\bibnamefont {Ruggiero}},\ }\href
	{\doibase 10.1088/0264-9381/32/3/035011} {\bibfield  {journal} {\bibinfo
			{journal} {Class.Quant.Grav.}\ }\textbf {\bibinfo {volume} {32}},\ \bibinfo
		{pages} {035011} (\bibinfo {year} {2015})},\ \Eprint
	{http://arxiv.org/abs/1408.0121} {arXiv:1408.0121 [gr-qc]} \BibitemShut
	{NoStop}%
	\bibitem [{\citenamefont {Ghosh}\ and\ \citenamefont
		{Pranzetti}(2014)}]{Ghosh:2014rra}%
	\BibitemOpen
	\bibfield  {author} {\bibinfo {author} {\bibfnamefont {A.}~\bibnamefont
			{Ghosh}}\ and\ \bibinfo {author} {\bibfnamefont {D.}~\bibnamefont
			{Pranzetti}},\ }\href {\doibase 10.1016/j.nuclphysb.2014.10.002} {\bibfield
		{journal} {\bibinfo  {journal} {Nucl. Phys.}\ }\textbf {\bibinfo {volume}
			{B889}},\ \bibinfo {pages} {1} (\bibinfo {year} {2014})},\ \Eprint
	{http://arxiv.org/abs/1405.7056} {arXiv:1405.7056 [gr-qc]} \BibitemShut
	{NoStop}%
	\bibitem [{\citenamefont {Bonzom}\ and\ \citenamefont
		{Dittrich}(2016)}]{Bonzom:2015ans}%
	\BibitemOpen
	\bibfield  {author} {\bibinfo {author} {\bibfnamefont {V.}~\bibnamefont
			{Bonzom}}\ and\ \bibinfo {author} {\bibfnamefont {B.}~\bibnamefont
			{Dittrich}},\ }\href {\doibase 10.1007/JHEP03(2016)208} {\bibfield  {journal}
		{\bibinfo  {journal} {JHEP}\ }\textbf {\bibinfo {volume} {03}},\ \bibinfo
		{pages} {208} (\bibinfo {year} {2016})},\ \Eprint
	{http://arxiv.org/abs/1511.05441} {arXiv:1511.05441 [hep-th]} \BibitemShut
	{NoStop}%
	\bibitem [{\citenamefont {Hamma}\ \emph {et~al.}(2015)\citenamefont {Hamma},
		\citenamefont {Hung}, \citenamefont {Marciano},\ and\ \citenamefont
		{Zhang}}]{Hamma:2015xla}%
	\BibitemOpen
	\bibfield  {author} {\bibinfo {author} {\bibfnamefont {A.}~\bibnamefont
			{Hamma}}, \bibinfo {author} {\bibfnamefont {L.-Y.}\ \bibnamefont {Hung}},
		\bibinfo {author} {\bibfnamefont {A.}~\bibnamefont {Marciano}}, \ and\
		\bibinfo {author} {\bibfnamefont {M.}~\bibnamefont {Zhang}},\ }\href@noop {}
	{\  (\bibinfo {year} {2015})},\ \Eprint {http://arxiv.org/abs/1506.01623}
	{arXiv:1506.01623 [gr-qc]} \BibitemShut {NoStop}%
	\bibitem [{\citenamefont {Bianchi}\ \emph {et~al.}(2015)\citenamefont
		{Bianchi}, \citenamefont {Hackl},\ and\ \citenamefont
		{Yokomizo}}]{Bianchi:2015fra}%
	\BibitemOpen
	\bibfield  {author} {\bibinfo {author} {\bibfnamefont {E.}~\bibnamefont
			{Bianchi}}, \bibinfo {author} {\bibfnamefont {L.}~\bibnamefont {Hackl}}, \
		and\ \bibinfo {author} {\bibfnamefont {N.}~\bibnamefont {Yokomizo}},\ }\href
	{\doibase 10.1103/PhysRevD.92.085045} {\bibfield  {journal} {\bibinfo
			{journal} {Phys. Rev.}\ }\textbf {\bibinfo {volume} {D92}},\ \bibinfo {pages}
		{085045} (\bibinfo {year} {2015})},\ \Eprint
	{http://arxiv.org/abs/1507.01567} {arXiv:1507.01567 [hep-th]} \BibitemShut
	{NoStop}%
	\bibitem [{\citenamefont {Han}(2014{\natexlab{a}})}]{Han:2014xna}%
	\BibitemOpen
	\bibfield  {author} {\bibinfo {author} {\bibfnamefont {M.}~\bibnamefont
			{Han}},\ }\href@noop {} {\  (\bibinfo {year} {2014}{\natexlab{a}})},\ \Eprint
	{http://arxiv.org/abs/1402.2084} {arXiv:1402.2084 [gr-qc]} \BibitemShut
	{NoStop}%
	\bibitem [{\citenamefont {Han}(2016)}]{Han:2015gma}%
	\BibitemOpen
	\bibfield  {author} {\bibinfo {author} {\bibfnamefont {M.}~\bibnamefont
			{Han}},\ }\href {\doibase 10.1007/JHEP01(2016)065} {\bibfield  {journal}
		{\bibinfo  {journal} {JHEP}\ }\textbf {\bibinfo {volume} {01}},\ \bibinfo
		{pages} {065} (\bibinfo {year} {2016})},\ \Eprint
	{http://arxiv.org/abs/1509.00466} {arXiv:1509.00466 [hep-th]} \BibitemShut
	{NoStop}%
	\bibitem [{\citenamefont {Oriti}\ \emph {et~al.}(2016)\citenamefont {Oriti},
		\citenamefont {Pranzetti},\ and\ \citenamefont {Sindoni}}]{Oriti:2015rwa}%
	\BibitemOpen
	\bibfield  {author} {\bibinfo {author} {\bibfnamefont {D.}~\bibnamefont
			{Oriti}}, \bibinfo {author} {\bibfnamefont {D.}~\bibnamefont {Pranzetti}}, \
		and\ \bibinfo {author} {\bibfnamefont {L.}~\bibnamefont {Sindoni}},\ }\href
	{\doibase 10.1103/PhysRevLett.116.211301} {\bibfield  {journal} {\bibinfo
			{journal} {Phys. Rev. Lett.}\ }\textbf {\bibinfo {volume} {116}},\ \bibinfo
		{pages} {211301} (\bibinfo {year} {2016})},\ \Eprint
	{http://arxiv.org/abs/1510.06991} {arXiv:1510.06991 [gr-qc]} \BibitemShut
	{NoStop}%
	\bibitem [{\citenamefont {Baratin}\ and\ \citenamefont
		{Oriti}(2010)}]{Baratin:2010wi}%
	\BibitemOpen
	\bibfield  {author} {\bibinfo {author} {\bibfnamefont {A.}~\bibnamefont
			{Baratin}}\ and\ \bibinfo {author} {\bibfnamefont {D.}~\bibnamefont
			{Oriti}},\ }\href {\doibase 10.1103/PhysRevLett.105.221302} {\bibfield
		{journal} {\bibinfo  {journal} {Phys. Rev. Lett.}\ }\textbf {\bibinfo
			{volume} {105}},\ \bibinfo {pages} {221302} (\bibinfo {year} {2010})},\
	\Eprint {http://arxiv.org/abs/1002.4723} {arXiv:1002.4723 [hep-th]}
	\BibitemShut {NoStop}%
	\bibitem [{\citenamefont {Baratin}\ and\ \citenamefont
		{Oriti}(2011)}]{Baratin:2011tx}%
	\BibitemOpen
	\bibfield  {author} {\bibinfo {author} {\bibfnamefont {A.}~\bibnamefont
			{Baratin}}\ and\ \bibinfo {author} {\bibfnamefont {D.}~\bibnamefont
			{Oriti}},\ }\href {\doibase 10.1088/1367-2630/13/12/125011} {\bibfield
		{journal} {\bibinfo  {journal} {New J. Phys.}\ }\textbf {\bibinfo {volume}
			{13}},\ \bibinfo {pages} {125011} (\bibinfo {year} {2011})},\ \Eprint
	{http://arxiv.org/abs/1108.1178} {arXiv:1108.1178 [gr-qc]} \BibitemShut
	{NoStop}%
	\bibitem [{\citenamefont {Baratin}\ and\ \citenamefont
		{Oriti}(2012{\natexlab{b}})}]{Baratin:2011hp}%
	\BibitemOpen
	\bibfield  {author} {\bibinfo {author} {\bibfnamefont {A.}~\bibnamefont
			{Baratin}}\ and\ \bibinfo {author} {\bibfnamefont {D.}~\bibnamefont
			{Oriti}},\ }\href {\doibase 10.1103/PhysRevD.85.044003} {\bibfield  {journal}
		{\bibinfo  {journal} {Phys. Rev.}\ }\textbf {\bibinfo {volume} {D85}},\
		\bibinfo {pages} {044003} (\bibinfo {year} {2012}{\natexlab{b}})},\ \Eprint
	{http://arxiv.org/abs/1111.5842} {arXiv:1111.5842 [hep-th]} \BibitemShut
	{NoStop}%
	\bibitem [{\citenamefont {De~Pietri}(2001)}]{DePietri:2000ke}%
	\BibitemOpen
	\bibfield  {author} {\bibinfo {author} {\bibfnamefont {R.}~\bibnamefont
			{De~Pietri}},\ }\bibfield  {booktitle} {\emph {\bibinfo {booktitle} {{Lattice
					field theory. Proceedings, 18th International Symposium, Lattice 2000,
					Bangalore, India, August 17-22, 2000}}},\ }\href {\doibase
		10.1016/S0920-5632(01)00880-5} {\bibfield  {journal} {\bibinfo  {journal}
			{Nucl. Phys. Proc. Suppl.}\ }\textbf {\bibinfo {volume} {94}},\ \bibinfo
		{pages} {697} (\bibinfo {year} {2001})},\ \bibinfo {note} {[,697(2000)]},\
	\Eprint {http://arxiv.org/abs/hep-lat/0011033} {arXiv:hep-lat/0011033
		[hep-lat]} \BibitemShut {NoStop}%
	\bibitem [{\citenamefont {De~Pietri}\ and\ \citenamefont
		{Petronio}(2000)}]{DePietri:2000ii}%
	\BibitemOpen
	\bibfield  {author} {\bibinfo {author} {\bibfnamefont {R.}~\bibnamefont
			{De~Pietri}}\ and\ \bibinfo {author} {\bibfnamefont {C.}~\bibnamefont
			{Petronio}},\ }\href {\doibase 10.1063/1.1290053} {\bibfield  {journal}
		{\bibinfo  {journal} {J. Math. Phys.}\ }\textbf {\bibinfo {volume} {41}},\
		\bibinfo {pages} {6671} (\bibinfo {year} {2000})},\ \Eprint
	{http://arxiv.org/abs/gr-qc/0004045} {arXiv:gr-qc/0004045 [gr-qc]}
	\BibitemShut {NoStop}%
	\bibitem [{\citenamefont {De~Pietri}\ \emph {et~al.}(2000)\citenamefont
		{De~Pietri}, \citenamefont {Freidel}, \citenamefont {Krasnov},\ and\
		\citenamefont {Rovelli}}]{DePietri:1999bx}%
	\BibitemOpen
	\bibfield  {author} {\bibinfo {author} {\bibfnamefont {R.}~\bibnamefont
			{De~Pietri}}, \bibinfo {author} {\bibfnamefont {L.}~\bibnamefont {Freidel}},
		\bibinfo {author} {\bibfnamefont {K.}~\bibnamefont {Krasnov}}, \ and\
		\bibinfo {author} {\bibfnamefont {C.}~\bibnamefont {Rovelli}},\ }\href
	{\doibase 10.1016/S0550-3213(00)00005-5} {\bibfield  {journal} {\bibinfo
			{journal} {Nucl. Phys.}\ }\textbf {\bibinfo {volume} {B574}},\ \bibinfo
		{pages} {785} (\bibinfo {year} {2000})},\ \Eprint
	{http://arxiv.org/abs/hep-th/9907154} {arXiv:hep-th/9907154} \BibitemShut
	{NoStop}%
	\bibitem [{\citenamefont {Oriti}\ \emph {et~al.}(2015)\citenamefont {Oriti},
		\citenamefont {Pranzetti}, \citenamefont {Ryan},\ and\ \citenamefont
		{Sindoni}}]{Oriti:2015qva}%
	\BibitemOpen
	\bibfield  {author} {\bibinfo {author} {\bibfnamefont {D.}~\bibnamefont
			{Oriti}}, \bibinfo {author} {\bibfnamefont {D.}~\bibnamefont {Pranzetti}},
		\bibinfo {author} {\bibfnamefont {J.~P.}\ \bibnamefont {Ryan}}, \ and\
		\bibinfo {author} {\bibfnamefont {L.}~\bibnamefont {Sindoni}},\ }\href
	{\doibase 10.1088/0264-9381/32/23/235016} {\bibfield  {journal} {\bibinfo
			{journal} {Class. Quant. Grav.}\ }\textbf {\bibinfo {volume} {32}},\ \bibinfo
		{pages} {235016} (\bibinfo {year} {2015})},\ \Eprint
	{http://arxiv.org/abs/1501.00936} {arXiv:1501.00936 [gr-qc]} \BibitemShut
	{NoStop}%
	\bibitem [{\citenamefont {Oriti}(2016{\natexlab{a}})}]{Oriti:2016acw}%
	\BibitemOpen
	\bibfield  {author} {\bibinfo {author} {\bibfnamefont {D.}~\bibnamefont
			{Oriti}},\ }\href@noop {} {\  (\bibinfo {year} {2016}{\natexlab{a}})},\
	\Eprint {http://arxiv.org/abs/1612.09521} {arXiv:1612.09521 [gr-qc]}
	\BibitemShut {NoStop}%
	\bibitem [{\citenamefont {Oriti}(2016{\natexlab{b}})}]{Oriti:2013aqa}%
	\BibitemOpen
	\bibfield  {author} {\bibinfo {author} {\bibfnamefont {D.}~\bibnamefont
			{Oriti}},\ }\href {\doibase 10.1088/0264-9381/33/8/085005} {\bibfield
		{journal} {\bibinfo  {journal} {Class. Quant. Grav.}\ }\textbf {\bibinfo
			{volume} {33}},\ \bibinfo {pages} {085005} (\bibinfo {year}
		{2016}{\natexlab{b}})},\ \Eprint {http://arxiv.org/abs/1310.7786}
	{arXiv:1310.7786 [gr-qc]} \BibitemShut {NoStop}%
	\bibitem [{\citenamefont {Singh}\ \emph {et~al.}(2011)\citenamefont {Singh},
		\citenamefont {Pfeifer},\ and\ \citenamefont {Vidal}}]{singh2011tensor}%
	\BibitemOpen
	\bibfield  {author} {\bibinfo {author} {\bibfnamefont {S.}~\bibnamefont
			{Singh}}, \bibinfo {author} {\bibfnamefont {R.~N.}\ \bibnamefont {Pfeifer}},
		\ and\ \bibinfo {author} {\bibfnamefont {G.}~\bibnamefont {Vidal}},\
	}\href@noop {} {\bibfield  {journal} {\bibinfo  {journal} {Physical Review
				B}\ }\textbf {\bibinfo {volume} {83}},\ \bibinfo {pages} {115125} (\bibinfo
	{year} {2011})}\BibitemShut {NoStop}%
	
	\bibitem{swi} B. Swingle, Phys. Rev. D 86, 065007 (2012), arXiv:0905.1317 [cond-mat.str-el].
	
	\bibitem [{\citenamefont {Oriti}(2014{\natexlab{b}})}]{Oriti:2013jga}%
	\BibitemOpen
	\bibfield  {author} {\bibinfo {author} {\bibfnamefont {D.}~\bibnamefont
			{Oriti}},\ }\href {\doibase 10.1016/j.shpsb.2013.10.006} {\bibfield
		{journal} {\bibinfo  {journal} {Stud. Hist. Phil. Sci.}\ }\textbf {\bibinfo
			{volume} {B46}},\ \bibinfo {pages} {186} (\bibinfo {year}
		{2014}{\natexlab{b}})},\ \Eprint {http://arxiv.org/abs/1302.2849}
	{arXiv:1302.2849 [physics.hist-ph]} \BibitemShut {NoStop}%
	\bibitem [{\citenamefont {Cao}\ \emph {et~al.}(2016)\citenamefont {Cao},
		\citenamefont {Carroll},\ and\ \citenamefont {Michalakis}}]{Cao:2016mst}%
	\BibitemOpen
	\bibfield  {author} {\bibinfo {author} {\bibfnamefont {C.}~\bibnamefont
			{Cao}}, \bibinfo {author} {\bibfnamefont {S.~M.}\ \bibnamefont {Carroll}}, \
		and\ \bibinfo {author} {\bibfnamefont {S.}~\bibnamefont {Michalakis}},\
	}\href@noop {} {\  (\bibinfo {year} {2016})},\ \Eprint
	{http://arxiv.org/abs/1606.08444} {arXiv:1606.08444 [hep-th]} \BibitemShut
	{NoStop}%
	\bibitem [{\citenamefont {Seiberg}(2006)}]{Seiberg:2006wf}%
	\BibitemOpen
	\bibfield  {author} {\bibinfo {author} {\bibfnamefont {N.}~\bibnamefont
			{Seiberg}},\ }in\ \href@noop {} {\emph {\bibinfo {booktitle} {{The Quantum
					Structure of Space and Time: Proceedings of the 23rd Solvay Conference on
					Physics. Brussels, Belgium. 1 - 3 December 2005}}}}\ (\bibinfo {year}
	{2006})\ pp.\ \bibinfo {pages} {163--178},\ \Eprint
	{http://arxiv.org/abs/hep-th/0601234} {arXiv:hep-th/0601234 [hep-th]}
	\BibitemShut {NoStop}%
	\bibitem [{\citenamefont {Koslowski}(2007)}]{Koslowski:2007kh}%
	\BibitemOpen
	\bibfield  {author} {\bibinfo {author} {\bibfnamefont {T.~A.}\ \bibnamefont
			{Koslowski}},\ }\href@noop {} {\  (\bibinfo {year} {2007})},\ \Eprint
	{http://arxiv.org/abs/0709.3465} {arXiv:0709.3465 [gr-qc]} \BibitemShut
	{NoStop}%
	\bibitem [{\citenamefont {Oriti}(2007)}]{Oriti:2007qd}%
	\BibitemOpen
	\bibfield  {author} {\bibinfo {author} {\bibfnamefont {D.}~\bibnamefont
			{Oriti}},\ }\href@noop {} {\bibfield  {journal} {\bibinfo  {journal} {PoS}\
		}\textbf {\bibinfo {volume} {QG-PH}},\ \bibinfo {pages} {030} (\bibinfo
		{year} {2007})},\ \Eprint {http://arxiv.org/abs/0710.3276} {arXiv:0710.3276
		[gr-qc]} \BibitemShut {NoStop}%
	\bibitem [{\citenamefont {Konopka}\ \emph {et~al.}(2006)\citenamefont
		{Konopka}, \citenamefont {Markopoulou},\ and\ \citenamefont
		{Smolin}}]{Konopka:2006hu}%
	\BibitemOpen
	\bibfield  {author} {\bibinfo {author} {\bibfnamefont {T.}~\bibnamefont
			{Konopka}}, \bibinfo {author} {\bibfnamefont {F.}~\bibnamefont
			{Markopoulou}}, \ and\ \bibinfo {author} {\bibfnamefont {L.}~\bibnamefont
			{Smolin}},\ }\href {http://arXiv.org/abs/hep-th/0611197} {\enquote {\bibinfo
			{title} {{Quantum Graphity}},}\ } (\bibinfo {year} {2006}),\ \Eprint
	{http://arxiv.org/abs/hep-th/0611197} {arXiv:hep-th/0611197 [hep-th]}
	\BibitemShut {NoStop}%
	\bibitem [{\citenamefont {Rivasseau}(2011)}]{Rivasseau:2011hm}%
	\BibitemOpen
	\bibfield  {author} {\bibinfo {author} {\bibfnamefont {V.}~\bibnamefont
			{Rivasseau}},\ }\bibfield  {booktitle} {\emph {\bibinfo {booktitle}
			{{Proceedings, 8th International Conference on Progress in theoretical
					physics (ICPTP 2011): Constantine, Algeria, October 23-25, 2011}}},\ }\href
	{\doibase 10.1063/1.4715396} {\bibfield  {journal} {\bibinfo  {journal} {AIP
				Conf. Proc.}\ }\textbf {\bibinfo {volume} {1444}},\ \bibinfo {pages} {18}
		(\bibinfo {year} {2011})},\ \Eprint {http://arxiv.org/abs/1112.5104}
	{arXiv:1112.5104 [hep-th]} \BibitemShut {NoStop}%
	\bibitem [{\citenamefont {Steinacker}(2010)}]{Steinacker:2010rh}%
	\BibitemOpen
	\bibfield  {author} {\bibinfo {author} {\bibfnamefont {H.}~\bibnamefont
			{Steinacker}},\ }\href {\doibase 10.1088/0264-9381/27/13/133001} {\bibfield
		{journal} {\bibinfo  {journal} {Class. Quant. Grav.}\ }\textbf {\bibinfo
			{volume} {27}},\ \bibinfo {pages} {133001} (\bibinfo {year} {2010})},\
	\Eprint {http://arxiv.org/abs/1003.4134} {arXiv:1003.4134 [hep-th]}
	\BibitemShut {NoStop}%
	\bibitem{hay2} P. Hayden, D. Leung, and A. Winter,
		Commun. Math. Phys. (2006) 265: 95. https://doi.org10.1007s00220-006-1535-6
		
	\bibitem [{\citenamefont {Chirco}\ \emph {et~al.}({\natexlab{b}})\citenamefont
		{Chirco}, \citenamefont {Oriti},\ and\ \citenamefont {Zhang}}]{ChircoRT}%
	\BibitemOpen
	\bibinfo {author} {\bibfnamefont {G.}~\bibnamefont {Chirco}}, \bibinfo
	{author} {\bibfnamefont {D.}~\bibnamefont {Oriti}}, \ and\ \bibinfo {author}
	{\bibfnamefont {M.}~\bibnamefont {Zhang}}\BibitemShut {NoStop}%
	\bibitem [{\citenamefont {Freidel}\ \emph {et~al.}(2009)\citenamefont
		{Freidel}, \citenamefont {Gurau},\ and\ \citenamefont
		{Oriti}}]{Freidel:2009hd}%
	\BibitemOpen
	\bibfield  {author} {  }\bibfield  {author} {\bibinfo {author} {\bibfnamefont
			{L.}~\bibnamefont {Freidel}}, \bibinfo {author} {\bibfnamefont
			{R.}~\bibnamefont {Gurau}}, \ and\ \bibinfo {author} {\bibfnamefont
			{D.}~\bibnamefont {Oriti}},\ }\href {\doibase 10.1103/PhysRevD.80.044007}
	{\bibfield  {journal} {\bibinfo  {journal} {Phys. Rev.}\ }\textbf {\bibinfo
			{volume} {D80}},\ \bibinfo {pages} {044007} (\bibinfo {year} {2009})},\
	\Eprint {http://arxiv.org/abs/0905.3772} {arXiv:0905.3772 [hep-th]}
	\BibitemShut {NoStop}%
	\bibitem [{\citenamefont {Magnen}\ \emph {et~al.}(2009)\citenamefont {Magnen},
		\citenamefont {Noui}, \citenamefont {Rivasseau},\ and\ \citenamefont
		{Smerlak}}]{Magnen:2009at}%
	\BibitemOpen
	\bibfield  {author} {\bibinfo {author} {\bibfnamefont {J.}~\bibnamefont
			{Magnen}}, \bibinfo {author} {\bibfnamefont {K.}~\bibnamefont {Noui}},
		\bibinfo {author} {\bibfnamefont {V.}~\bibnamefont {Rivasseau}}, \ and\
		\bibinfo {author} {\bibfnamefont {M.}~\bibnamefont {Smerlak}},\ }\href
	{\doibase 10.1088/0264-9381/26/18/185012} {\bibfield  {journal} {\bibinfo
			{journal} {Class. Quant. Grav.}\ }\textbf {\bibinfo {volume} {26}},\ \bibinfo
		{pages} {185012} (\bibinfo {year} {2009})},\ \Eprint
	{http://arxiv.org/abs/0906.5477} {arXiv:0906.5477 [hep-th]} \BibitemShut
	{NoStop}%
	\bibitem [{\citenamefont {Barrett}\ and\ \citenamefont
		{Naish-Guzman}(2006)}]{Barrett:2006ru}%
	\BibitemOpen
	\bibfield  {author} {\bibinfo {author} {\bibfnamefont {J.~W.}\ \bibnamefont
			{Barrett}}\ and\ \bibinfo {author} {\bibfnamefont {I.}~\bibnamefont
			{Naish-Guzman}},\ }in\ \href@noop {} {\emph {\bibinfo {booktitle} {{Recent
					developments in theoretical and experimental general relativity, gravitation
					and relativistic field theories. Proceedings, 11th Marcel Grossmann Meeting,
					MG11, Berlin, Germany, July 23-29, 2006. Pt. A-C}}}}\ (\bibinfo {year}
	{2006})\ pp.\ \bibinfo {pages} {2782--2784},\ \Eprint
	{http://arxiv.org/abs/gr-qc/0612170} {arXiv:gr-qc/0612170 [gr-qc]}
	\BibitemShut {NoStop}%
	\bibitem [{\citenamefont {Bonzom}\ and\ \citenamefont
		{Smerlak}(2010)}]{Bonzom:2010ar}%
	\BibitemOpen
	\bibfield  {author} {\bibinfo {author} {\bibfnamefont {V.}~\bibnamefont
			{Bonzom}}\ and\ \bibinfo {author} {\bibfnamefont {M.}~\bibnamefont
			{Smerlak}},\ }\href {\doibase 10.1007/s11005-010-0414-4} {\bibfield
		{journal} {\bibinfo  {journal} {Lett. Math. Phys.}\ }\textbf {\bibinfo
			{volume} {93}},\ \bibinfo {pages} {295} (\bibinfo {year} {2010})},\ \Eprint
	{http://arxiv.org/abs/1004.5196} {arXiv:1004.5196 [gr-qc]} \BibitemShut
	{NoStop}%
	\bibitem [{\citenamefont {Bonzom}\ and\ \citenamefont
		{Smerlak}(2012{\natexlab{a}})}]{Bonzom:2010zh}%
	\BibitemOpen
	\bibfield  {author} {\bibinfo {author} {\bibfnamefont {V.}~\bibnamefont
			{Bonzom}}\ and\ \bibinfo {author} {\bibfnamefont {M.}~\bibnamefont
			{Smerlak}},\ }\href {\doibase 10.1007/s00220-012-1477-0} {\bibfield
		{journal} {\bibinfo  {journal} {Commun. Math. Phys.}\ }\textbf {\bibinfo
			{volume} {312}},\ \bibinfo {pages} {399} (\bibinfo {year}
		{2012}{\natexlab{a}})},\ \Eprint {http://arxiv.org/abs/1008.1476}
	{arXiv:1008.1476 [math-ph]} \BibitemShut {NoStop}%
	\bibitem [{\citenamefont {Bonzom}\ and\ \citenamefont
		{Smerlak}(2012{\natexlab{b}})}]{Bonzom:2011br}%
	\BibitemOpen
	\bibfield  {author} {\bibinfo {author} {\bibfnamefont {V.}~\bibnamefont
			{Bonzom}}\ and\ \bibinfo {author} {\bibfnamefont {M.}~\bibnamefont
			{Smerlak}},\ }\href {\doibase 10.1007/s00023-011-0127-y} {\bibfield
		{journal} {\bibinfo  {journal} {Annales Henri Poincare}\ }\textbf {\bibinfo
			{volume} {13}},\ \bibinfo {pages} {185} (\bibinfo {year}
		{2012}{\natexlab{b}})},\ \Eprint {http://arxiv.org/abs/1103.3961}
	{arXiv:1103.3961 [gr-qc]} \BibitemShut {NoStop}%
	\bibitem [{\citenamefont {Han}(2013)}]{Han:2013ina}%
	\BibitemOpen
	\bibfield  {author} {\bibinfo {author} {\bibfnamefont {M.}~\bibnamefont
			{Han}},\ }\href {\doibase 10.1103/PhysRevD.88.044051} {\bibfield  {journal}
		{\bibinfo  {journal} {Phys. Rev.}\ }\textbf {\bibinfo {volume} {D88}},\
		\bibinfo {pages} {044051} (\bibinfo {year} {2013})},\ \Eprint
	{http://arxiv.org/abs/1304.5628} {arXiv:1304.5628 [gr-qc]} \BibitemShut
	{NoStop}%
	\bibitem [{\citenamefont {Han}(2014{\natexlab{b}})}]{Han:2013hna}%
	\BibitemOpen
	\bibfield  {author} {\bibinfo {author} {\bibfnamefont {M.}~\bibnamefont
			{Han}},\ }\href {\doibase 10.1088/0264-9381/31/1/015004} {\bibfield
		{journal} {\bibinfo  {journal} {Class.Quant.Grav.}\ }\textbf {\bibinfo
			{volume} {31}},\ \bibinfo {pages} {015004} (\bibinfo {year}
		{2014}{\natexlab{b}})},\ \Eprint {http://arxiv.org/abs/1304.5627}
	{arXiv:1304.5627 [gr-qc]} \BibitemShut {NoStop}%
	\bibitem [{\citenamefont {Han}(2014{\natexlab{c}})}]{Han:2013tap}%
	\BibitemOpen
	\bibfield  {author} {\bibinfo {author} {\bibfnamefont {M.}~\bibnamefont
			{Han}},\ }\href {\doibase 10.1103/PhysRevD.89.124001} {\bibfield  {journal}
		{\bibinfo  {journal} {Phys.Rev.}\ }\textbf {\bibinfo {volume} {D89}},\
		\bibinfo {pages} {124001} (\bibinfo {year} {2014}{\natexlab{c}})},\ \Eprint
	{http://arxiv.org/abs/1308.4063} {arXiv:1308.4063 [gr-qc]} \BibitemShut
	{NoStop}%
	\bibitem [{\citenamefont {Gurau}(2014)}]{Gurau:2011kk}%
	\BibitemOpen
	\bibfield  {author} {\bibinfo {author} {\bibfnamefont {R.}~\bibnamefont
			{Gurau}},\ }\href {\doibase 10.1214/13-AIHP567} {\bibfield  {journal}
		{\bibinfo  {journal} {Ann. Inst. H. Poincare Probab. Statist.}\ }\textbf
		{\bibinfo {volume} {50}},\ \bibinfo {pages} {1474} (\bibinfo {year}
		{2014})},\ \Eprint {http://arxiv.org/abs/1111.0519} {arXiv:1111.0519
		[math.PR]} \BibitemShut {NoStop}%
\end{thebibliography}

%

\end{document}